\documentclass[preprint]{ptephy_v1}

\preprintnumber{XXXX-XXXX} 

\usepackage{amsmath,amssymb}
\usepackage{graphicx}
\usepackage{subfig} 

\usepackage{mathrsfs}
\usepackage{url}
\usepackage{braket}
\usepackage{bm}
\usepackage{color}
\usepackage{lineno}
\newcommand*\patchAmsMathEnvironmentForLineno[1]{
  \expandafter\let\csname old#1\expandafter\endcsname\csname #1\endcsname
  \expandafter\let\csname oldend#1\expandafter\endcsname\csname end#1\endcsname
  \renewenvironment{#1}
     {\linenomath\csname old#1\endcsname}
     {\csname oldend#1\endcsname\endlinenomath}}
\newcommand*\patchBothAmsMathEnvironmentsForLineno[1]{
  \patchAmsMathEnvironmentForLineno{#1}
  \patchAmsMathEnvironmentForLineno{#1*}}
\AtBeginDocument{
\patchBothAmsMathEnvironmentsForLineno{equation}
\patchBothAmsMathEnvironmentsForLineno{align}
\patchBothAmsMathEnvironmentsForLineno{flalign}
\patchBothAmsMathEnvironmentsForLineno{alignat}
\patchBothAmsMathEnvironmentsForLineno{gather}
\patchBothAmsMathEnvironmentsForLineno{multline}
}

\newcommand{\red}[1]{#1}
\newcommand{\commentout}[1]{}

\newcommand{\Fig}[1]{Fig.~\ref{#1}}
\newcommand{\Figure}[1]{Figure~\ref{#1}}

\begin{document}

\title{Measurement of differential cross sections for $\Sigma^{+}p$ elastic scattering in the momentum range \mbox{0.44 -- 0.80 GeV/$c$}} 


\author[1,2]{T.~Nanamura}
\author[3,5]{K.~Miwa}
\author[4]{J.~K.~Ahn}
\author[5]{Y.~Akazawa}
\author[3]{T.~Aramaki}
\author[1]{S.~Ashikaga}
\author[6]{S.~Callier}
\author[3]{N.~Chiga}
\author[4]{S.~W.~Choi}
\author[7]{H.~Ekawa}
\author[8]{P.~Evtoukhovitch}
\author[3]{N.~Fujioka}
\author[2]{M.~Fujita}
\author[1]{T.~Gogami}
\author[1,2]{T.~K.~Harada}
\author[2]{S.~Hasegawa}
\author[3]{S.~H.~Hayakawa}
\author[5]{R.~Honda}
\author[9]{S.~Hoshino}
\author[2]{K.~Hosomi}
\author[1,2]{M.~Ichikawa}
\author[2]{Y.~Ichikawa}
\author[5]{M.~Ieiri}
\author[3]{M.~Ikeda}
\author[2]{K.~Imai}
\author[3]{Y.~Ishikawa}
\author[5]{S.~Ishimoto}
\author[4]{W.~S.~Jung}
\author[3]{S.~Kajikawa}
\author[3]{H.~Kanauchi}
\author[10]{H.~Kanda}
\author[3]{T.~Kitaoka}
\author[4]{B.~M.~Kang}
\author[11]{H.~Kawai}
\author[4]{S.~H.~Kim}
\author[9]{K.~Kobayashi}
\author[3]{T.~Koike}
\author[3]{K.~Matsuda}
\author[3]{Y.~Matsumoto}
\author[3]{S.~Nagao}
\author[9]{R.~Nagatomi}
\author[9]{Y.~Nakada}
\author[7]{M.~Nakagawa}
\author[5]{I.~Nakamura}
\author[1,2]{M.~Naruki}
\author[3]{S.~Ozawa}
\author[6]{L.~Raux}
\author[3]{T.~G.~Rogers}
\author[9]{A.~Sakaguchi}
\author[3]{T.~Sakao}
\author[2]{H.~Sako}
\author[2]{S.~Sato}
\author[3]{T.~Shiozaki}
\author[10]{K.~Shirotori}
\author[1]{K.~N.~Suzuki}
\author[5]{S.~Suzuki}
\author[11]{M.~Tabata}
\author[6]{C.~d.~L.~Taille}
\author[5]{H.~Takahashi}
\author[5]{T.~Takahashi}
\author[12]{T.~N.~Takahashi}
\author[2,3]{H.~Tamura}
\author[5]{M.~Tanaka}
\author[2]{K.~Tanida}
\author[8,13]{Z.~Tsamalaidze}
\author[3,5]{M.~Ukai}
\author[3]{H.~Umetsu}
\author[3]{S.~Wada}
\author[2]{T.~O.~Yamamoto}
\author[3]{J.~Yoshida}
\author[14]{K.~Yoshimura}

\affil[1]{Department of Physics, Kyoto University, Kyoto 606-8502, Japan}
\affil[2]{Advanced Science Research Center (ASRC), Japan Atomic Energy Agency (JAEA), Tokai, Ibaraki 319-1195, Japan}
\affil[3]{Department of Physics, Tohoku University, Sendai 980-8578, Japan}
\affil[4]{Department of Physics, Korea University, Seoul 02841, Korea}
\affil[5]{Institute of Particle and Nuclear Studies (IPNS), High Energy Accelerator Research Organization (KEK), Tsukuba 305-0801, Japan}
\affil[6]{OMEGA Ecole Polytechnique-CNRS/IN2P3, 3 rue Michel-Ange, 75794 Paris 16, France}
\affil[7]{High Energy Nuclear Physics Laboratory, RIKEN, Wako, 351-0198, Japan}
\affil[8]{Joint Institute for Nuclear Research (JINR), Dubna, Moscow Region 141980, Russia}
\affil[9]{Department of Physics, Osaka University, Toyonaka 560-0043, Japan}
\affil[10]{Research Center for Nuclear Physics (RCNP), Osaka University, Ibaraki 567-0047, Japan}
\affil[11]{Department of Physics, Chiba University, Chiba 263-8522, Japan}
\affil[12]{Nishina Center for Accelerator-based Science, RIKEN, Wako, 351-0198, Japan}
\affil[13]{Georgian Technical University (GTU), Tbilisi, Georgia}
\affil[14]{Department of Physics, Okayama University, Okayama 700-8530, Japan \email{nanamura@scphys.kyoto-u.ac.jp}}



\begin{abstract}%
We performed a novel $\Sigma^+ p$ scattering experiment at the J-PARC Hadron Experimental Facility.
Approximately 2400 $\Sigma^+ p$ elastic scattering events were identified from $4.9 \times 10^7$ tagged $\Sigma^+$ particles in the $\Sigma^+$ momentum range 0.44 -- 0.80 GeV/$c$. The differential cross sections of the $\Sigma^+ p$ elastic scattering were derived with much better precision than in previous experiments. The obtained differential cross sections were approximately 2 mb/sr or less, which were not as large as those predicted by the fss2 and FSS models based on the quark cluster model in the short-range region. By performing  phase-shift analyses for the obtained differential cross sections, we experimentally derived the phase shifts of the $^3 S_1$ and $^1 P_1$ channels for the first time. The phase shift of the $^3 S_1$ channel, where a large repulsive core was predicted owing to the Pauli effect between quarks, was evaluated \red{as} $20^\circ<|\delta_{^3S_1}|<35^\circ$. If the sign of $\delta_{^3S_1}$ is assumed to be negative, the interaction in this channel is moderately repulsive, as the Nijmegen extended-sort-core models predicted.
\end{abstract}

\subjectindex{xxxx, xxx}

\maketitle

\section{Introduction}
Understanding the origin of the short-range repulsion in the nuclear force is a central challenge in nuclear physics.
In pioneering nuclear force research, the short-range repulsion was treated phenomenologically \cite{Hamada:1962} or was attributed to the $\omega$ exchange in a boson-exchange picture \cite{Machleidt:1987}.
The effect of antisymmetrization among quarks on short-range interactions was first considered using the quark cluster model (QCM) by Oka and Yazaki \cite{Oka:1981_1, Oka:1981_2}.
They reported that the short-range repulsion could be explained by the Pauli principle between quarks and the color-magnetic interactions.
In the nuclear force, a detailed study of the Pauli principle at the quark level is impossible because quark Pauli-repulsive spin-isospin configurations 
are excluded as a result of the Pauli principle at the baryon level. 
However, by extending the nucleon-nucleon ($NN$) interaction to the baryon-baryon ($BB$) interaction between octet baryons, it is possible to investigate the distinct quark-Pauli forbidden states, which are labeled as 8s-plet and 10-plet in the flavor SU(3) symmetry representations for completely and partially Pauli-forbidden states, respectively \cite{Oka:1986}.
Table \ref{tab1} shows the relationship between the isospin and flavor SU(3) bases for the $BB$ interaction for strangeness $0$ and $-1$ sectors.
In particular, the $\Sigma N$ ($I=3/2$) channel is one of the best channels for studying the repulsive nature in the 10-plet because the $\Sigma N$ ($I=3/2$)  channel is simply represented by the 27-plet and 10-plet.
Here, the channels represented by 27-plet, such as the $^1S_0$ channel in the $\Sigma^{+} p$ system are less uncertain because 27-plet is well-estimated from the $NN$ ($I=1$) interaction based on the flavor SU(3) symmetry.
Therefore, the nature of the 10-plet can be extracted from information on the $\Sigma N$ ($I=3/2$) system.
\begin{table}[!h]
\begin{center}
\caption{Relationship between the isospin and flavor SU(3) bases for the $BB$ interaction channels. Spin states $s$ and a parity of the orbital angular momentum $L$ are denoted as $^{2s+1}(-1)^L$, such as singlet-even ($^1\text{even}$) and triplet-odd ($^3 \text{odd}$).}
\label{tab1} 
\begin{tabular}{cccc}
\hline\noalign{\smallskip}
strangeness & $BB$ channel ($I$) & $^1$even or $^3$odd & $^3$even or $^1$odd \\
\noalign{\smallskip}\hline\noalign{\smallskip}
0 & $NN(I=0)$ & -- & ($\bm{10^*}$) \\
 & $NN(I=1)$ & ($\bm{27}$) & --  \\
\noalign{\smallskip}\hline\noalign{\smallskip}
  & $\Lambda N(I=\frac{1}{2})$ & $\frac{1}{\sqrt{10}}[ (\bm{8_s})+3(\bm{27})] $& $\frac{1}{\sqrt{2}}[ -(\bm{8_a})+(\bm{10^*})]$\\
$-1$  & $\Sigma N(I=\frac{1}{2})$ & $\frac{1}{\sqrt{10}}[ 3(\bm{8_s})-(\bm{27})]$ & $\frac{1}{\sqrt{2}}[ (\bm{8_a})+(\bm{10^*})]$\\
 & $\Sigma N(I=\frac{3}{2})$ & $(\bm{27})$ & ($\bm{10}$)\\
\noalign{\smallskip}\hline
\end{tabular}
\end{center}
\end{table}

Theoretical treatments of the short-range $BB$ interactions have led to relatively different results for the $\Sigma N$ ($I=3/2$, $^3S_1$) interaction.
The fss2 model, which includes the QCM in the short-range region and empirical meson-exchange potential in the middle- and long-range regions, naturally predicts a repulsive interaction in the $\Sigma N$ ($I=3/2$, $^3S_1$) channel \cite{Fujiwara:2007}.
However, the meson-exchange models such as Nijmegen soft core (NSC) model \cite{Maessen:1989, Rijken:1999} and J\"{u}lich hyperon-nucleon ($YN$) model A \cite{Holzenkamp:1989}, which represent \red{short-range repulsion} from a heavy vector meson exchange, cannot predict such a large repulsive force.
The NSC97 model, whose $\Lambda N$ interaction is extensively used in $\Lambda$ hypernuclear studies, predicts an attractive interaction for the $\Sigma N$ ($I=3/2$, $^3S_1$) channel \cite{Rijken:1999}.
Experimental information on the $\Sigma N$ interaction in the nuclear core region was limited owing to the lack of  observations of $\Sigma$ hypernuclei, except for $^{4}_{\Sigma}$He \cite{Nagae:1998}.
However, the quasi-free $\Sigma^{-}$ production spectra in medium nuclei obtained at KEK-PS revealed that the spin-isospin-averaged $\Sigma$ potential had \red{a strong repulsion and a sizable absorption} \cite{Noumi:2002, Saha:2004}. This was confirmed even for the $\Sigma^{-}$ + $^{5}$He system \cite{Harada:2018, Honda:2017}.
Based on these experimental results,  the $\Sigma N$ ($I=3/2$, $^3S_1$) channel is believed to be repulsive.
In the Nijmegen extended-soft-core (ESC) model, an additional short-range interaction owing to Pomeron exchange is included, to explain the repulsive nature of the $\Sigma N$ ($I=3/2$, $^3S_1$) channel \cite{Rijken:2010, Nagels:2019}.
The potentials of the $BB$ $S$-wave interaction in the flavor-irreducible representation were calculated from a lattice QCD simulation in the flavor SU(3) limit, and the potential shapes agreed with QCM predictions \cite{Inoue:2019}.
Furthermore, the $\Sigma N$ ($I=3/2$, $^3S_1$) potential derived from the lattice QCD simulation in almost physical quark masses demonstrated a repulsive core in the short-range region, without any attractive pocket in the middle-range region \cite{Nemura:2018}.
Recent chiral effective field theory ($\chi$EFT) calculations, extended to the $YN$ sectors, also predicted repulsive interactions for this channel \cite{Haidenbauer:2013, Haidenbauer:2020}.
In $\chi$EFT calculations, short-range interactions were included as contact interactions represented by low- energy constants (LECs).

Currently, all theoretical calculations predict the repulsive interactions in the $\Sigma N$ ($I=3/2$) channel.
However, the predicted strength of the repulsive interaction, that is, the phase shift of the $^{3}S_{1}$ channel, are  different from each other; therefore, this strength should be experimentally determined.
The theoretical predictions of differential cross sections for $\Sigma^+ p$ scattering in the intermediate momentum region (above 400 MeV/$c$) vary depending on the size of the repulsion of the $^{3}S_{1}$ channel \cite{Rijken:2010, Nagels:2019, Fujiwara:2007, Haidenbauer:2013, Haidenbauer:2020}.
Therefore, an accurate determination of the differential cross sections of $\Sigma^{+} p$ plays a crucial role in determining the strength of the repulsive interaction.
Moreover, owing to the simple representation of the multiplet, one can experimentally derive the phase shift of the 10-plet from a numerical phase shift analysis of the differential cross sections with some assumptions, as explained in section \ref{discussionsect}.
Until now, none of the phase-shift values of the $YN$ and $YY$ channels have been determined experimentally.
This is in contrast to the $NN$ interaction, in which the phase shifts have been precisely determined from the scattering observables of $pp$ and $np$ scatterings \cite{Arndt:2000}.

The $BB$ interactions provide essential information for predicting the onset of hyperons in neutron stars \cite{Vidana:2013}.
Recently, the particle composition in the high-density region of the inner core of neutron stars has been extensively discussed to consider the mechanism that supports massive neutron stars with two solar masses.
Because the onset of $\Sigma^{-}$ permits the appearance of protons owing to charge neutrality, the $\Sigma^{-}$'s impact on the particle composition is significant.
Although the $\Sigma$ potential in symmetric nuclear matter is estimated from $\Sigma^{-}$ quasi-free production data in medium-heavy nuclei, the $\Sigma^{-}$ potential in neutron matter provides important information regarding neutron stars.
The $\Sigma^{-}n$ interaction, which is equal to the $\Sigma^{+}p$ interaction due to isospin symmetry, is an important input for obtaining the $\Sigma^{-}$ potential in neutron matter through G-matrix calculations \cite{Yamamoto:2014}.
Therefore, investigation of the $\Sigma^{+}p$ interaction is also essential to determine the nature of $\Sigma^{-}$ in neutron stars.

Experimental data of the $YN$ scatterings are, historically, rare owing to the experimental difficulties regarding the short lifetime of hyperons \cite{Sechi-zorn:1968, Alexander:1968, Kadyk:1971, Hauptman:1977, Engelmann:1966, Eisele:1971, Stephen:1970, Kondo:2000, Kanda:2005}.
For the intermediate energy, two experiments to measure the differential cross sections of the $\Sigma^{+}p$ channel were conducted \cite{Goto:1999, Kanda:2005}. 
However,  no conclusions could be drawn on the repulsive nature owing to the insufficient \red{precision} of the studies stemming from low statistics.
Although there is one spin-observable measurement on the $\Sigma^{+}p$ channel \cite{Kurosawa:2006},  information on $\Sigma^{+} p$ scattering is relatively limited.
However, we (J-PARC E40 collaboration) have recently succeeded in the systematic measurements of $\Sigma^{\pm}p$ scatterings with high statistics at the J-PARC Hadron Experimental Facility.
$\Sigma^{-}p$  results can be found in \cite{Miwa:2021, Miwa:2021_2}.
In this article, we report the results of the differential cross section measurements of $\Sigma^{+}p$ elastic scattering in the $\Sigma^{+}$ momentum range from 0.44 to 0.80 GeV/$c$.

\section{Experiment}
We performed the  $\Sigma^+ p$ scattering experiment, J-PARC E40, at the K1.8 beam line \cite{Agari:2012} in the J-PARC Hadron Experimental Facility. Secondary \red{particles} were produced by exposing a primary Au target to a \mbox{30-GeV} primary proton beam from the J-PARC main ring, and were delivered to the experimental area. \red{Undesired} beam particles were deflected from the central beam orbit by the electric field of the two stages of the electric separators, with the aid of correction magnets. A mass-separated $\pi^+$ beam was used within the experiment. The spill cycle and beam duration during accelerator operation were 5.2 and 2~s, respectively. 

\begin{figure}[!h]
\begin{center}
\includegraphics[width=5.0in]{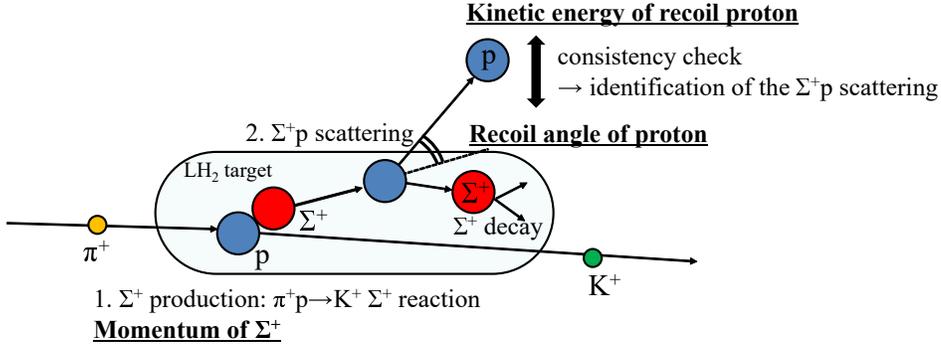}
\end{center}
\caption{Conceptual drawing of the $\Sigma^+ p$ scattering caused by the initial $\Sigma^+$ traveling in the $\text{LH}_2$ target. The initial $\Sigma^+$ is produced by the $\pi^+ p \to K^+ \Sigma^+$ reaction. The $\Sigma^+ p$ scattering can be kinematically identified by measuring the initial $\Sigma^+$'s momentum, the recoil proton's kinetic energy, and recoil angle.}
\label{fig:1}       
\end{figure}

\begin{figure}[!h]
\begin{center}
\includegraphics[width=5.0in]{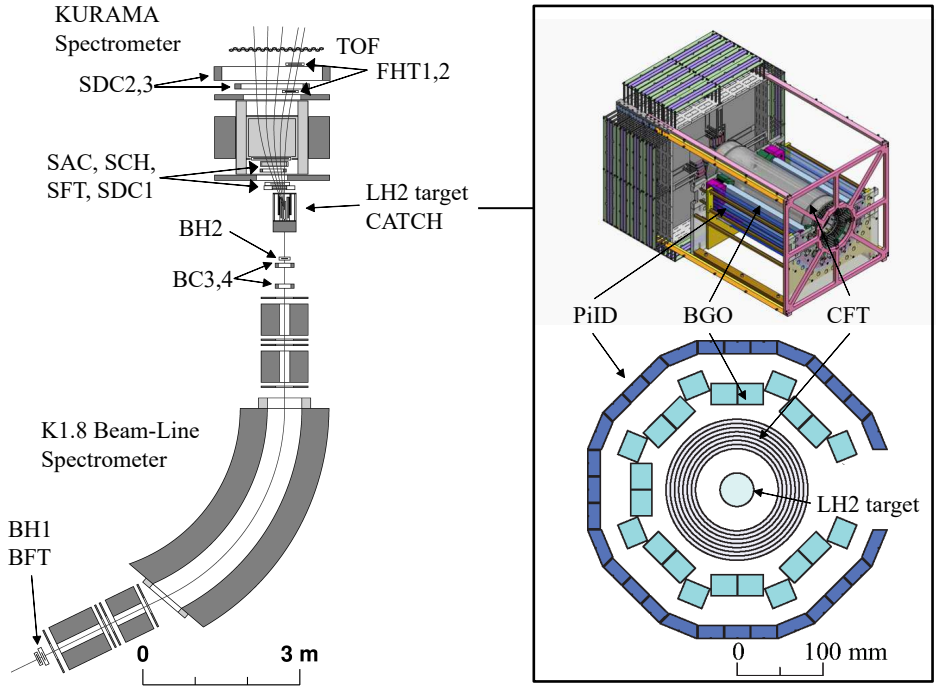}
\end{center}
\caption{(Left) Experimental setup of the J-PARC E40 experiment. The K1.8 beam-line spectrometer comprises five analyzer magnets in a QQDQQ configuration, two hodoscopes (BH1 and BH2), and three tracking detectors (BFT, BC3, and BC4). In the KURAMA spectrometer, seven tracking detectors (SFT, SCH, SDC1, SDC2, SDC3, FHT1, and FHT2) and two counters (SAC and TOF) are used. (Right) Perspective and cross sectional views of  CATCH.  CATCH comprises a cylindrical fiber tracker (CFT), BGO calorimeter (BGO), and plastic scintillator hodoscope (PiID)\red{;} and surrounds the $\text{LH}_2$ target.  The BGO and PiID segments overlapping the $K^+$ path to the KURAMA spectrometer were removed.}
\label{setup}       
\end{figure}

A conceptual drawing of the $\Sigma^+ p$ scattering identification is shown in \Fig{fig:1}, and the experimental setup used to realize this concept is shown in \Fig{setup}. A high-intensity $\pi^+$ beam of approximately $2\times 10^7$/spill was used to produce many $\Sigma^+$ particles inside a liquid hydrogen ($\text{LH}_2$) target via the $\pi^+p\to K^+\Sigma^+$ reaction. The central momentum of the beam was 1.41~GeV/$c$. The produced $\Sigma^+$ particles traveled in the $\text{LH}_2$ target within their lifetimes. Such $\Sigma^+$ particles are regarded as ``incident $\Sigma^+$'' for $\Sigma^+ p$ scattering. $\Sigma^+ p$ scattering can occur during the $\Sigma^+$ flight in the $\text{LH}_2$ target. The momentum of each $\Sigma^+$ particle can be calculated as the missing momentum of the $\pi^+$ beam and scattered $K^+$, analyzed using the K1.8 beam-line spectrometer and forward magnetic spectrometer (KURAMA spectrometer), as shown on the left-hand side of \Fig{setup}. The $\text{LH}_2$ target was surrounded by the so-called CATCH system, which comprises a cylindrical fiber tracker  (CFT), BGO calorimeter (BGO), and plastic scintillator hodoscope (PiID) \cite{Akazawa:2021}, as shown on the right-hand side of \Fig{setup}. The momentum vector\red{, which means both the momentum amplitude and direction,} of the recoil proton was determined using CATCH. Subsequently, the $\Sigma^+ p$ scattering events can be kinematically identified from the momentum vectors of the incident $\Sigma^+$ and recoil proton in the $\Sigma^+ p$ scattering. Spectrometers and CATCH are described in detail in the following paragraphs.


The $\pi^+$ beam was focused on the center of the $\text{LH}_2$ target via a set of QQDQQ magnets  downstream of the K1.8 beam line. These magnets form the K1.8 beam-line spectrometer together with detectors  upstream and downstream of the magnets. A \red{plastic scintillator} hodoscope (BH1) and \red{plastic scintillating fiber detector} (BFT \cite{BFT}) were placed upstream of the magnets. In contrast, the drift chambers (BC3 and BC4) and \red{another plastic scintillator} hodoscope (BH2) were placed downstream of these magnets. BH2 determines the origin of the timing for all detectors. The $\pi^+$ beam momentum was reconstructed event-by-event using the spatial information at BFT, BC3, BC4, and the third-order transfer matrix for the spectrometer. 

$\text{LH}_2$ was filled in the target cylindrical container of diameter 40~mm and length 300~mm with half-sphere end-caps at both edges. A vacuum window around the target region was created using a CFRP cylinder of diameter 80~mm and thickness 1~mm.

The outgoing particles produced at the $\text{LH}_2$ target by the $\pi^+ p$ reaction were analyzed using the KURAMA spectrometer downstream of the $\text{LH}_2$ target. The KURAMA spectrometer comprises a dipole magnet (KURAMA magnet), \red{plastic scintillating} fiber tracker (SFT), fine segmented \red{plastic scintillator} hodoscopes  (SCH, FHT1, and FHT2), three drift chambers (SDC1, SDC2, and SDC3), \red{a plastic scintillator wall (TOF), and an aerogel Cherenkov counter (SAC)}. The KURAMA magnet was excited to 0.78~T at the central position. SFT, SDC1, SAC, and SCH were placed either at the entrance or inside the KURAMA magnet gap. SDC2, SDC3, FHT1, FHT2, and TOF were installed downstream of the KURAMA magnet. The trajectories of the charged particles in the magnetic field were reconstructed using the Runge-Kutta method \cite{RungeKutta}. Their momenta were obtained to reproduce the hit positions measured at the tracking detectors. The time-of-flight of the outgoing particle along a flight path of approximately 3-m distance was measured using TOF. The typical time resolution was 300~ps. The spectrometer acceptance for $K^+$ in the $\pi^+ p \to K^+ \Sigma^+$ reaction was approximately 6.7\%, and the survival ratio of $K^+$ was 65\%. The large acceptance and short flight length are advantages of the KURAMA spectrometer for accumulating many $\Sigma^+$ particles.

Charged particles involved in $\Sigma^+ p$ scattering, such as the recoil proton and decay products of $\Sigma^+$, were detected using CATCH \cite{Akazawa:2021}. CATCH comprised CFT, BGO, and PiID. CFT along the beam axis is 400~mm long. It comprises eight cylindrical layers of \red{plastic scintillating} fibers. The fibers were placed parallel to the beam axis in four layers, called $\phi$ layers. In the other four layers, called $uv$ layers, fibers were arranged in a spiral shape. This configuration enabled us to reconstruct the trajectories of the charged particles in three dimensions. The BGO calorimeter was placed around CFT, and designed to measure the kinetic energy of the recoil proton from $\Sigma^+ p$ scattering by stopping it in the calorimeter.
The size of each BGO crystal was 400~mm $(l) \times 30~$mm $(w) \times 25~$mm $(t)$.
PiID was placed outside BGO to determine whether the charged particles penetrated BGO.

The experiment was performed in April 2019 and May-June 2020. In each period, we collected $\Sigma^+ p$ scattering data for approximately 10 days of beam time. Additionally, $pp$ scattering data using proton beams with various momenta between 0.45 and 0.85~GeV/$c$ were collected. The $pp$ scattering data were used for energy calibration and estimation of CATCH detection efficiency. 

\section{Analysis I: \red{Identification of the $\Sigma^+$ production events}}
\label{Sec3}
The analysis of the $\Sigma^+ p$ scattering events consists of three components. First, $\Sigma^+$ production events were identified, and the momentum of each $\Sigma^+$ was tagged from the analyses of the K1.8 beam-line and KURAMA spectrometers. Second, the $\Sigma^+ p$ scattering events were identified by requiring  kinematical consistency for the recoil proton, which was detected using CATCH. Finally, differential cross sections were derived. In this section, we explain the first component by detailing the analysis of the two spectrometers. The identification of the $\Sigma^+ p$ scattering events and the derivation of the differential cross sections for $\Sigma^+ p$ scattering are described in Sections \ref{Sec4} and \ref{Sec5}, respectively.

\subsection{$\pi^+$ analysis using the K1.8 beam-line spectrometer}
\begin{figure}[!h]
\begin{center}
\includegraphics[width=4.0in]{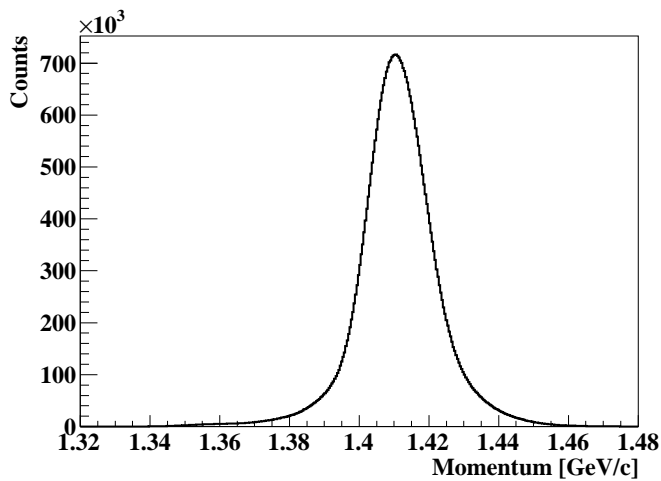}
\end{center}
\caption{Momentum distribution of the $\pi^+$ beam analyzed by the K1.8 beam line spectrometer.}
\label{BeamMom}       
\end{figure}

The momenta of incoming $\pi^+$ particles were analyzed event by event using the K1.8 beam-line spectrometer.
The position and angle downstream of the spectrometer magnet were reconstructed using BC3 and BC4. Additionally, the one-dimensional hit position upstream was measured using BFT. The beam momentum was reconstructed by connecting them with a third-order transfer matrix. Details are described in \cite{Honda:2017}. 
The distribution of the reconstructed momentum of the $\pi^+$ beam is shown in \Fig{BeamMom}.
The momentum resolution of the K1.8 beam-line spectrometer \red{$\sigma_p/p$ is better than $3 \times 10^{-3}$}  \cite{Takahashi:2012}.
\subsection{$K^+$ analysis using the KURAMA spectrometer}
\begin{figure}[!h]
\begin{center}
\includegraphics[width=4.0in]{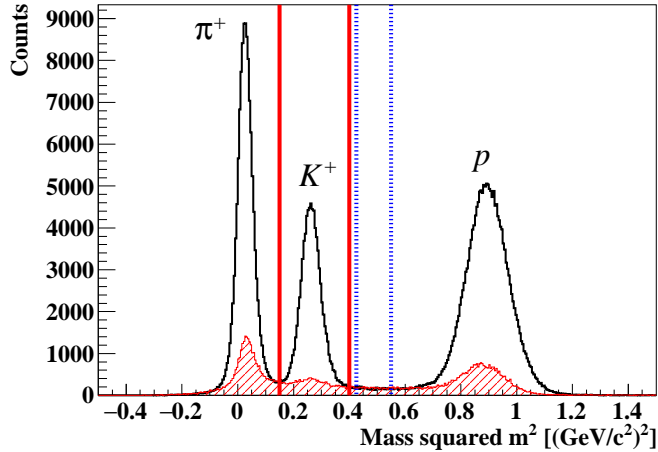}
\end{center}
\caption{Reconstructed $m^2$ distribution after the momentum selection of $0.65<p [\text{GeV}/c]<1.05$. The red shaded spectrum shows the estimated background due to the miscalculated time-of-flight for multi-beam events. The $K^+$ and side-band regions are indicated by the red solid and blue dashed lines, respectively. The side-band region was used to estimate the contribution of the miscalculated events in further analyses.}
\label{m2}    
\end{figure}

The outgoing particles produced at the $\text{LH}_2$ target by the $\pi^+ p$ reaction were analyzed using the KURAMA spectrometer. The trajectories of the outgoing particles in the magnetic field were traced using the Runge-Kutta method \cite{RungeKutta}, based on the equation of motion defined by the initial parameters, namely, the momentum vector and the position at TOF. The initial parameters were determined using a set of hit positions measured by the tracking detectors. The velocity $\beta$ of the outgoing particle can be calculated using the path length $L_{\text{track}}$ of the reconstructed trajectory and the time-of-flight between the target and TOF, $t$. The outgoing $K^+$ was identified by calculating the mass squared $m^2$ of the outgoing particles as follows:
\begin{align}
m^2 &= \Bigl{(}\frac{p}{\beta c}\Bigr{)}^2 (1-\beta^2), \\ 
\beta&=\frac{L_{\text{track}}}{ct}. \notag 
\end{align}
$K^+$ was selected from the $m^2$ gate of $0.15<m^2 [\text{GeV}^2/c^4]<0.40$, as indicated by the red solid lines in \Fig{m2} with an additional momentum gate of $0.65<p [\text{GeV}/c]<1.05$. There was a \red{background contamination} in the $m^2$ spectrum between the $\pi^+$ and the proton peaks. These background events were caused by the high-intensity $\pi^+$ beam, which is explained as follows.
When multiple particles pass the same segment of BH2 within a shorter time interval than its pulse shape, BH2 sometimes \red{failed to record the timing of the later pulse. In case that the later beam reacted at the target, the recorded timing for the former pulse was regarded as the BH2 timing for such events. These fake BH2 timings} irrelevant to the reaction results in the time-of-flight to be miscalculated, where the hits of BH2 and TOF were attributed to mismatched events. This is why a constant background exists in the $m^2$ spectrum. These background events were partially rejected using information on the energy deposit in TOF, as described in \cite{Miwa:2021}. The background structure was estimated as the shaded spectrum in \Fig{m2} by selecting the non-$K^+$ region in the TOF $dE/dx$ analysis for multi-beam events. Although some $K^+$ events are present in the spectrum, the background is expected to have a smooth structure. Finally, the contamination fraction of the miscalculated events was estimated to be 10.0\% for the selected $K^+$ events, by fitting the $m^2$ spectrum with the $K^+$ peak and  background contribution.

The momentum resolution of the KURAMA spectrometer was evaluated as \red{$\sigma_p/p=2.5 \times 10^{-2}$} for 1.37~GeV/c $\pi^+$ by analyzing the $\pi^+p$ elastic scattering reaction. This resolution was insufficient for identifying the $\Sigma^+ p$ scattering event. However, once $K^+$ is identified from $m^2$, the $K^+$ momentum can be calculated from the scattering angle $\theta_{K^+}$, based on the two-body kinematics of the $\pi^+p \to K^+ \Sigma^+$ reaction. By applying this analysis method, the momentum resolution for $K^+$ was improved to \red{$\sigma_p/p=6.5 \times 10^{-3}$}.


\subsection{$\Sigma^+$ identification with $(\pi^+, K^+)$ analysis}
\begin{figure}[!h]
\centerline{\subfloat[$z$ distribution]{\includegraphics[height=11\baselineskip]{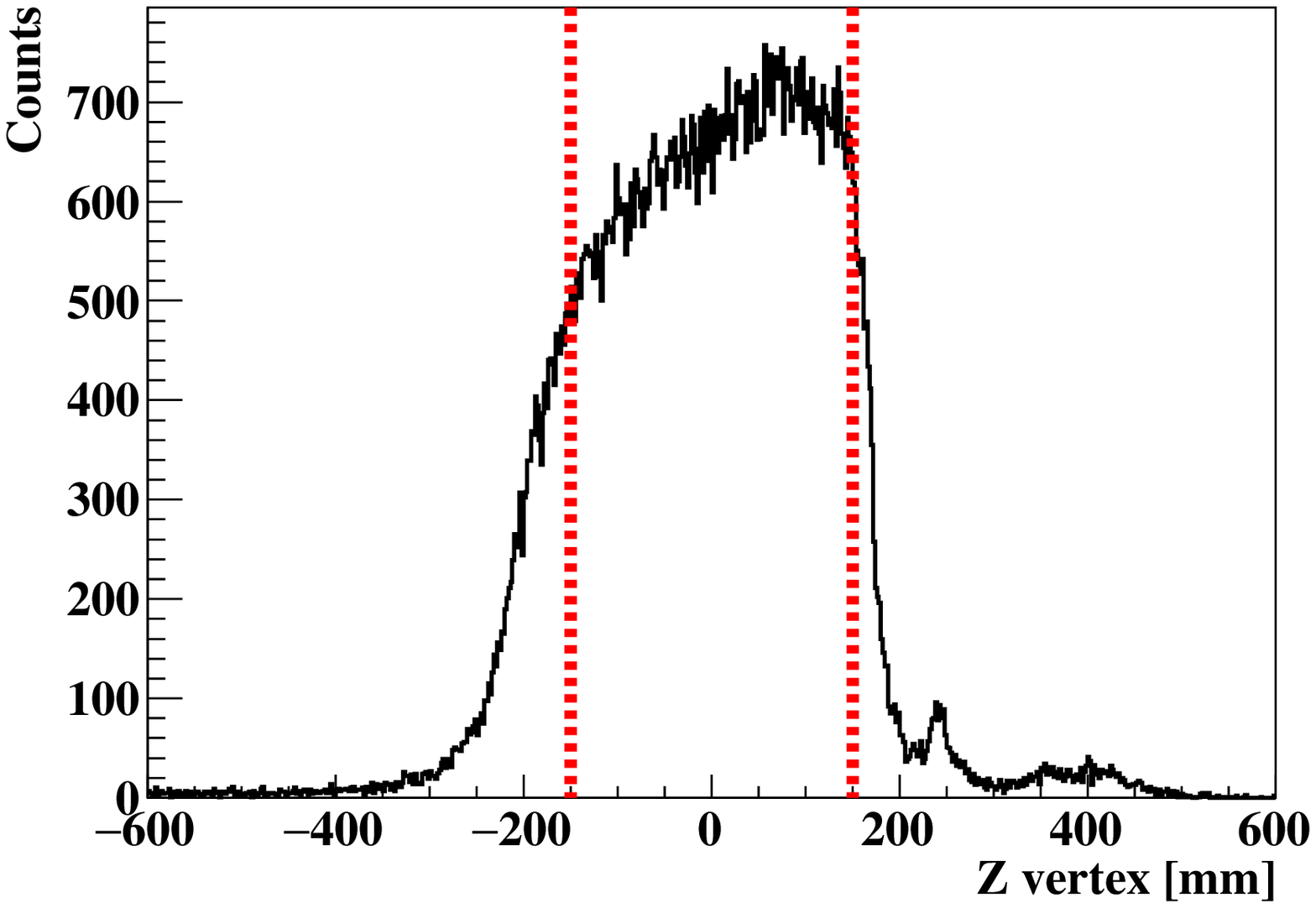}
\label{vtz_KURAMA}
}
\hfil
\subfloat[$xy$ distribution]{
\includegraphics[height=11\baselineskip]{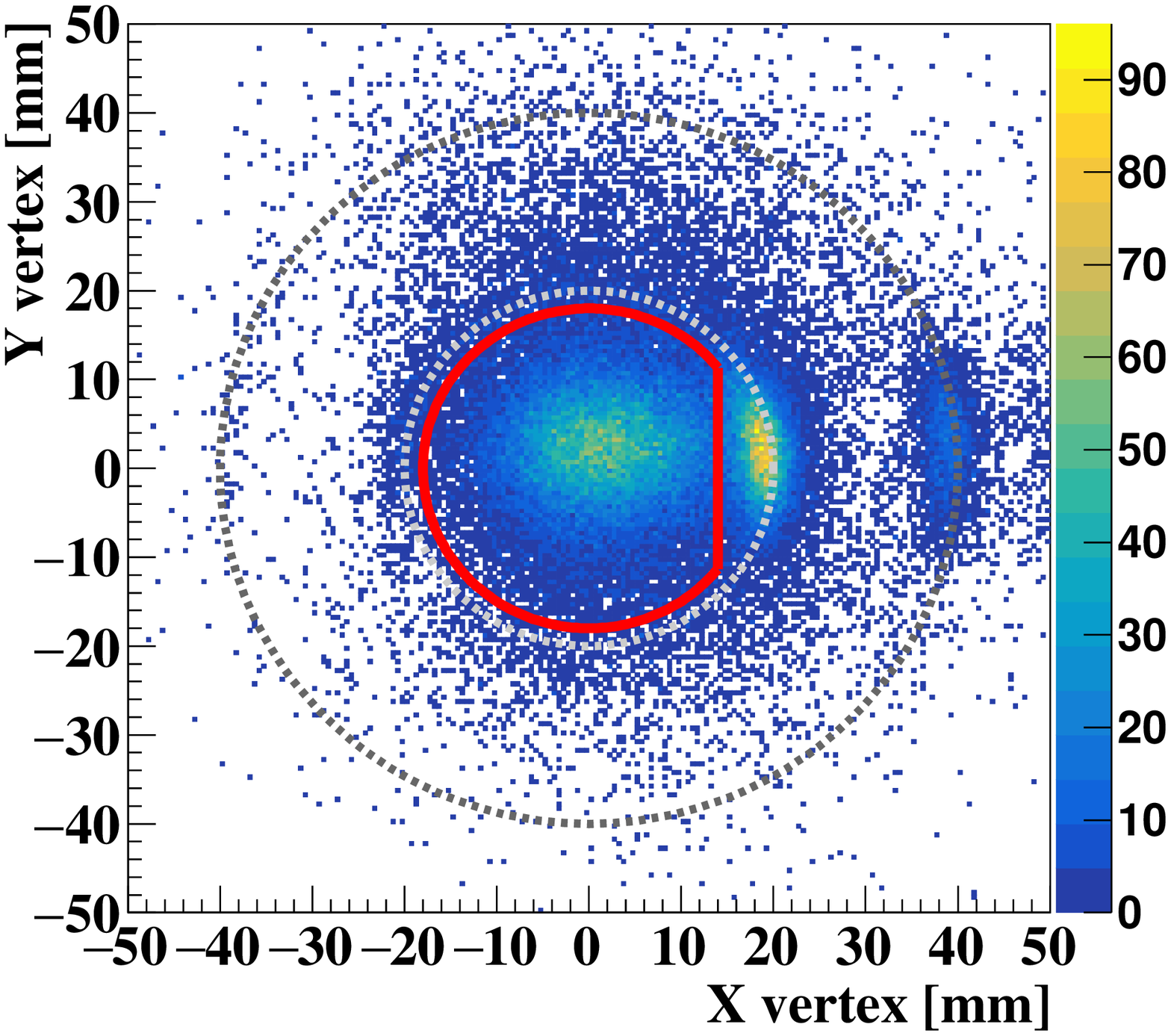}
\label{vtxy_KURAMA}
}}
\caption{Vertex distributions of the ($\pi^+,K^+$) reaction. (a) $z$-vertex distribution. (b) Correlation between the $x$- and $y$-vertices for events in which two protons were detected with CATCH. The gray dotted lines show the envelopes of the target container and vacuum window. The events inside the red-line region were selected to suppress the contamination of the reaction at the target container.}
\label{vertex}       
\end{figure}

$\Sigma^+$ particles were identified from the missing mass spectrum of the $\pi^+ p \to K^+ X$ reaction using the reconstructed \red{momentum} of the $\pi^+$ beam and the outgoing $K^+$. To select the reactions occurring in the $\text{LH}_2$ target, information regarding the vertex of the $\pi^+ p \to K^+ X$ reaction was used. This was determined to be the closest point between the $\pi^+$ and $K^+$ tracks.
\Figure{vertex} (a) shows the $z$-vertex distribution. The $\text{LH}_2$ target can be identified from $-200$~mm to 150~mm from the vertex image. Reflecting the differential cross section with a forward peak, the vertex distribution was flatter than that in $\Sigma^-$ production \cite{Miwa:2021}. For the $\Sigma^+ p $ scattering analysis, the $-150 <z [\text{mm}] <150$ region shown by the dotted lines in \Fig{vertex} (a) was selected, considering CATCH acceptance. \Figure{vertex} (b) shows the correlation between the $x$- and $y$-vertices. In this plot, the detection of two protons with CATCH was required to \red{enhance} the background events, owing to interactions between the $\pi^+$ beam and the target vessel. A horizontally wider beam caused such background events.
To suppress them, $x$- and $y$-vertices are required within the red line in \Fig{vertex} (b). The vertex resolution of the spectrometers was evaluated using multi-particle events, such as the $\pi^+ p \to \pi^+\pi^+\pi^- p$ reaction, by comparing the vertex obtained from the spectrometer analysis with that obtained from the two other tracks measured by CATCH from the same reaction vertex. The $z$-vertex resolution depends on the scattering angle $\theta_{K^+}$, and typical resolutions are $\sigma_z=16$~mm and $10$~mm for $\theta_{K^+}=10^\circ$ and $20^\circ$, respectively. The $x$- and $y$-vertex resolutions were evaluated as $\sigma_x=2.6$~mm and $\sigma_y=3.5$~mm, respectively, with a negligibly small angular dependence. These vertex resolutions were considered in the simulation study to estimate the analysis cut efficiency. 

\begin{figure}[!h]
\begin{center}
\includegraphics[width=4.0in]{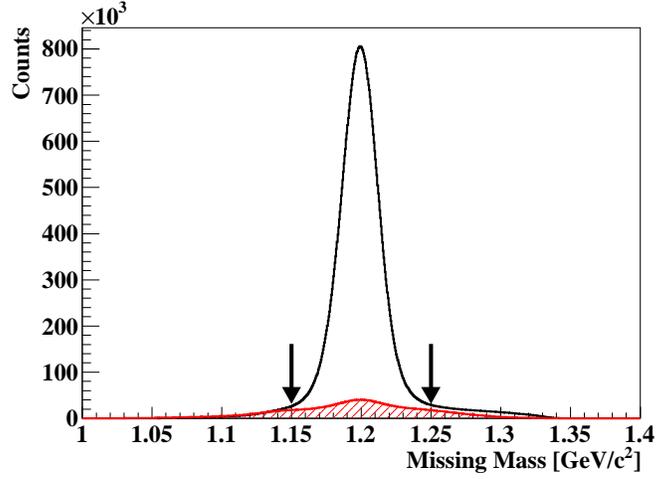}
\end{center}
\caption[Missing mass spectrum of the $\pi^+p \to K^+ X$ reaction]{Missing mass spectrum of the $\pi^+p \to K^+ X$ reaction for the $K^+$ events (black open histogram) and side-band events of $K^+$ (red shaded histogram) to estimate the effect of the contamination of the miscalculated events under the $K^+$ region in the $m^2$ spectrum. $\Sigma^+$ events were selected by the $1.15<M_X [\text{GeV}/c^2]<1.25$ gate, represented by the arrows.}
\label{MissingMass}       
\end{figure}

The missing mass spectrum of the $\pi^+p \to K^+ X$ reaction is shown in \Fig{MissingMass}. A clear peak corresponding to $\Sigma^+$ was identified. As mentioned in the previous subsection, there were misidentification backgrounds in the $K^+$ selection owing to multiple beam events. Their contribution was examined by selecting the sideband region of $K^+$, as shown by the blue dashed lines in \Fig{m2} with a momentum gate of $0.65<p [\text{GeV}/c]<1.05$. The red shaded histogram in \Fig{MissingMass} shows the missing mass spectrum for the sideband contribution. We selected $\Sigma^+$ particles from 1.15 to 1.25 GeV/$c^2$, as shown by the arrows in \Fig{MissingMass}. Contamination was estimated to be 8.6\% for $\Sigma^+$ selection. In total, 4.9 $\times 10^7$ $\Sigma^+$ particles were accumulated after subtracting the contamination.

\begin{figure}[!h]
\begin{center}
\includegraphics[width=4.0in]{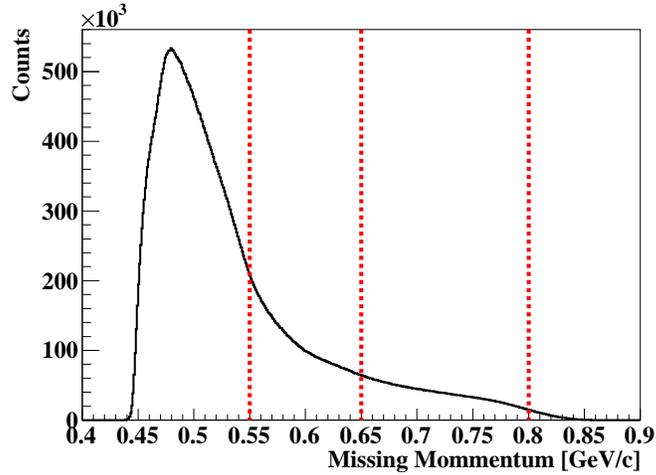}
\end{center}
\caption{$\Sigma^+$ momentum reconstructed as the missing momentum of the $\pi^+ p \to K^+ \Sigma^+$ reaction. The red dotted lines show the boundaries of the three momentum regions: low- ($0.44<p_\Sigma [\text{GeV}/c]< 0.55$), middle- ($0.55 <p_\Sigma [\text{GeV}/c]< 0.65$), and high-momentum ($0.65 <p_\Sigma [\text{GeV}/c]< 0.80$) regions.}
\label{MissingMom}       
\end{figure}

The reconstructed $\Sigma^+$ momentum, as the missing momentum of the $\pi^+p \to K^+ X$ reaction, is shown in \Fig{MissingMom}. This ranges from 0.44 to 0.85 GeV/$c$. In the $\Sigma^+ p$ scattering analysis, the $\Sigma^+$ events were categorized into three momentum ranges: the low- ($0.44<p_\Sigma [\text{GeV}/c]< 0.55$), middle- ($0.55 <p_\Sigma [\text{GeV}/c]< 0.65$), and high-momentum ($0.65 <p_\Sigma [\text{GeV}/c]< 0.80$) regions. The resolution of the $\Sigma^+$ momentum was $6 \times 10^{-3}$~GeV/$c$ in $\sigma$, which was  determined predominantly by the momentum resolution of the KURAMA spectrometer.

\section{Analysis II: Identification of the $\Sigma^+p$ scattering events}
\label{Sec4}
As explained in Section \ref{Sec3}, the momentum vector of the incident $\Sigma^+$ is reconstructed from the spectrometer information. In addition, the momentum vector of the recoil proton was measured using CATCH. 
Combining these momentum vectors enables us to identify the $\Sigma^+ p$ scattering events by checking the kinematical consistency of the recoil proton
between the measured energy $E_{\text{meas}}$ and the calculated energy $E_{\text{cal}}$ from the recoil angle. This section describes the analysis of the $\Sigma^+ p$ identification, with an emphasis on background suppression and derivation of the numbers of scattering events.

\subsection{Analysis for recoil/decay protons in $\Sigma^+$ production events using CATCH}
\label{CATCHtext}
The charged particles involved in the $\Sigma^+ p$ scattering, that is, the recoil proton and decay product of $\Sigma^+$, were detected using CATCH. The trajectory was reconstructed using CFT with an angular resolution of $1.5^\circ$ in $\sigma$. The kinetic energy was measured by summing the energy deposits in CFT ($dE_{\text{CFT}}$) and BGO ($E_{\text{BGO}}$) on the trajectory. This measured energy is denoted as $E_{\text{meas}}$ and its resolution was evaluated \red{as} 6~MeV in $\sigma$ for a 100~MeV proton. 

\begin{figure}[h]
\begin{center}
\includegraphics[width=4.0in]{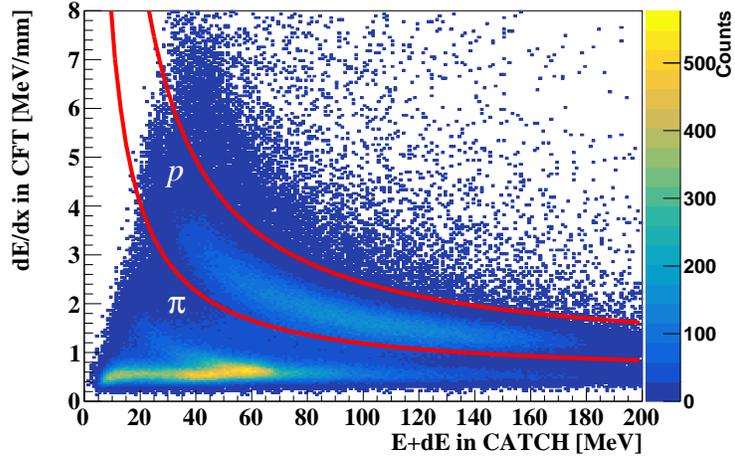}
\end{center}
\caption{$dE$-$E$ correlation between the energy loss in CFT and summed energy deposit in CATCH. The two red lines show the selection region for protons.}
\label{CATCHPID}       
\end{figure}

Particle identification in CATCH was performed using the so-called $dE$-$E$ method between $dE_{\text{CFT}}$, corrected by the path length in CFT ($dE/dx$), and $E_{\text{meas}}$ for each track. \Figure{CATCHPID} shows the $dE$-$E$ plot for the $\Sigma^+$ production events. The locus defined by the two lines corresponds to the protons. The typical purity of a proton was 90\% for the selection gate. The other locus shows an approximately constant $dE/dx$ distribution with a branch toward the higher $dE/dx$ value mainly corresponding to $\pi^+$s. Most of $\pi^+$s penetrated BGO by losing only a part of the kinetic energy. Therefore, the only available information for $\pi^+$ is the tracking information.

Low-energy protons, which \red{stopped} in CFT before arriving at BGO, were also identified by setting a $dE/dx$ value larger than 2.7~MeV/mm in CFT. Additionally, these protons were used for the $\Sigma^+ p$ scattering analysis.

\subsection{Kinematical consistency check for recoil proton}
\begin{figure}[!h]
\begin{center}
\includegraphics[width=3.0in]{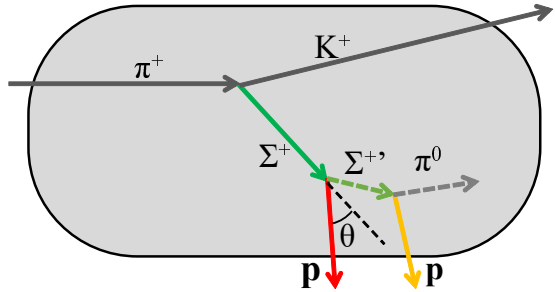}
\end{center}
\caption[Schematic of the $\Sigma^+ p$ scattering (following $\Sigma^+ \to p \pi^0$ decay.)]{Schematic of the $\Sigma^+ p$ scattering followed by the $\Sigma^+ \to p \pi^0$ decay. $\theta$ indicates the opening angle between the $\Sigma^+$ track, which is \red{reconstructed from the incoming $\pi^+$ and outroing $K^+$ using the spectrometers}, and the recoil proton track measured with CATCH.}
\label{SigmaPfig}       
\end{figure}

A schematic of $\Sigma^+ p$ scattering in the $\text{LH}_2$ target is shown in \Fig{SigmaPfig}.
The identification of the $\Sigma^+ p$ scattering event was performed by a kinematical consistency check for the recoil proton, as indicated by the red arrow in \Fig{SigmaPfig}. From the kinematic relation of the $\Sigma^+ p$ scattering, the kinetic energy of the recoil proton  $E_{\text{cal}}$ can be calculated from the momentum of the incident $\Sigma^+$ and the recoil angle of the recoil proton $\theta$. 
The kinetic energy of the recoil proton was also measured using CATCH, and is denoted by $E_{\text{meas}}$.
We then defined the difference between the two measurements $\Delta E$ as $\Delta E=E_{\text{meas}}-E_{\text{cal}}$.
If the proton recoils in $\Sigma^+ p$ scattering, such events would produce a peak around $\Delta E=0$.

\begin{figure}[!h]
\centerline{\subfloat[Data without cuts]{
\includegraphics[width=3in]{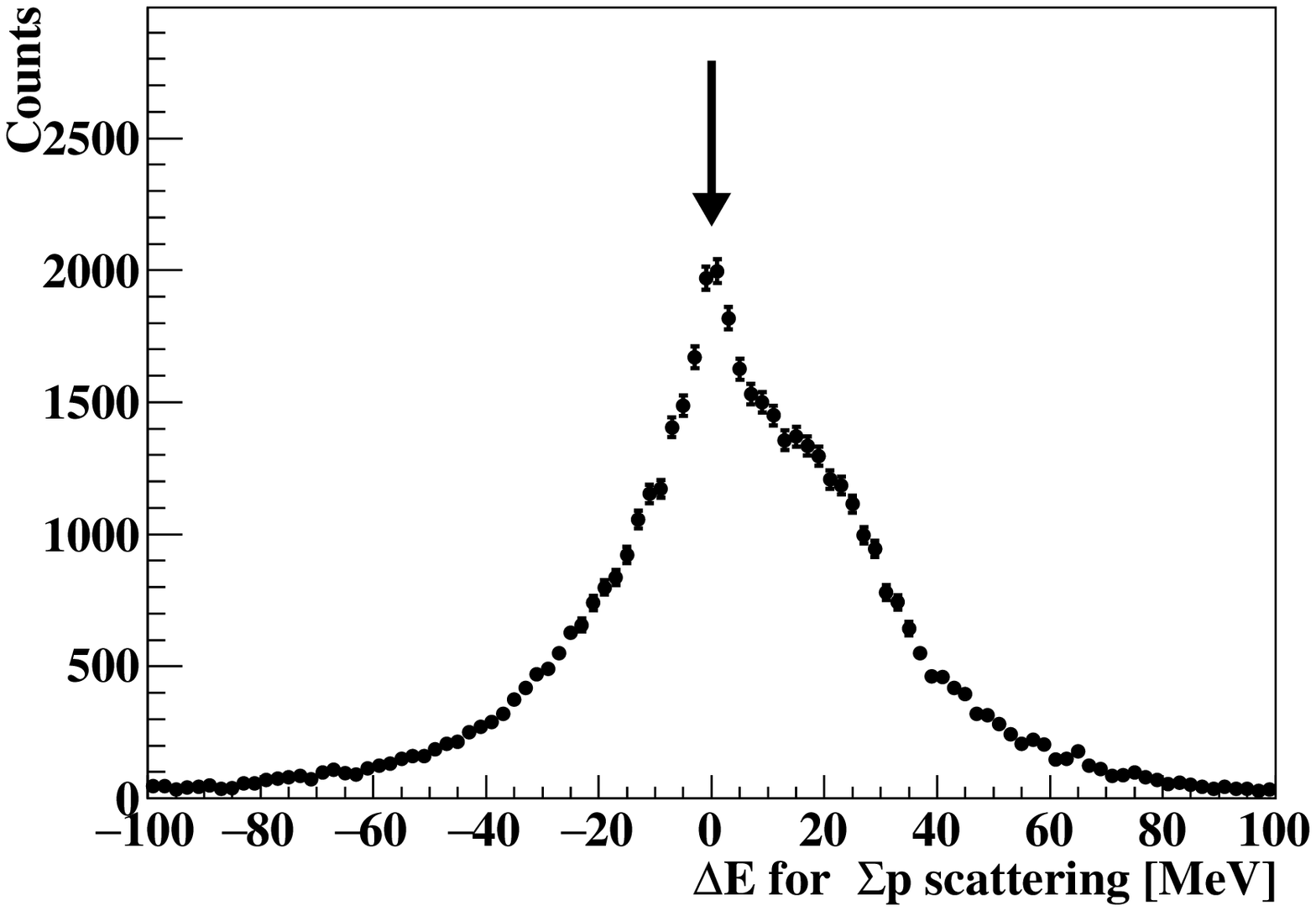}
\label{DeltaESigmaPScat2ppiBeforeData}
}\hfil
\subfloat[Simulation without cuts]{
\includegraphics[width=3in]{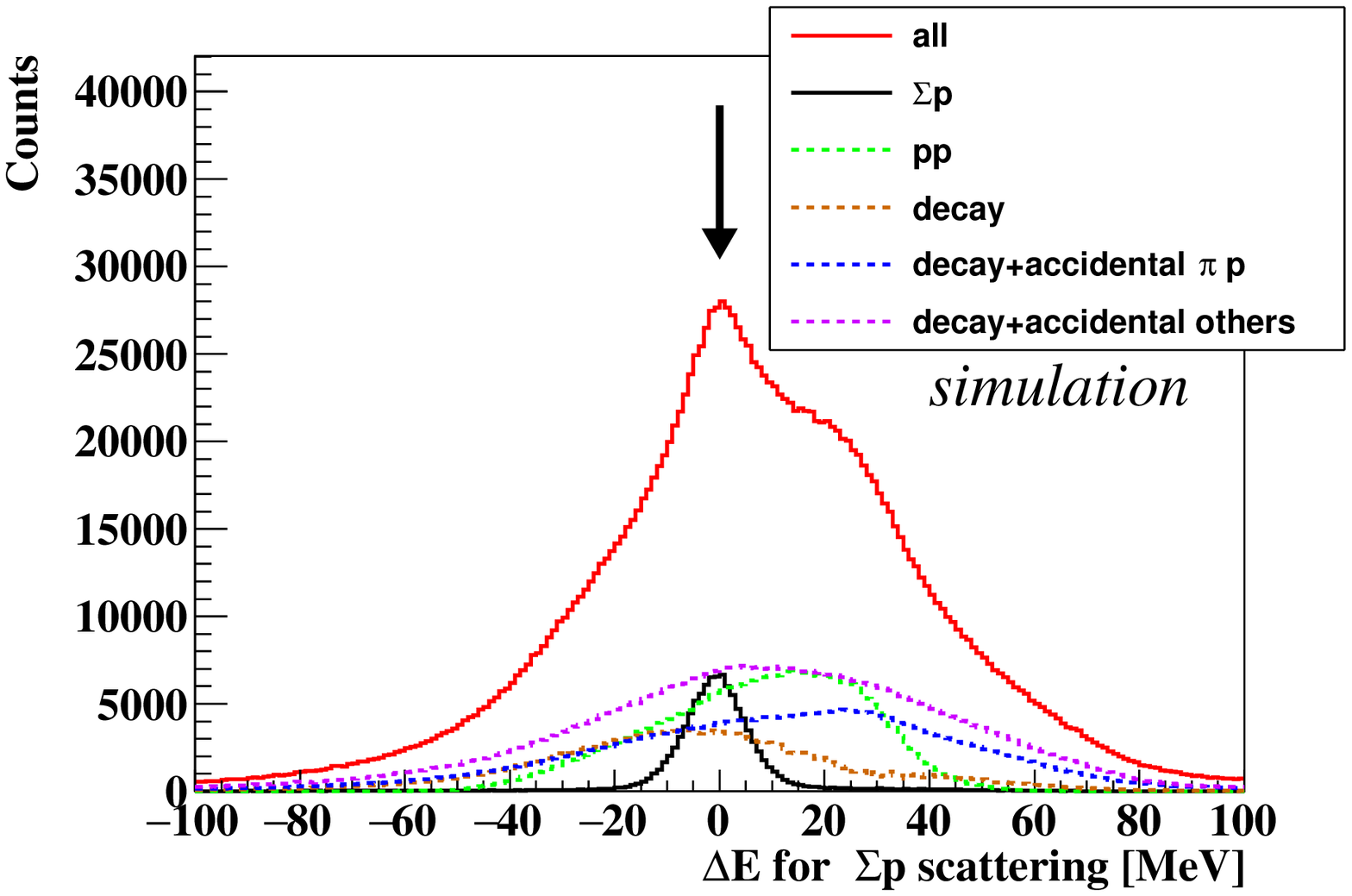}
\label{DeltaESigmaPScat2ppiBeforeSim}
}}
\caption{$\Delta E$ spectra for the two-proton events without cuts to select the $\Sigma^+ p$ scattering events for (a) data and (b) simulation. As shown by the arrows, a peak structure was identified around $\Delta E =0$. In the simulation (b), the contributions of the assumed reactions are additionally shown, reaction-by-reaction.}
\label{DeltaESigmaPScat2ppiBefore}       
\end{figure}

After $\Sigma^+ p$ scattering, the scattered $\Sigma^+$ decays mainly into $n \pi^+$ or $p \pi^0$. \red{For $\Sigma^+ p$ scattering followed by $\Sigma^+ \to n\pi^+$ decay, one proton, that is, the recoil proton is in the final state. However, this recoil proton is severely contaminated by the mere $\Sigma^+ \to p \pi^0$ decay and other secondary background events. Therefore, we focus on $\Sigma^+ p$ scattering followed by $\Sigma^+ \to p \pi^0$ decay, in which two protons can be observed with CATCH.
In the following analysis, the detection of two protons with CATCH is required and these events are called ``two-proton events''.}
The $\Delta E$ spectrum for two-proton events is shown in \Fig{DeltaESigmaPScat2ppiBefore} (a). In the analysis of two-proton events, the proton with the smaller $|\Delta E|$ value was regarded as the recoil proton.  In \Fig{DeltaESigmaPScat2ppiBefore} (a), a peak structure can be identified around $\Delta E=0$ without any further selection of the $\Sigma^+p$ scattering.

\begin{figure}[!h]
\begin{center}
\includegraphics[width=4in]{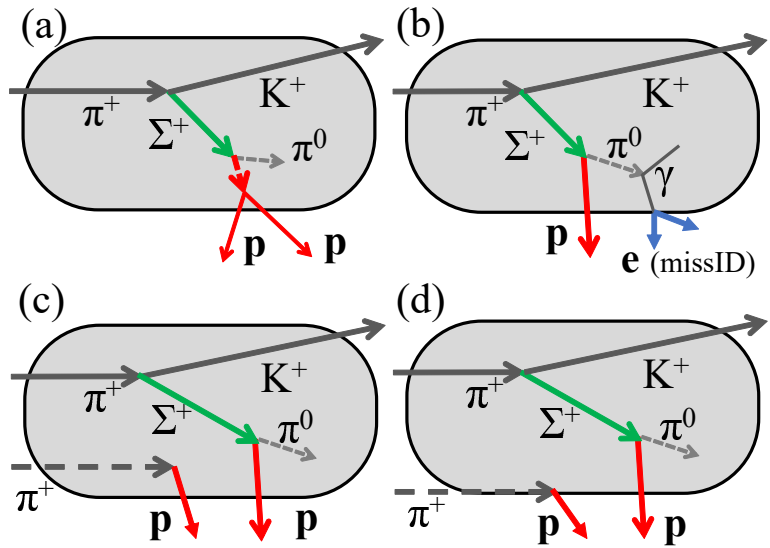}
\end{center}
\caption[Schematics of four types of backgrounds in the two-proton events.]{Schematics of four types of backgrounds in the two-proton events, (a) the $pp$ scattering following the $\Sigma^+ \to p \pi^0$ decay, (b) the combination of a proton from the $\Sigma^+ \to p \pi^0$ decay and misidentified $e^\pm$, (c) the combination of a proton from the $\Sigma^+ \to p \pi^0$ decay and a proton from the elastic $\pi^+ p$ scattering in the $\text{LH}_2$ target caused by \red{the accidental $\pi^+$}, and (d) the combination of a proton from the $\Sigma^+ \to p \pi^0$ decay and a proton from the scattering between \red{the accidental $\pi^+$} and the target vessel.}
\label{BGforSP}       
\end{figure}

The $\Delta E$ spectrum was compared with that of a Monte Carlo simulation. For the \red{background contamination} in the two-proton events, the four cases shown in \Fig{BGforSP} were considered within the simulation. \Figure{BGforSP} (a) shows the $pp$ scattering following $\Sigma^+ \to p \pi^0$ decay, generated based on the cross section. In the case shown in \Fig{BGforSP} (b), the $\Sigma^+ \to p \pi^0$ decay finally produces a proton and $e^+e^-$ pair, where the $e^+$ or $e^-$ are misidentified as a proton by CATCH after losing a large energy deposit in CFT, as mentioned in Subsection \ref{CATCHtext}. This misidentification was reproduced by the simulation. Therefore, we can estimate this background contribution by generating $\Sigma^+$ events within the simulation. The two other reactions are accidental coincidences of $\Sigma^+$ production and different reactions of the $\text{LH}_2$ target or the target vessel, shown in \Fig{BGforSP} (c) and (d), respectively, induced by the accidental $\pi^+$ beam. To reproduce accidental backgrounds, \red{the probability of accidental coincidence} and the distributions of the energy and angle of \red{the accidental protons} must be estimated from the real data. In real data, the $\pi^+ p$ elastic scattering events can be identified by detecting the recoil proton using the KURAMA spectrometer. For these events, an additional proton was searched for with CATCH, because this additional proton was attributed to \red{accidental coincidence}. From this analysis, the probabilities for types (c) and (d) were obtained as approximately 0.8\% and 1.2\% for the number of $\Sigma^+$ production events, respectively. These probabilities are considered in the simulations. \Figure{DeltaESigmaPScat2ppiBefore} (b) shows the $\Delta E$ spectrum obtained by analyzing the simulation, considering these backgrounds. The simulated spectrum consistently reproduced that of the real data.

\subsection{Cut conditions to select $\Sigma^+ p$ scattering events}
\label{Cutconditiontext}
To reduce the background in the $\Delta E$ spectrum, shown in \Fig{DeltaESigmaPScat2ppiBefore} (a), additional cuts regarding the spatial and kinematical information obtained from the two detected protons were applied. The detailed procedure is described in the following subsections. Table \ref{tab:cutcondition} summarizes the survival ratios for $\Sigma^+ p$ scattering events and the considered four types of background events. Finally, more than 90\% of the background events were eliminated, while maintaining approximately half of the $\Sigma^+ p$ scattering events.

\begin{table}[!h]
\begin{center}
\caption{Survival ratios of the $\Sigma^+ p$ scattering and background events after applying cuts as described in each subsection. (a)-(d) types correspond to the background types in \Fig{BGforSP}.}
\begin{tabular}{cccccc}\hline
\label{tab:cutcondition} 
 &$\Sigma^+p$ & (a) type & (b) type & (c) type & (d) type \\ \hline
Cuts in \ref{cut1}& 89.9\% & 69.3\% &74.8\% & 55.7\% & 53.5\% \\ 
Cuts in \ref{cut2}& 86.7\% & 68.2\% &8.3\% & 13.0\% & 5.7\% \\ 
Cuts in \ref{cut3}& 69.8\% & 12.8\% &7.6\% & 11.3\% & 4.7\% \\ 
Cuts in \ref{cut4} (All cuts)& 54.5\% & 7.1\% &5.5\% & 0.4\% & 1.6\% \\ \hline
\end{tabular}
\end{center}
\end{table}

\subsubsection{Vertex cut and closest distance cut}
\label{cut1}
\begin{figure}[!h]
\centerline{\subfloat[At the scattering point]{
\includegraphics[width=3in]{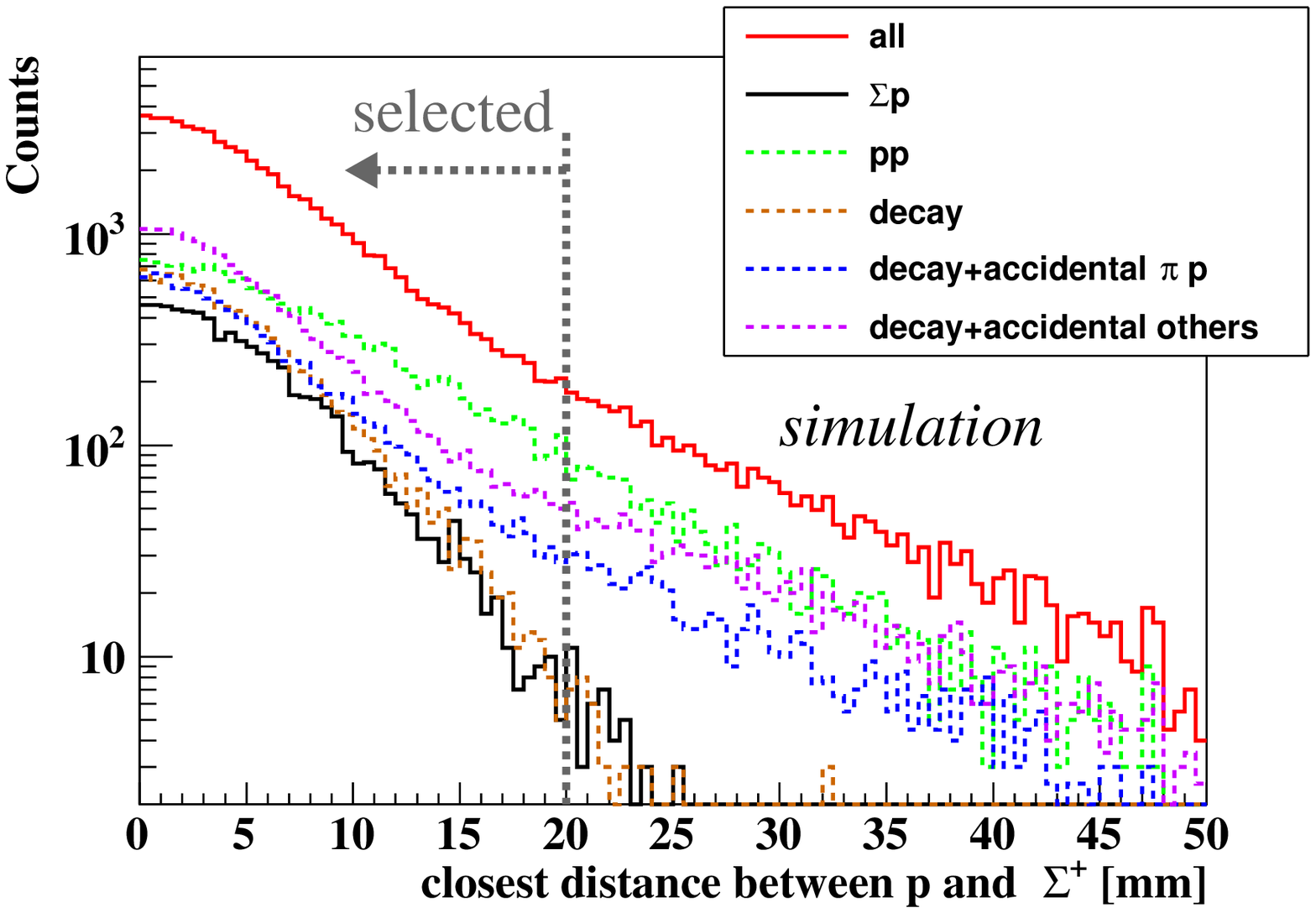}
\label{cdistscat}
}\hfil
\subfloat[At the decay point]{
\includegraphics[width=3in]{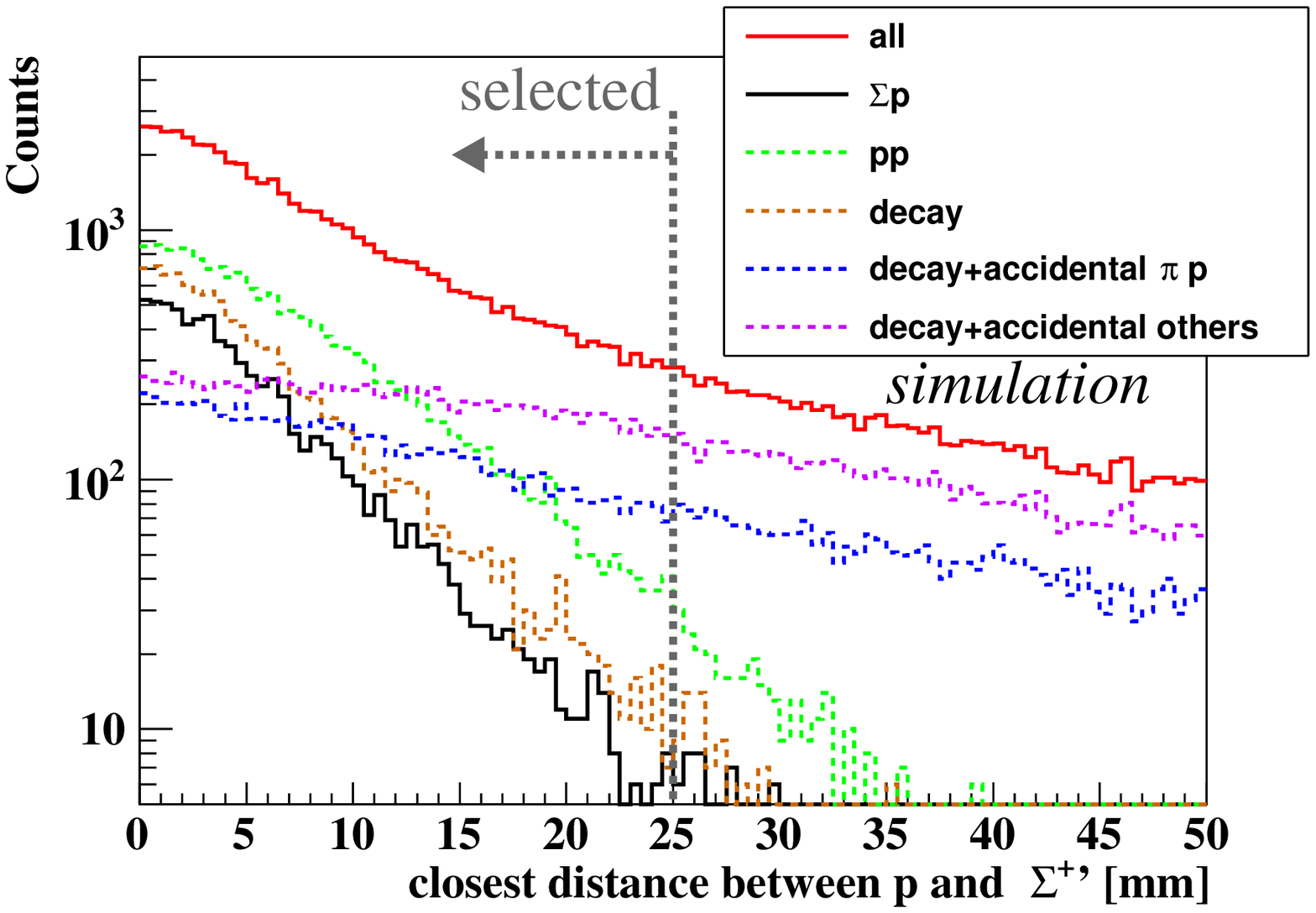}
\label{cdistdecay}
}}
\caption{Simulated distributions of the closest distances at the scattering \red{(a)} and decay points \red{(b)}. The dotted lines indicate the boundaries of selected regions: 20 and 25 mm \red{in (a) and (b)}, respectively.}
\label{cdist}       
\end{figure}

The spatial consistency between two proton tracks can be used to select $\Sigma^+ p$ scattering events. The vertex of $\Sigma^+ p$ scattering, which was calculated as the closest point between the incident $\Sigma^+$ and recoil proton tracks, should be inside the target. Accordingly, the scattering vertex ($x_{\text{scat}},y_{\text{scat}}, z_{\text{scat}}$) must be $x_{\text{scat}}^2+y_{\text{scat}}^2<25^2$ mm$^2$, and $|z_{\text{scat}}|<170$ mm. Similarly, the decay vertex ($x_{\text{decay}},y_{\text{decay}}, z_{\text{decay}}$), which was obtained from the decay proton and scattered $\Sigma^+$, was required to be $-30<x_{\text{decay}}<25$ mm, $|y_{\text{decay}}|<30$ mm, and $|z_{\text{decay}}|<180$ mm. The momentum vector of the scattered $\Sigma^+$, denoted as $\Sigma^{+'}$, can be kinematically calculated from the recoil angle of the proton. The trajectory of $\Sigma^{+'}$ is reconstructed from the momentum vector and the scattering vertex. The closest distances between $\Sigma^+$ and the proton tracks at the scattering and decay points also reflect spatial consistency. The simulated distributions of the closest distances at the two vertices are shown in \Fig{cdist}. The closest distances at the scattering and decay points were required to be less than 20 mm and 25 mm, respectively. These cuts can reduce the background events by \red{30--50\%} while maintaining approximately 90\% of the $\Sigma^+ p$ scattering events, as shown in the second row of Table \ref{tab:cutcondition}.

\subsubsection{Missing mass cut to select the scattered $\Sigma^+$ decay }
\label{cut2}
\begin{figure}[!h]
\begin{center}
\includegraphics[width=4.0in]{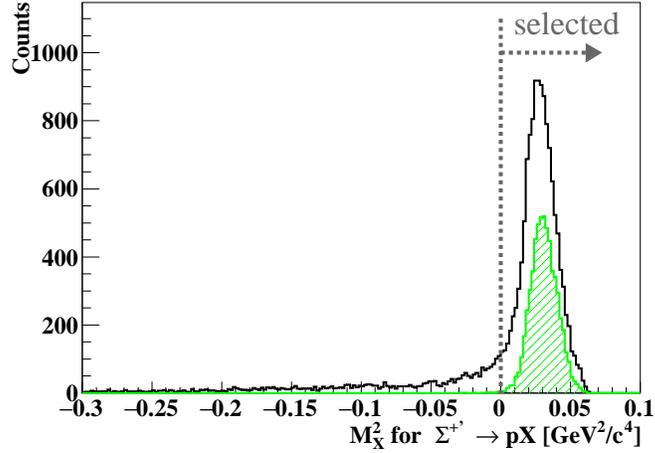}
\end{center}
\caption{Squared missing mass distribution of the $\Sigma^{+'} \to pX$ reaction. The green shaded spectrum shows the $pp$ scattering following the $\Sigma^+ \to p \pi^0$ decay, which can be selected using the kinematical consistency and opening angle of the two detected protons, as explained in sub-subsection \ref{cut3}. Here, $M_X^2>0$ is selected.}
\label{MMcut}       
\end{figure}

Assuming that $\Sigma^+ p$ scattering is followed by $\Sigma^{+'} \to p \pi^0$ decay, the missing mass of the $\Sigma^{+'} \to pX$ reaction, $M_X$, should be the mass of $\pi^0$. The squared missing mass ($M_X^2$) distribution shows a $\pi^0$ peak and a broad distribution toward the negative region, as shown in \Fig{MMcut}. The broad distribution in the negative region was mainly attributed to accidental backgrounds. \red{The event forming the $\pi^0$ peak comes not only from $\Sigma^+ p$ scattering, but also from $pp$ scattering following the $\Sigma^+ \to p \pi^0$ decay}, because both reactions have the same final state of $pp \pi^0$, originating from the initial $\Sigma^+$. The green shaded spectrum in \Fig{MMcut} shows the $M_X^2$ distribution for $pp$ scattering,  identified with the cuts described in sub-subsection \ref{cut3}. The cut condition $M_X^2>0$, shown by the dotted line in \Fig{MMcut}, was determined to include the green spectrum. The survival ratios estimated from the Monte Carlo simulation are listed in the third row of Table \ref{tab:cutcondition}.

\subsubsection{Kinematical cut for secondary $pp$ scattering}
\label{cut3}
The $pp$ scattering following the $\Sigma^+ \to p \pi^0$ decay, shown in \Fig{BGforSP} (a), can be identified by the kinematical consistency check for the decay proton, as is the case for $\Sigma^+p$ scattering.

\begin{figure}[!h]
\begin{center}
\includegraphics[width=4.0in]{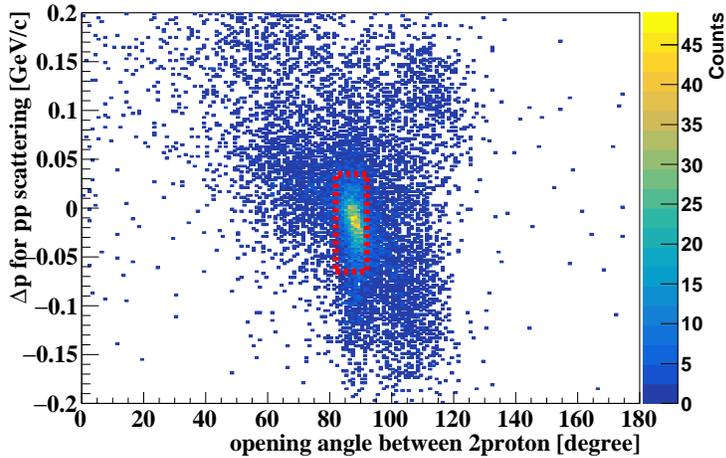}
\end{center}
\caption{Correlation between $\Delta p$ and the opening angle of two detected protons from real data. The events inside the red-line region were rejected as the secondary $pp$ scattering background.}
\label{DeltaPcor}       
\end{figure}

Assuming that the two protons originate from $pp$ scattering, the incident proton's momentum can be reconstructed from the momenta of the two protons. Because the incident proton in this $pp$ scattering is also the decay proton from the $\Sigma^+ \to p\pi^0$ decay, the reconstructed momentum should satisfy the $\Sigma^+$ decay kinematics for a real $pp$ scattering event. This kinematical check was performed by evaluating the consistency between the reconstructed proton momentum $p_{\text{meas}}$ and the calculated momentum $p_{\text{cal}}$, from the emission angle of the decay proton from the initial $\Sigma^+$. The difference between $p_{\text{meas}}$ and $p_{\text{cal}}$, $\Delta p$, was evaluated to identify the $pp$ scattering event from the peak near $\Delta p =0$. \red{The opening angle of the two protons, $\alpha$, was also verified} because $\alpha$ is kinematically constrained to be approximately $90^\circ$ for the $pp$ scattering event. The correlation between $\Delta p$ and $\alpha$ is illustrated in \Fig{DeltaPcor}, where a clear event concentration \red{owing to $pp$ scattering} can be confirmed within the expected region. To reject \red{the identified $pp$ scattering events}, the cut region defined as the $\pm 2 \sigma$ areas for $\Delta p$ and $\alpha$ was defined, as shown by the red box in \Fig{DeltaPcor}. Although this cut condition rejected approximately 20\% of the $\Sigma^+ p$ scattering events, the signal-to-noise (S/N) ratio was improved by cutting approximately 80\% of the secondary $pp$ scattering events.

\subsubsection{Kinematical cut for $\pi^+ p$ elastic scattering}
\label{cut4}
\begin{figure}[!h]
\begin{center}
\includegraphics[width=4.0in]{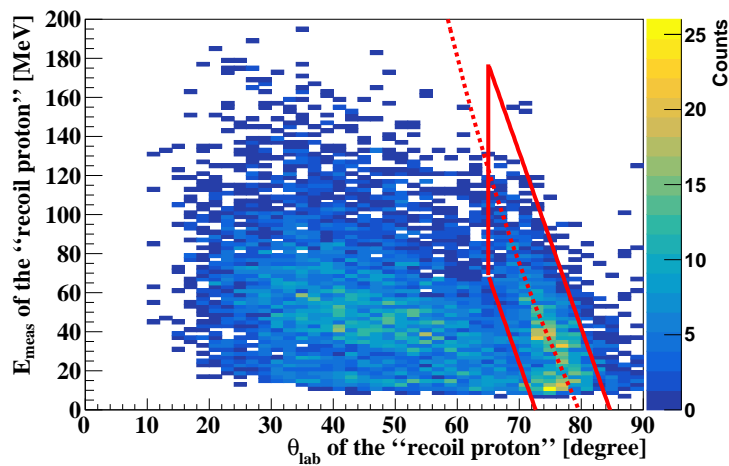}
\end{center}
\caption{Correlation between the angle of protons \red{with respect to} the central axis of CFT, $\theta_{\text{lab}}$, and $E_{meas}$ of the recoil proton. The red dotted line indicates the kinematics of $\pi^+ p$ elastic scattering induced by the \mbox{1.41-GeV/$c$} $\pi^+$ beam. When calculating the kinematics, the energy loss in the $\text{LH}_2$ target was considered. The events inside the red-line region were rejected as the protons from elastic $\pi^+ p$ scattering in response to the accidental $\pi^+$.}
\label{pipcut}       
\end{figure}

Finally, we describe the rejection of the accidental coincidence of $\pi^+ p$ elastic scattering. The recoil proton by $\pi^+p$ elastic scattering shows a kinematical correlation between the recoil angle and energy. \Figure{pipcut} shows the correlation between the angle of protons \red{with respect to} the central axis of CFT and the energy. The locus corresponding to the $\pi^+ p$ kinematics, indicated by the red dotted line, can be determined. \red{The events inside the red-line region in \Fig{pipcut} were rejected as the protons from elastic $\pi^+ p$ scattering.} This cut reduced almost all of the type (c) background in \Fig{BGforSP}.

\subsubsection{$\Sigma^+ p$ scattering identification after all cuts}
\begin{figure}[!h]
\centerline{\subfloat[Data with cuts]{
\includegraphics[width=3in]{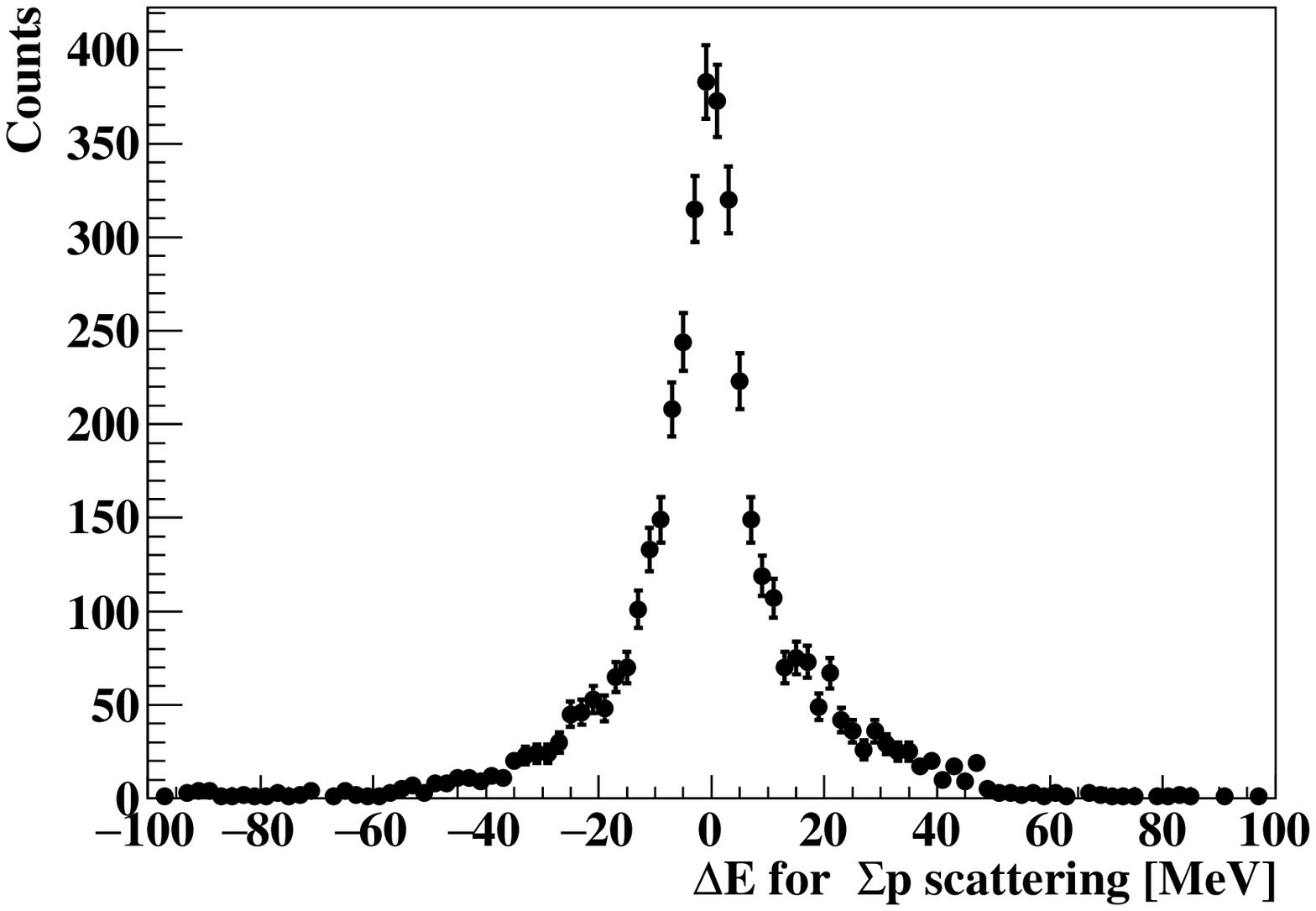}
\label{DeltaESigmaPScat2ppiAfterData}
}\hfil
\subfloat[Simulation with cuts]{
\includegraphics[width=3in]{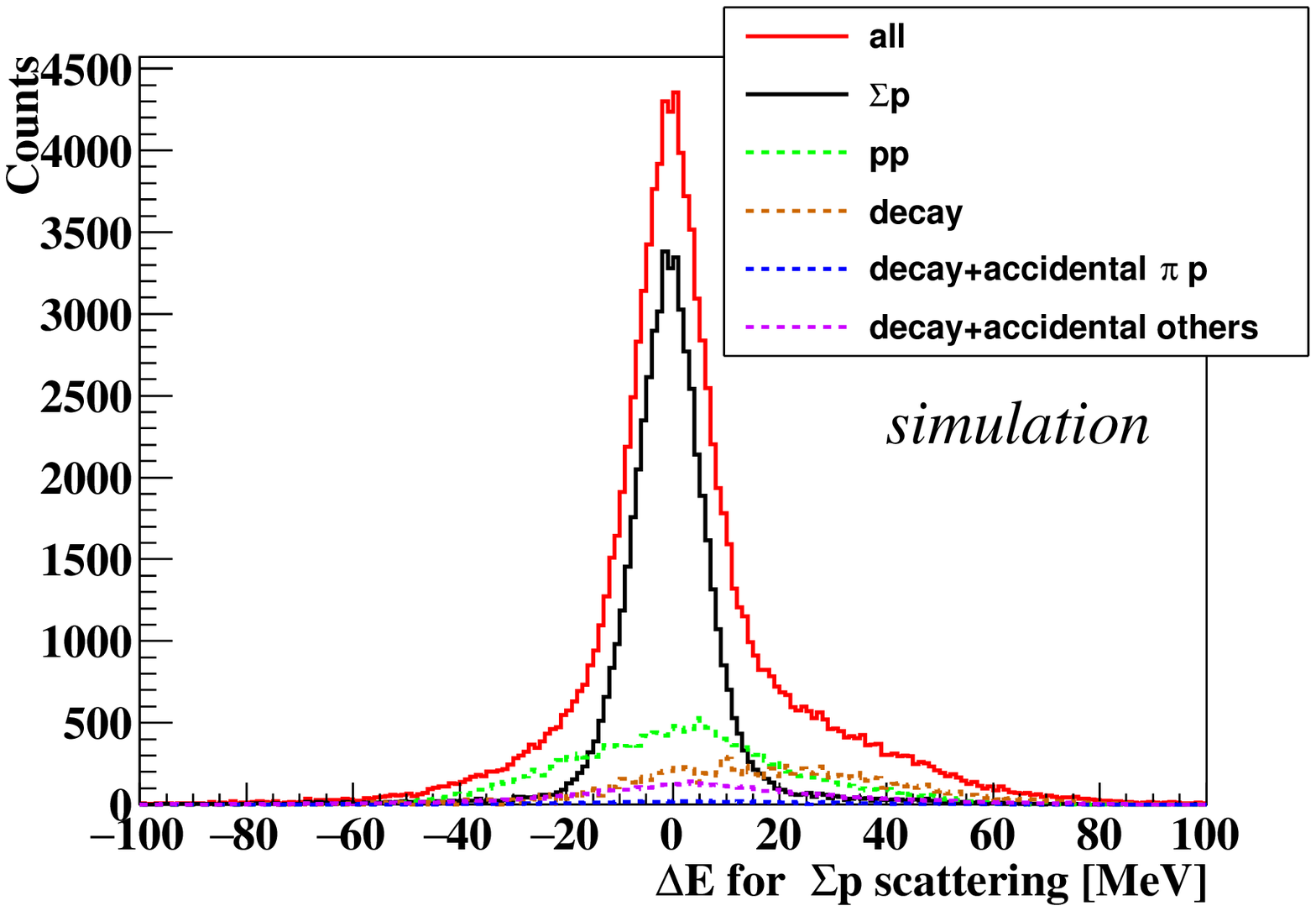}
\label{DeltaESigmaPScat2ppiAfterSim}
}}
\caption{$\Delta E$ spectra for the two-proton events with cuts to select the $\Sigma^+ p$ scattering events for  (a) data and (b) simulation. In the simulation (b), the contributions of the assumed reactions are shown reaction-by-reaction.}
\label{DeltaESigmaPScat2ppiAfter}       
\end{figure}

The $\Delta E$ spectrum for two-proton events after applying all cuts is shown in \Fig{DeltaESigmaPScat2ppiAfter} (a). The S/N ratio in the peak region of $-20<\Delta E [\text{MeV}]<20$ was significantly improved to 1.78. The evaluation of the S/N ratio is based on the fitting results of the $\Delta E$ spectra explained in the next subsection. The simulated spectrum after the same cuts was in agreement with the data shown in \Fig{DeltaESigmaPScat2ppiAfter} (b).
The analysis efficiency for the $\Sigma^+ p$ scattering events was estimated to be 54.5\%, and the rejection \red{factors} of the four background sources were greater than 90\%, as summarized in Table \ref{tab:cutcondition}.

\subsection{Estimation of the number of $\Sigma^+ p$ scattering events}
\label{BGtext}
\begin{figure}[!h]
\centerline{\subfloat[$\Delta E$ spectrum for $\Sigma^+ p$ scattering]{
\includegraphics[width=3in]{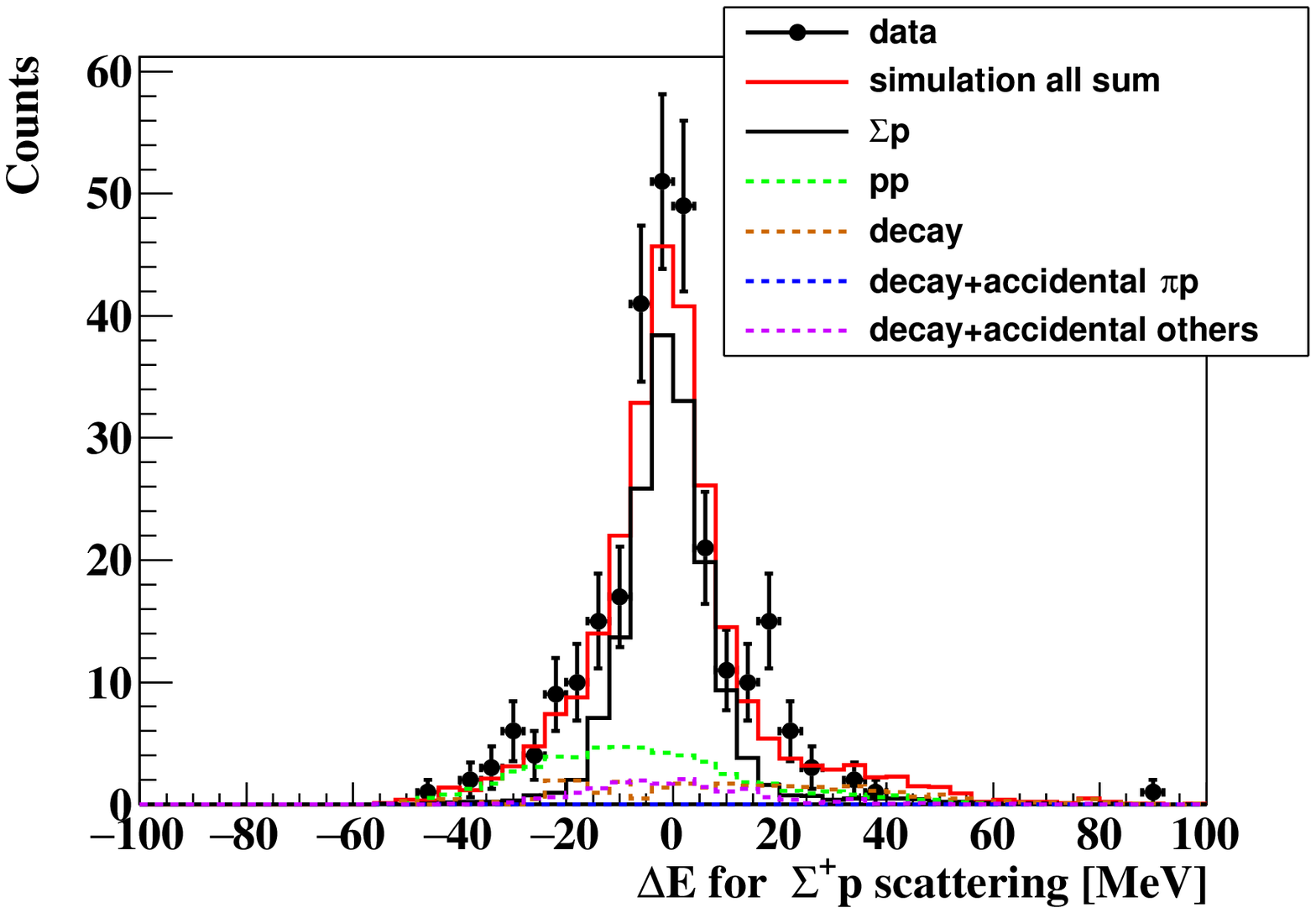}
\label{DeltaEfittingintext}
}\hfil
\subfloat[$\Delta p$ spectrum for $pp$ scattering]{
\includegraphics[width=3in]{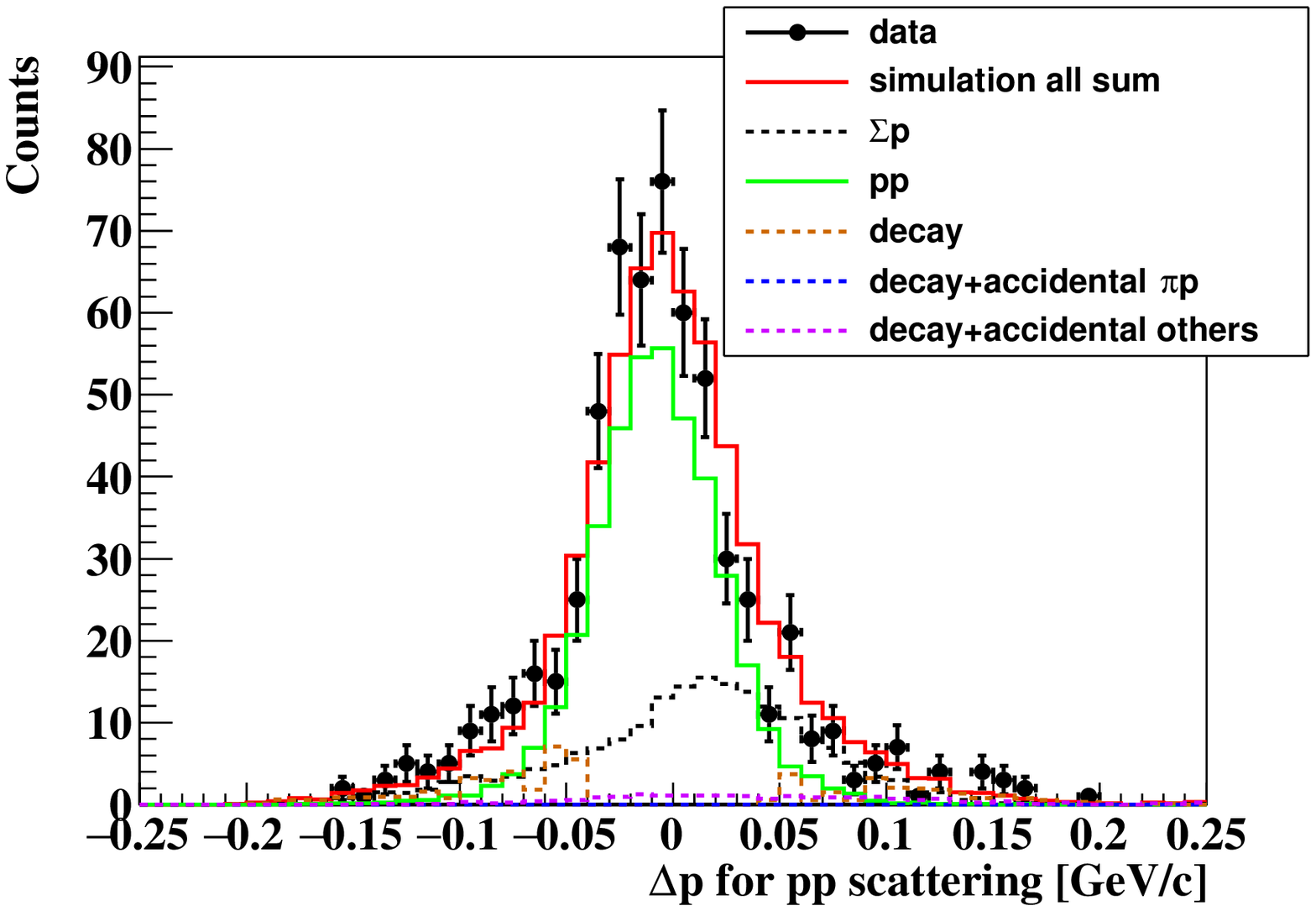}
\label{DeltaPfittingintext}
}}
\caption{Fitted $\Delta E$ and $\Delta p$ spectra for $\Sigma^+ p$ scattering and  $pp$ scattering following the $\Sigma^+ \to p \pi^0$ decay at the scattering angle of $-0.4<\cos \theta_{\text{CM}}<-0.3$ in the momentum range of $0.44<p_\Sigma [\text{GeV/}c]<0.55$. In the $\Delta p$ spectra (b), the $pp$ scattering rejection cut was not applied.}
\label{fittingintext}       
\end{figure}

To estimate the number of $\Sigma^+ p$ scattering events and survival background events, the $\Delta E$ spectrum for the $\Sigma^+ p$ scattering was fitted with the sum of the simulated spectra for both the $\Sigma^+ p$ scattering and background reactions. Fitting was performed for the $\Delta E$ spectrum at each scattering angle independently to correctly reproduce the angular dependence of the background contribution. \Figure{fittingintext} (a) shows a typical fitting result of the $\Delta E$ spectrum for the scattering angle of $-0.4<\cos \theta_{\text{CM}}<-0.3$ in the momentum range of $0.44<p_\Sigma [\text{GeV/}c]<0.55$. To constrain the contamination of the $pp$ scattering events in this fitting, the $\Delta p$ spectrum for the $pp$ scattering kinematics was simultaneously fitted with the simulated spectra, as shown in \Fig{fittingintext} (b). The cut condition for the $\Delta p$ spectrum, where the $pp$ scattering rejection cut described in sub-subsection \ref{cut3} was not applied, was different from that for the $\Delta E$ spectrum. However, the same scale parameters were used for both the $\Delta E$ and $\Delta p$ spectra for each reaction, and these parameters were obtained by simultaneous fitting. We also examined the fitting of only the $\Delta E$ spectrum with the simulated spectra in order to study the systematic differences due to the background estimation and fitting procedure.
The difference in the estimated $\Sigma^+ p$ scattering events in these fittings is considered \red{as} the systematic uncertainty.
The uncertainty due to the bin size of the spectra was also estimated by iterating the same procedure for the spectra with different sets of bin sizes. This uncertainty was also included as a systematic error, although it was smaller than that owing to the fitting condition mentioned earlier.

\begin{figure}[!h]
\begin{center}
\includegraphics[width=6.0in]{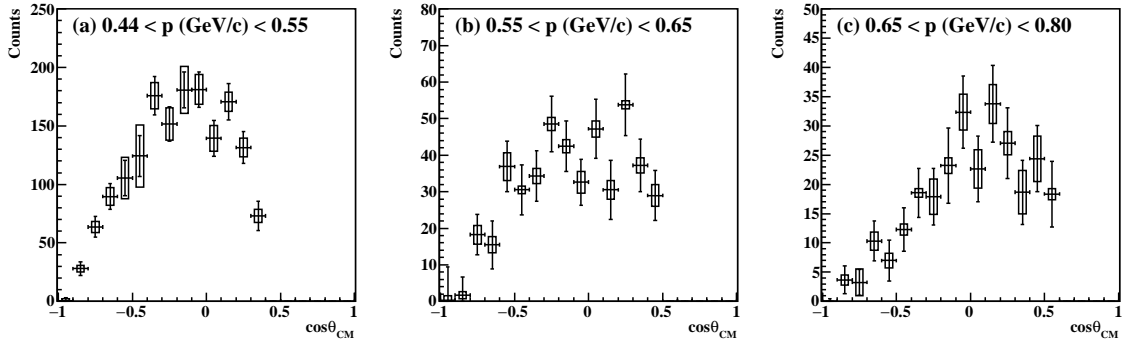}
\end{center}
\caption{The estimated numbers of the $\Sigma^+ p$ scattering events for each scattering angle and momentum region of $\Sigma^+$: (a) the low momentum region ($0.44<p_\Sigma [\text{GeV}/c]<0.55$),  (b) the middle momentum region ($0.55<p_\Sigma [\text{GeV}/c]<0.65$), and (c) the high momentum region ($0.65<p_\Sigma [\text{GeV}/c]<0.80$). The error bars and boxes indicate the statistical and systematic errors, respectively.}
\label{SPcounts}       
\end{figure}

The fitting results of $\Delta E$ for all measurements within detector acceptance are shown in \Fig{fitLow}, \ref{fitMid}, and \ref{fitHigh} in Appendix \ref{Fittext}. As shown in these spectra, the $S/N$ ratio of the $\Delta E$ spectra worsened \red{in} the forward angular region because of the limited acceptance of the low-energy recoil proton.
Therefore, we set the maximum scattering angle for each incident $\Sigma^+$ momentum region. The obtained number of $\Sigma^+ p$ scattering events is shown in \Fig{SPcounts} as a function of $\cos \theta_{\text{CM}}$, for each incident $\Sigma^+$ momentum region. The error bars and boxes represent statistical and systematic errors, respectively. The systematic error includes the uncertainty in the background estimation, as previously discussed. A total of approximately 2400 $\Sigma^+ p$ scattering events were identified. 

\section{Analysis III: Derivation of differential cross sections}
\label{Sec5}
For the analysis deriving the differential cross sections, several values should be \red{evaluated} for each scattering angle and $\Sigma^+$ momentum region. Therefore, these values are denoted as a function of $p_\Sigma$ and $\cos \theta_{\text{CM}}$, such that $N(p_\Sigma,\cos\theta_{\text{CM}})$ represents the number of scattering events.
The differential cross section was calculated as follows:
\begin{equation}
\label{DCScalcdata}
\frac{d\sigma}{d\Omega}=\frac{N(p_\Sigma, \cos \theta_{\text{CM}})}{\rho \cdot N_A \cdot L(p_\Sigma) \cdot \bar{\varepsilon}(p_\Sigma, \cos \theta_{\text{CM}}) \cdot \Delta\Omega},
\end{equation}
where $\rho$ and $N_A$ represent the density of the $\text{LH}_2$ target, 0.071~g$/\text{cm}^3$, and Avogadro's number, respectively.
$L(p_\Sigma)$ is the total flight length of the incident $\Sigma^+$ in the $\text{LH}_2$ target. $\bar{\varepsilon}$ represents the efficiency of the $\Sigma^+ p$ scattering event averaged for the vertex position. $\Delta \Omega$ represents a constant solid angle of $\Delta \Omega=2\pi \Delta \cos \theta_{\text{CM}}$.
\red{The following subsections describe} the evaluation of each factor, where finally, the differential cross sections were derived.
\subsection{Total track length of the incident $\Sigma^+$ in the $\text{LH}_2$ target}
\label{TotalLengthtext}
\begin{figure}[!h]
\begin{center}
\includegraphics[width=4in]{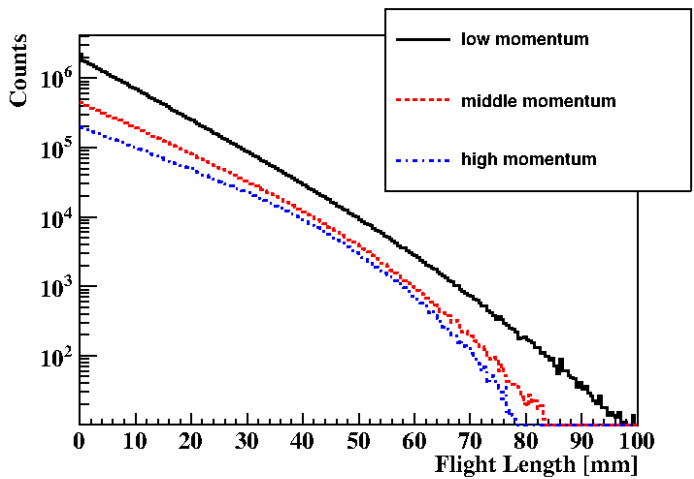}
\end{center}
\caption[Distribution of estimated $\Sigma^+$ track lengths.]{Distribution of estimated $\Sigma^+$ \red{track lengths} for the three momentum regions.}
\label{TotalLength}       
\end{figure}

In ordinary scattering experiments, the expression $\rho \cdot N_{Avo}\cdot t \cdot N_{\text{beam}}$ is used for the luminosity, where $t$ and $N_{\text{beam}}$ represent a target thickness and number of beam particles, respectively. However, this evaluation was inappropriate in this experiment because the incident $\Sigma^+$ was produced in the $\text{LH}_2$ target, and primarily decayed inside the target. The direct measurement of the $\Sigma^+$ track length, event-by-event, is also difficult because of the limited acceptance of the decay proton. However, the total track length of the incident $\Sigma^+$ can be reliably evaluated using a Monte Carlo simulation. Information regarding the production vertices and momentum vectors of all identified incident $\Sigma^+$ particles was obtained from the spectrometer analysis. $\Sigma^+$ particles with measured momenta were generated at the production points in the simulation. The flight length of $\Sigma^+$ was subsequently summed until $\Sigma^+$ decayed or exited the target. \Figure{TotalLength} shows the estimated $\Sigma^+$ track length distribution. The total track lengths for each momentum region $L(p_\Sigma)$ were obtained by integrals of these histograms. The background contribution in the $\Sigma^+$ identification was also estimated from the sideband event in the $m^2$ distribution.
Table \ref{tab:TotalLength} summarizes the estimated total track lengths after subtracting background contributions.
\begin{table}[!h]
\begin{center}
\caption{Estimated $\Sigma^+$ total track lengths for the three momentum regions: low-momentum region ($0.44<p_\Sigma [\text{GeV}/c]<0.55$), middle-momentum region ($0.55<p_\Sigma [\text{GeV}/c]<0.65$), and high-momentum region ($0.65<p_\Sigma [\text{GeV}/c]<0.80$). \red{The values} in the row of ``All events'' include the contributions from the miscalculated background events. The background contributions are listed in the row of ``Sideband BG''. By subtracting ``sideband BG'' from ``All events'', the $\Sigma^+$ track length was calculated and is listed in the row labeled ``$\Sigma^+$''.}
\begin{tabular}{cccc}\hline
\label{tab:TotalLength} 
 Region & Low &Middle&High \\ \hline
All events [cm] & $3.69\times 10^7 $ & $1.13\times 10^7 $ & $6.70 \times 10^6 $ \\
Sideband BG [cm] & $0.27\times 10^7 $ & $0.12\times 10^7 $ & $0.86 \times 10 ^6 $ \\ \hline \hline
$\Sigma^+$ [cm] & $3.42\times 10^7 $ & $1.00\times 10^7 $ & $5.84 \times 10^6 $ \\ \hline
\end{tabular}
\end{center}
\end{table}

In this procedure, the simulation inputs of the vertex point and momentum vector contained uncertainties owing to the resolution and systematic errors of the spectrometers. This may have caused uncertainties in the estimated track length.
Such uncertainties were estimated to be \red{3\% at maximum}, which is similar to that of the $\Sigma^-$ case \cite{Miwa:2021}.  This uncertainty is considerably smaller than other uncertainties, such as the statistical errors shown in \Fig{SPcounts}.
\subsection{Average efficiency of the $\Sigma^+ p$ scattering events including the detection and analysis efficiency}
The average efficiency $\bar{\varepsilon}$ for the $\Sigma^+ p$ scattering events, including the detection and analysis efficiencies, was evaluated by analyzing the simulated data with the same analyzer program for the real data. The difference in the detection efficiency of CATCH for protons between the simulation and real data should be considered. In the following, we first discuss the evaluation of the CATCH efficiency. \red{Then, the average efficiency $\bar{\varepsilon}$ is obtained by correcting the difference between the simulation and real data.}
\subsubsection{Detection efficiency of CATCH}
\label{CATCHefftext}
The detection efficiency of CATCH includes the geometrical acceptance, tracking efficiency of CFT, and the energy measurement efficiency for protons. They depend on the angle in the laboratory frame $\theta_{\text{lab}}$, the kinetic energy $E$, and \red{$z$-vertex} position $z$. The efficiencies were evaluated based on the $pp$ scattering data taken for calibration, by irradiating the $\text{LH}_2$ target with proton beams of seven momenta between $0.45$ and $0.85$~GeV/$c$. In parallel, we estimated the proton efficiency in a Monte Carlo simulation, where protons with arbitrary angles and energies can be generated. 

\begin{figure}[!h]
\centerline{\subfloat[Energy dependence of the energy measurement efficiency]{
\includegraphics[width=3in]{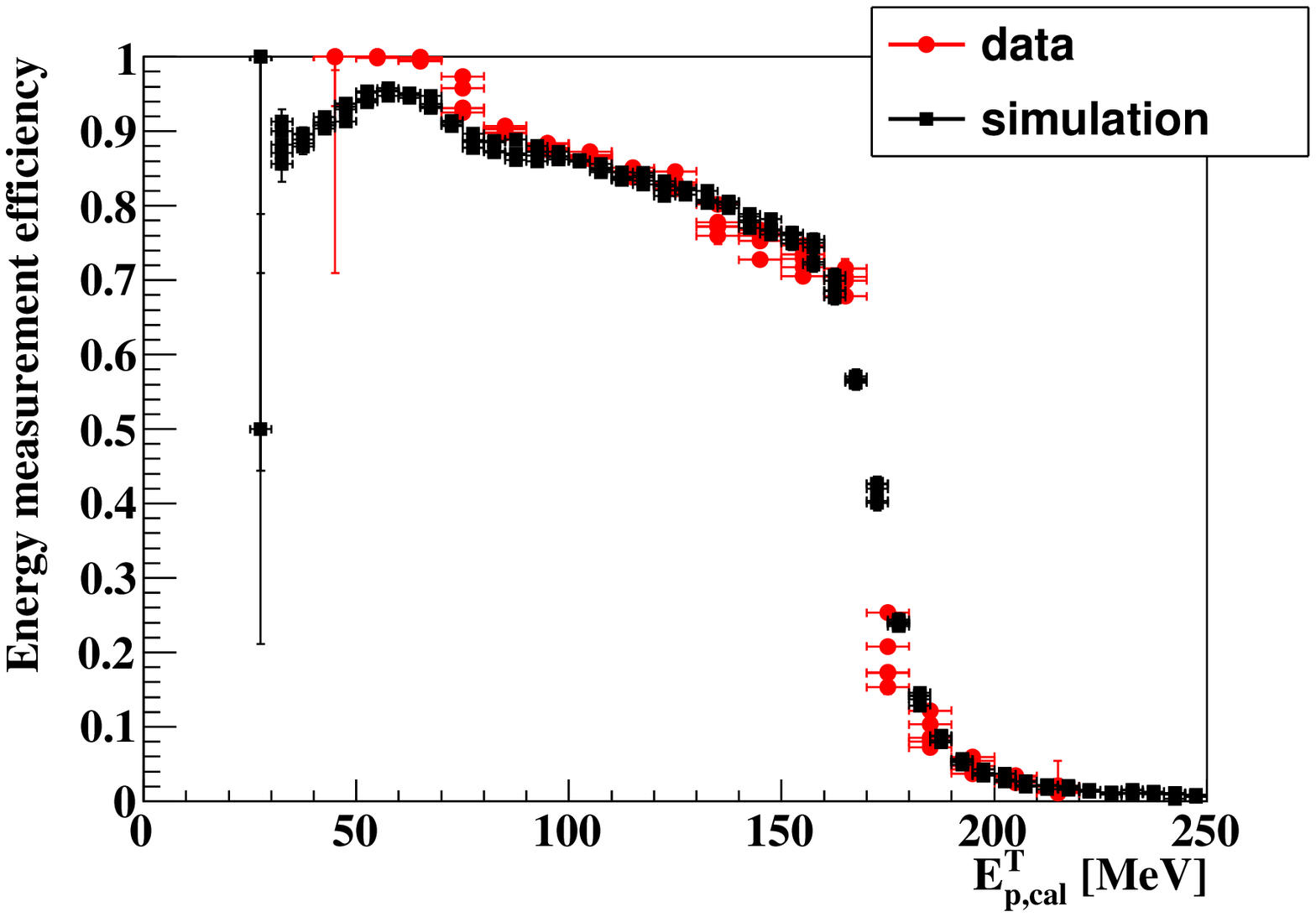}
\label{BGOeffproj}
}\hfil
\subfloat[Energy measurement efficiency]{
\includegraphics[width=3in]{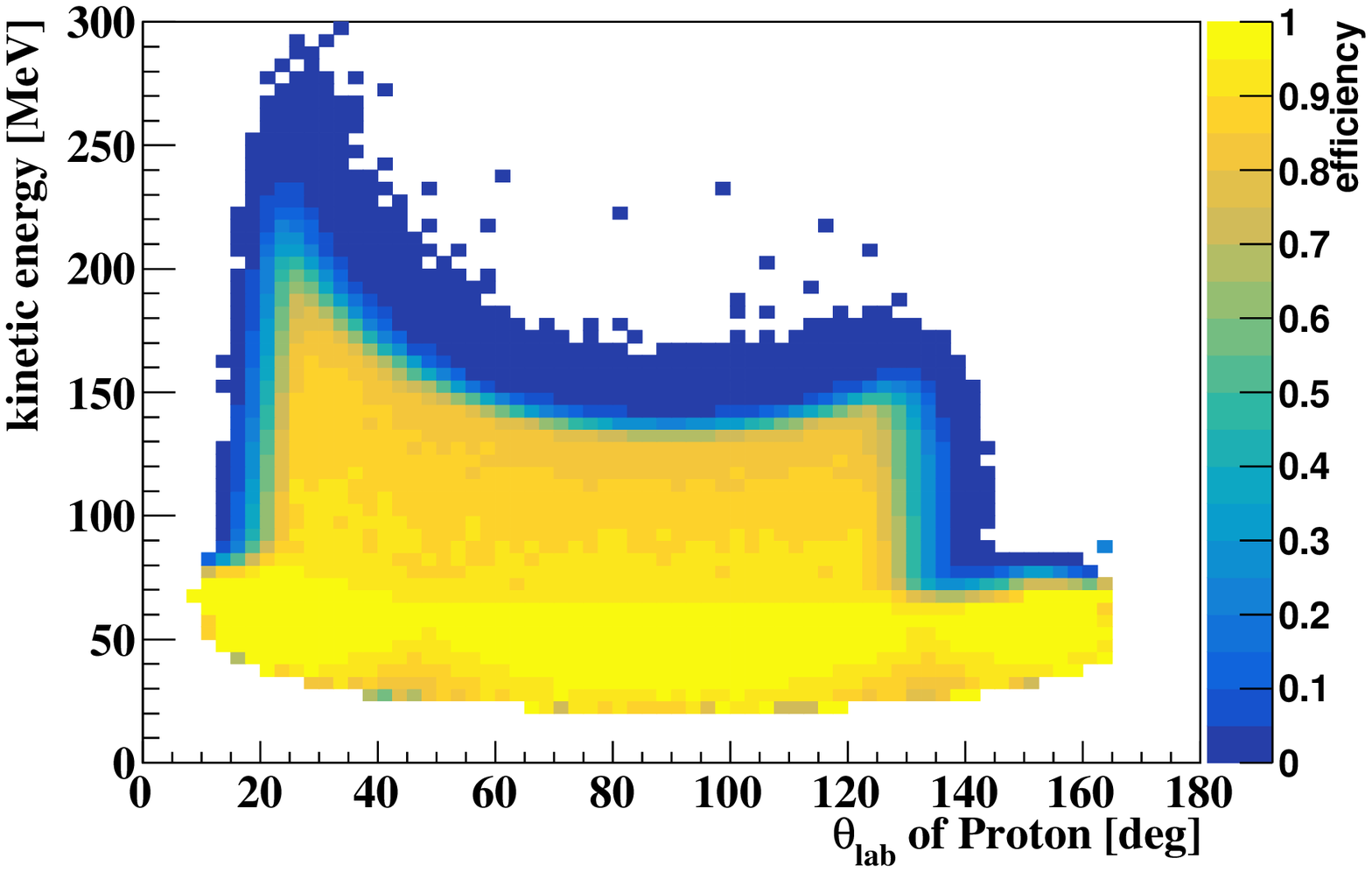}
\label{BGOeff}}}
\caption{ (a) \red{Energy measurement efficiency as a function of the kinetic energy at the scattering angle of $\theta_{\text{lab}}=37^\circ$}. The red circles and black squares represent the data and simulation, respectively.  The simulated efficiency reproduces the data well.
(b) Energy measurement efficiency map evaluated by the simulation.}
\label{BGOeffall}
\end{figure}

The procedures for the efficiency evaluation using the $pp$ scattering data are as follows:
\begin{enumerate}
\item For the efficiency estimation, at least one proton among the two protons in the final state must be detected by CATCH. From the kinematics, the momentum ($\bm{p}' (\theta')$) of the detected proton with the recoil angle $\theta'$ was calculated.
The scattering vertex $(x_{\text{scat}},y_{\text{scat}},z_{\text{scat}})$ was also reconstructed as the closest point between the beam and recoil proton tracks. 
\item The momentum vector of the other proton was obtained as $\bm{p}=\bm{p}_{\text{beam}}-\bm{p'}(\theta')$, where the $\bm{p}_{\text{beam}}$ was the proton beam momentum analyzed by the K1.8 beam-line spectrometer. From the momentum vector $\bm{p}$, the angle and kinetic energy of the second proton can be predicted; they are denoted as $\theta$ and $E$, respectively. 
\item The CATCH efficiency was estimated by checking whether the predicted track and energy were measured or not. The tracking and energy measurement efficiencies were derived separately.
\end{enumerate}
First, we explain energy measurement efficiency  $\varepsilon_{\text{BGO}}(\theta,E, z_{\text{scat}})$. In this case, we checked whether the measured energy for the predicted track agreed with the predicted $E$ within 40~MeV. The obtained $\varepsilon_{\text{BGO}}(\theta,E, z_{\text{scat}})$ \red{at} $\theta=37^\circ$  is shown as the red points in \Fig{BGOeffall} (a) as an example. This efficiency was compared with the efficiency estimated from the simulation. As shown in \Fig{BGOeffall} (a), the simulation-based efficiency accurately reproduced the data-based efficiency. Therefore, the simulation-based efficiency for the energy measurement was used for further analysis to cover the entire $\theta,E,$ and $z_{\text{scat}}$ regions, as shown in \Fig{BGOeffall} (b).

\begin{figure}
\centerline{\subfloat[Energy dependence of the tracking efficiency]{
\includegraphics[width=3in]{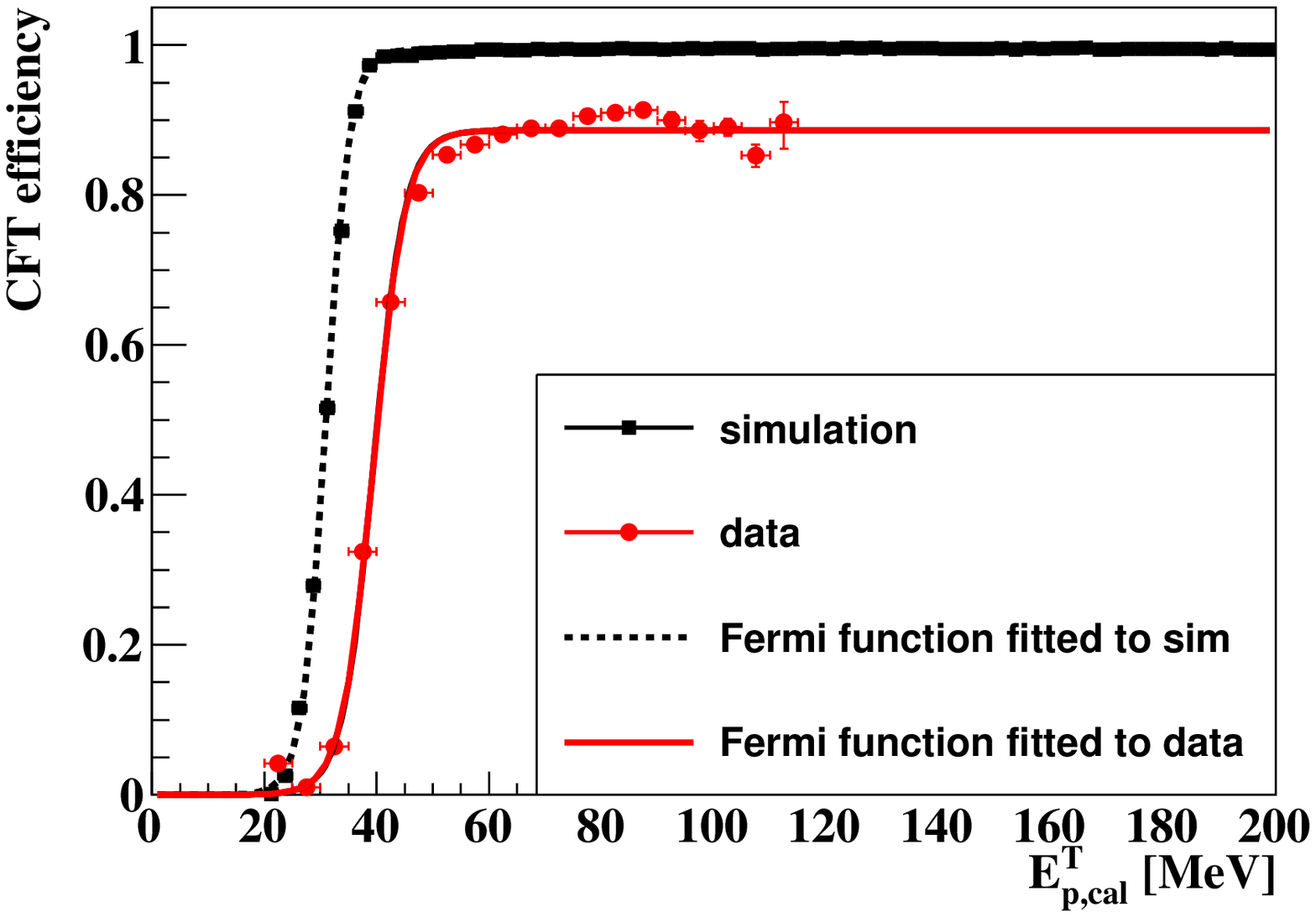}
\label{CFTeffproj}
}\hfil
\subfloat[Tracking efficiency including the geometrical acceptance]{
\includegraphics[width=3in]{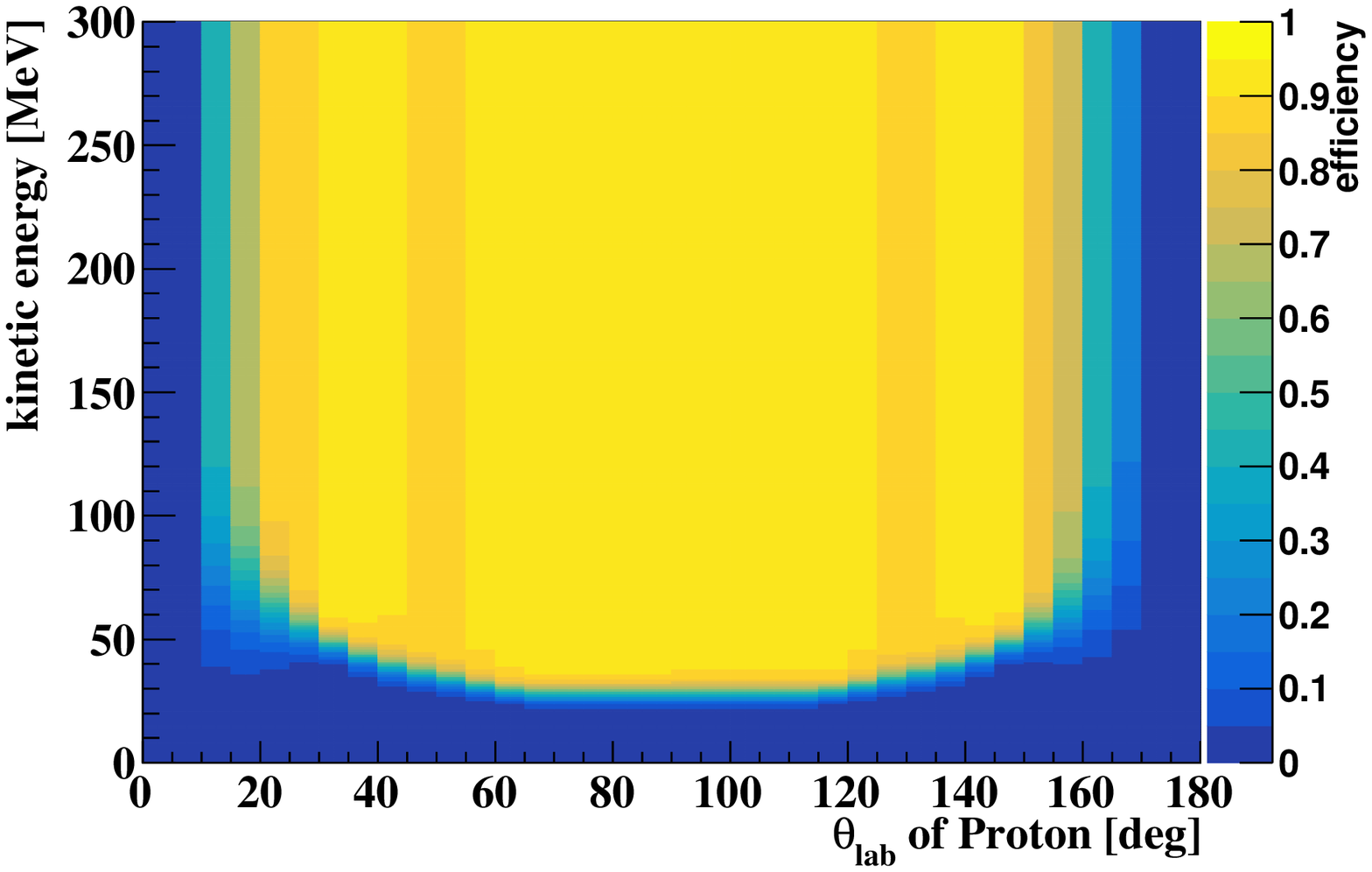}
\label{CFTeff}
}}
\caption{(a) \red{CFT tracking efficiency as a function of the kinetic energy at the scattering angle of $\theta_{\text{lab}}=54^\circ$.} The red circles and black squares represent the data and simulation, respectively. The red and black curves are the fit functions obtained for the data and simulation. (b) \red{Data-based} tracking efficiency map used for analysis, \red{which was calculated from Eq. (\ref{CFTformula}) and parameters were determined by the fitting to the real data.} }
\label{CFTeffall}
\end{figure}

Hereafter, we describe the tracking efficiency of CFT $\varepsilon_{\text{CFT}}(\theta,E,z)$. This was evaluated by checking whether a  track with the predicted direction was detected. Therefore, the effect of detector acceptance was also included in $\varepsilon_{\text{CFT}}$. The energy dependence of the tracking efficiencies estimated from the simulation and $pp$ scattering data is shown by the black and red points in \Fig{CFTeffall} (a), respectively. Because CFT tracking required at least six layer hits, the efficiency decreased sharply at low energies. This energy dependence of the efficiency can be phenomenologically represented by the Fermi function for both the data and simulation. The efficiency was then formulated as follows:
\begin{equation}
\varepsilon_{\text{CFT}}(\theta,E,z)=\frac{\varepsilon_{\text{max}}(\theta,z)}{1+\exp\Bigl{(}\frac{E-E_{\text{half}}(\theta)}{d(\theta)}\Bigr{)}}, \label{CFTformula}
\end{equation}
where $\varepsilon_{\text{max}}(\theta, z), d(\theta)$ and $E_{\text{half}}(\theta)$ are parameters representing the maximum efficiency, diffusion, and kinetic energy with half efficiency, respectively. These parameters were determined by fitting \red{Eq. (\ref{CFTformula}) to} the estimated efficiency, as indicated by the solid red line in \Fig{CFTeffall} (a).
 The realistic efficiency is slightly lower than that of the simulation, typically by 10\%. This difference is attributed to the geometrical effect of the fiber placement in the $uv$ layers of CFT. There are ineffective regions for tracks with scattering angles of approximately  $45^\circ$, owing to the zigzag fiber configuration in the $uv$ layers. This is illustrated in Figs. 42 and 43 in \cite{Akazawa:2021}. In addition, both the kinetic energy with half efficiency $E_{\text{half}}$ and diffusion parameter $d$ were slightly larger than those in the simulation. These differences indicated that the realistic amount of material in the experimental setup was larger than that considered in the simulation. It was difficult to incorporate the real spiral fiber configuration into the CFT $uv$ layers and the missing amount of  material within the simulation. Therefore, the data-based efficiency for CFT tracking was used for the analysis of the cross section. \Figure{CFTeffall} (b) shows the efficiency map as a function of $\theta$ and $E$.

\begin{figure}[!h]
\begin{center}
\includegraphics[width=4in]{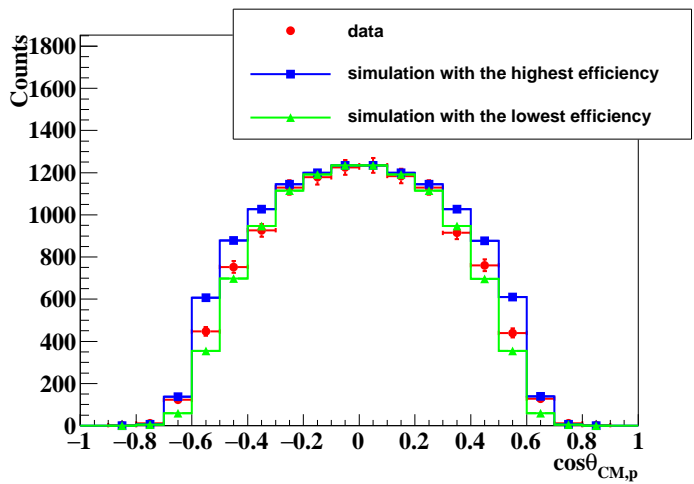}
\end{center}
\caption{Angular distribution of the protons from the $pp$ scattering following the $\Sigma^+ \to p \pi^0$ decay events. The red points represent the data. The green and blue points show the \red{simulation results} with the efficiency correction using the possible lowest and highest CFT tracking efficiencies, respectively. The simulations are normalized by the counts at $0<\cos \theta_{\text{CM},p}<0.1$.}
\label{ppangular}       
\end{figure}

The obtained efficiency map was checked by deriving the differential cross sections for the calibration $pp$ scattering data.
The derived values agreed with the reference values to within 5\%, except for the acceptance edge. 
To estimate the effect of the uncertainty of the efficiency in the acceptance edge for the $\Sigma^+p $ scattering analysis, the possible lowest and highest CFT tracking efficiencies were also estimated by changing the parameters within a reasonable range, i.e., \red{changing $d$ within 20\%} and $E_{\text{half}}$ within 4 MeV. The validity of the margin of efficiency defined by the lowest and highest cases was additionally verified using another calibration reaction. This calibration reaction was that of $pp$ scattering following the $\Sigma^+ \to p \pi^0$ decay, which has been described as the background event for the $\Sigma^+ p$ scattering up to now. From the data analysis, the angular distribution of the recoil proton was obtained, shown by the red points in \Fig{ppangular}. This angular distribution was compared with a Monte Carlo simulation, including the secondary $pp$ scattering process with a realistic angular distribution. While analyzing the simulated data, the data-based CFT tracking efficiencies for the lowest and highest cases were considered. We then confirmed that the angular distribution in the data was sandwiched \red{between the two distributions estimated using the highest and lowest efficiencies, as shown by the blue and green points in \Fig{ppangular}}. In the next sub-subsection, the detection efficiency for $\Sigma^+ p$ scattering was corrected using these two efficiencies, and the difference was considered to be the systematic uncertainty.

\subsubsection{Evaluation of average efficiency for the $\Sigma^+ p$ scattering events considering the real detection efficiency}
\label{Aveefftext}
\begin{figure}[!h]
\begin{center}
\includegraphics[width=6.0in]{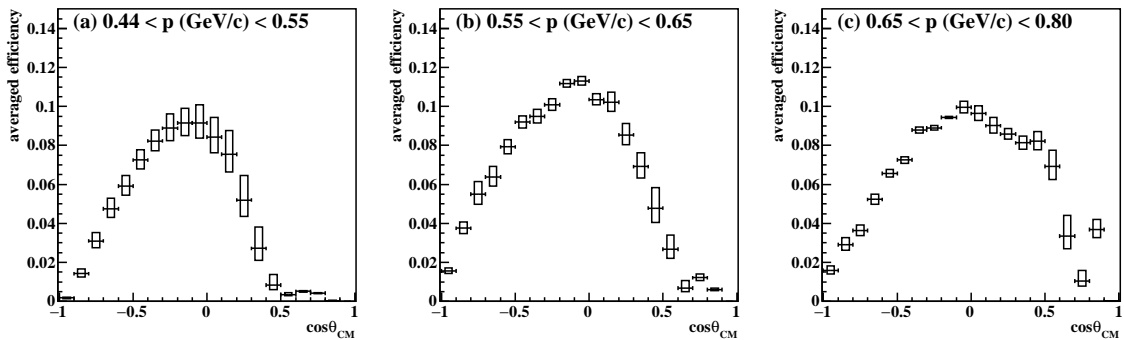}
\end{center}
\caption{Averaged efficiencies for the $\Sigma^+ p$ scattering for each scattering angle and momentum region. \red{The vertical error} represents the difference in the obtained efficiency using the two possible lowest and highest CFT tracking efficiencies. \red{The horizontal error corresponds to the angular step of $\Delta \cos \theta_{\text{CM}}=0.1$.}}
\label{SPeffs}       
\end{figure}

The average efficiency $\bar{\varepsilon}$ for the $\Sigma^+ p$ scattering events, including the detection and analysis efficiencies, was evaluated by analyzing the simulated data with the same analyzer program for the real data. $\bar{\varepsilon}$ is defined as follows:
\begin{align}
\bar{\varepsilon}(p_\Sigma, \cos \theta_{\text{CM}})&=\frac{N_{\text{analyzed}}(p_\Sigma, \cos \theta_{\text{CM}})}{N_{\text{generated}, \Sigma^+ \text{id}}(p_\Sigma, \cos \theta_{\text{CM}})} \label{eff1} \\
&=\frac{N_{\text{detected}}(p_\Sigma, \cos \theta_{\text{CM}})}{N_{\text{generated}, \Sigma^+ \text{id}}(p_\Sigma, \cos \theta_{\text{CM}})} \cdot \frac{N_{\text{analyzed}}(p_\Sigma, \cos \theta_{\text{CM}})}{N_{\text{detected}}(p_\Sigma, \cos \theta_{\text{CM}})} \notag \\
&=\bar{\varepsilon}_{\text{detect}}(p_\Sigma, \cos \theta_{\text{CM}}) \cdot \bar{\varepsilon}_{\text{ana}}(p_\Sigma, \cos \theta_{\text{CM}}).
\end{align}
$N_{\text{generated}, \Sigma^+ \text{id}}$ in Equation (\ref{eff1}) represents the number of generated $\Sigma^+ p$ scattering events in the simulation. $\Sigma^+$ identification from missing mass analysis in the analyzer program is required. $N_{\text{analyzed}}$ represents the number of identified $\Sigma^+ p$ scattering events that satisfy all cut conditions for $\Sigma^+ p$ scattering for the two-proton events. The effect of the branching ratio of the $\Sigma^+ \to p\pi^0$ decay is also included in $N_{\text{analyzed}}$.
$\bar{\varepsilon}$ \red{is factorized by} the detection efficiency in CATCH, $\varepsilon_{\text{detect}}$, and the analysis cut efficiency, $\varepsilon_{\text{ana}}$, from the second equation. 
 $N_{\text{detected}}$ represents the number of events in which the two protons were detected by CATCH for the tagged $\Sigma^+$ events. The difference in the CFT tracking efficiency between the data, $\varepsilon_{\text{CFT}}^{\text{data}}$, and simulation, $\varepsilon_{\text{CFT}}^{\text{sim}}$, was corrected by changing $\bar{\varepsilon}_{\text{detect}}$ as follows:
\begin{equation}
\bar{\varepsilon}_{\text{detect}} \mapsto \frac{1}{N_{\text{generated}, \Sigma^+ \text{id}}}\Bigl{(} \sum_{\text{events}}^{N_\text{detected}} \frac{\varepsilon_{\text{CFT}}^{\text{data}}(\theta_{p_1},E_{\text{cal},p_1}, z_{\text{scat}}) \cdot \varepsilon_{\text{CFT}}^{\text{data}}(\theta_{p_2},E_{\text{cal},p_2}, z_{\text{decay}})}{\varepsilon_{\text{CFT}}^{\text{sim}}(\theta_{p_1},E_{\text{cal},p_1}, z_{\text{scat}}) \cdot \varepsilon_{\text{CFT}}^{\text{sim}}(\theta_{p_2},E_{\text{cal},p_2}, z_{\text{decay}})} \Bigr{)}, 
\end{equation}
where the efficiency correction for both protons was considered. The analysis cuts are explained in subsection \ref{Cutconditiontext}; and the analysis efficiency for the entire angular region is summarized in Table \ref{tab:cutcondition}. The efficiency of the analysis was estimated for each angular region.
The efficiencies obtained are presented in \Fig{SPeffs}. \red{The vertical error} represents the difference between the lowest and highest possible CFT tracking efficiencies, as mentioned in the previous sub-subsection. The angular dependence of the efficiency can be understood from the kinetic energies of protons. The efficiency decreases for the forward scattering angle, because the tracking efficiency also decreases for recoil protons with a lower kinetic energy. Similarly, the kinetic energy of the decay proton decreases for the backward angle, and the efficiency therefore decreases for the backward angle. The errors in the efficiency were considered \red{as} systematic errors in the derivation of the differential cross sections.
\subsection{Differential cross sections}
\begin{figure}[!h]
\begin{center}
\includegraphics[width=6.0in]{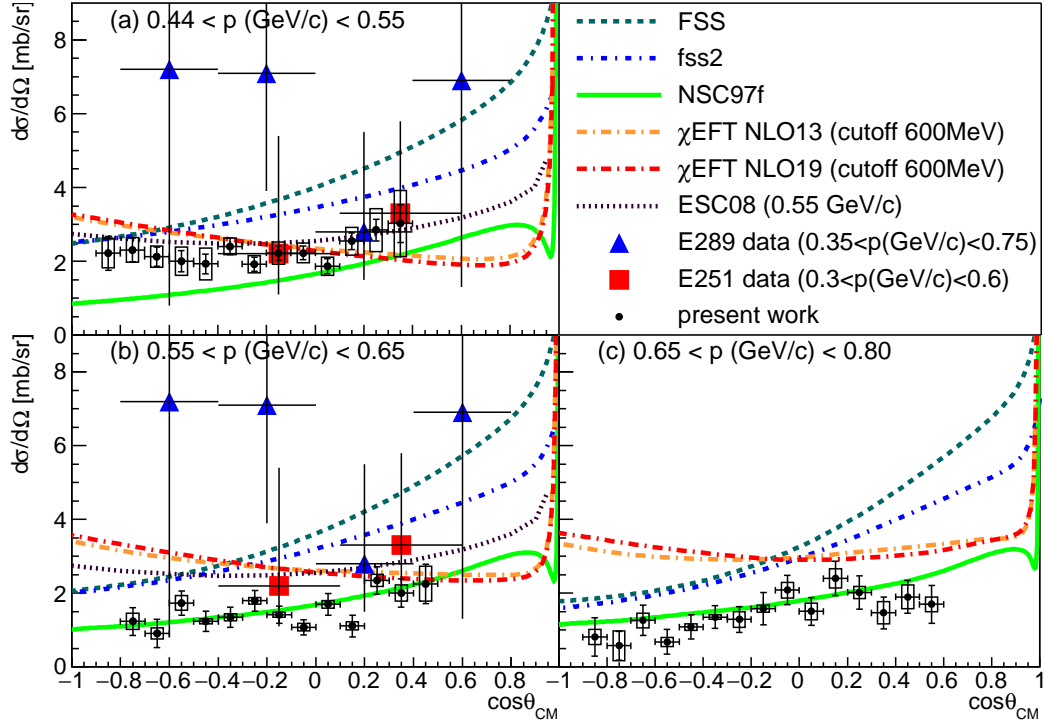}
\end{center}
\caption{Derived differential cross sections of the $\Sigma^+ p$ scattering for the three momentum regions. The error bars and boxes show the statistical and systematic uncertainties, respectively. The red boxes and blue triangles represent the data of past measurements, KEK E251 \cite{Goto:1999} and KEK E289 \cite{Kanda:2005}, respectively. 
The blue dotted and dot-dashed lines show the calculations from FSS and fss2 \cite{Fujiwara:2007}, respectively. The green-solid lines and black-dashed lines show the calculations from the Nijmegen NSC97f \cite{Rijken:1999} and ESC08 \cite{Rijken:2010} models, respectively. The orange and red dot-dashed lines show the calculations from the $\chi$EFT NLO models \cite{Haidenbauer:2013} \cite{Haidenbauer:2020}.}
\label{SPDCS}       
\end{figure}

The differential cross section was calculated using \red{Eq. (\ref{DCScalcdata})}. The obtained differential cross sections for the three incident $\Sigma^+$ momentum regions are shown as black circles in \Fig{SPDCS}. The mean momenta of the three momentum regions are 0.50, 0.59, and 0.71 GeV/$c$, respectively. The error bars and boxes of the data points represent statistical and systematic uncertainties, respectively. The systematic error was estimated as the quadratic sum of the error from the background estimation, average efficiency, and $\Sigma^+$ total flight length. The values of the differential cross sections and their uncertainties are summarized in Tables \ref{tab:DCSlow}, \ref{tab:DCSmid}, and \ref{tab:DCShigh} in Appendix \ref{Tablesdiff}. For the lower two momentum regions, past measurements at KEK PS are plotted in \Fig{SPDCS} \red{with} red boxes \cite{Goto:1999} and blue triangles \cite{Kanda:2005}. The data quality in the present experiment was improved significantly. Thus, a meaningful comparison with theories has become possible. The angular dependences are isotropic for the present angular regions, especially for low momentum. Moreover, the obtained values of the differential cross sections are not as large as those predicted by the fss2 and FSS models based on the QCM in the short-range region \cite{Fujiwara:2007}, as discussed in the next section.

\section{Discussion}
\label{discussionsect}
\subsection{Comparison with theoretical calculations}
\label{comparetext}
The obtained data were compared with the theoretical calculations, which are overlaid as lines in \Fig{SPDCS}. 

The blue dotted and dot-dashed lines show the calculations from the FSS and fss2 models, which include the QCM in the short-range region \cite{Fujiwara:2007}. Verification of the predicted large repulsive force originating from the quark Pauli effect is an important motivation for this experiment. The difference in the strength of the quark Pauli effect between the two models is attributed to the size parameter, which defines the size of the quark cluster in baryons. The FSS model, using the larger size parameter, predicts a repulsive interaction, which increases the differential cross section. However, the predictions by FSS and fss2 are much larger than the present data, indicating that the repulsive forces in FSS and fss2 are too large and unrealistic.

The green solid and black-dashed lines show the predictions from the Nijmegen NSC97f \cite{Rijken:1999} and ESC08 \cite{Rijken:2010} models, respectively, based on the boson-exchange picture. Historically, in the Nijmegen models, it has been difficult to describe the repulsive nature of the $\Sigma N (I=3/2, ^3S_1)$ channel. Although NSC97f agrees well with our data in terms of the differential cross sections, it predicts an attractive $\Sigma^+ p$ interaction, which does not agree with the current common understanding of the $\Sigma N$ interaction. In ESC08, additional repulsive effects, including the quark picture, are considered by making an effective Pomeron potential as the sum of a pure Pomeron exchange and a Pomeron-like representation of the Pauli repulsion. Subsequently, ESC08 predicts a moderate repulsive force within this channel. Although there were sizable discrepancies between our data and ESC08, especially in the middle-momentum region, ESC08 was closer to the data than fss2. This suggests that the size of the repulsive force used in ESC08 is reasonable.

The orange and red dot-dashed lines show the calculations using the $\chi$EFT models extended to the $YN$ sector (NLO13 \cite{Haidenbauer:2013} and NLO19 \cite{Haidenbauer:2020}, respectively), which use different sets of LECs. In both cases, a cutoff value of 600 MeV was used. The LECs are essential parameters of the $\chi$EFT models, representing the short-range part of the interaction, and should be determined from the experimental data. At present, the LECs for $S$ waves have been determined based on existing hyperon-proton scattering data in the low-momentum region. However, the LECs for $P$ waves have not been well-constrained owing to the lack of experimental data, especially for the momentum region around the present data. At present, $\chi$EFT predicts much larger cross sections, especially in the higher-momentum region. Our data are used to determine the LECs for $P$ waves in the $\chi$EFT models.

For the first time, we presented \red{precise} data for the $\Sigma^+ p$ channel in the higher-momentum range.
Currently, no theoretical model can reproduce our data consistently for the three momentum regions. This was mainly because of the lack of \red{precise} data. Therefore, our data are essential inputs for improving these theoretical calculations in order to become realistic $BB$ interaction models.

\subsection{Numerical phase-shift analysis}
To extract the phase shifts of the $\Sigma^+ p$ interaction, particularly for the $^3S_1$ channel, from the obtained differential cross sections, a phase-shift analysis was performed based on a general formulation of the scattering problem in quantum mechanics. This was the first application of hyperon-nucleon scattering data, whereas precise phase-shift analysis has been performed for the $NN$ scattering data to derive the phase shifts for each partial wave \cite{Arndt:2000}.
The differential cross sections can be represented as a function of the phase shifts for the 27-plet $\bm{\delta}_{[27]}$ and 10-plet $\bm{\delta}_{[10]}$, scattering angle $\theta_{\text{CM}}$, and momentum $p_{\text{CM}}$ in the CM system (see Appendix \ref{PAformula}). This function is denoted by $I_0(\theta_{\text{CM}},p_{\text{CM}},\bm{\delta}_{[27]}, \bm{\delta}_{[10]})$. In our analysis, partial waves up to $D$ are considered. The phase shifts for the 27-plet are taken up to five total spin states: $\bm{\delta}_{[27]}=\{\delta_{^1S_0}, \delta_{^3P_2}, \delta_{^3P_1}, \delta_{^3P_0}, \delta_{^1D_2}\} $. For the 10-plet, five phase shifts and a mixing parameter for $^3S_1$-$^3D_1$ mixing, that is, $\bm{\delta}_{[10]}=\{\delta_{^3S_1}, \delta_{^1P_1}, \delta_{^3D_3}, \delta_{^3D_2}, \delta_{^3D_1}, \epsilon_1\}$, are included. The function $I_0(\theta_{\text{CM}},p_{\text{CM}},\bm{\delta}_{[27]},\bm{\delta}_{[10]})$ has 11 phase-shift parameters. To extract meaningful information from the fitting of the differential cross sections, the number of phase shifts to be fitted should be reduced.

The phase shifts $\bm{\delta}_{[27]}$ can be constrained with reliability because $\bm{\delta}_{[27]}$ becomes identical to the phase shifts in the $NN  (I=1)$ channel in the limit of flavor SU(3) symmetry. In this limit, $\bm{\delta}_{[27]}$ can be obtained from the phase shifts of  $pp$ scattering for the corresponding momentum. However, in reality, $\bm{\delta}_{[27]}$ in the $\Sigma^+ p$ scattering should be slightly different from that in the $pp$ scattering, owing to the breaking of flavor symmetry. In fact, all theories (FSS, fss2, ESC, and NSC97f) predict smaller $^1S_0$ phase shifts in $\Sigma^+ p$ scattering than those in $pp$ scattering. However, the difference between the theoretical predictions of $\bm{\delta}_{[27]}$ is small because these models are also constrained by $pp$ scattering data. In this analysis, the effect of the uncertainty in $\bm{\delta}_{[27]}$ was examined using three different sets of $\bm{\delta}_{[27]}$. The phase-shift values of $pp$ scattering and theoretical predictions in ESC16 \cite{Nagels:2019} and NSC97f \cite{Rijken:1999} were used in this study.

In contrast, the phase shifts $\bm{\delta}_{[10]}$ are unique to the $\Sigma N (I=3/2)$ channel, and these phase shifts should be determined from the fitting. The theoretically uncertain two phase shifts $\delta_{^3S_1}$ and $\delta_{^1P_1}$, representing the short-range interaction, were regarded as free parameters. For the remaining phase shifts, namely $\delta_{^3D_3},\delta_{^3D_2}, \delta_{^3D_1}$, and $\epsilon_1$, the variation among the theoretical models is rather small because the pion-exchange mechanism is expected to be dominant for long-range interactions. Therefore, these phase shifts were fixed \red{at} the theoretical values as an approximation.

In summary, the two phase-shift parameters, $\delta_{^3S_1}$ and $\delta_{^1P_1}$, were obtained by fitting the differential cross sections with the function $I_0(\theta_{\text{CM}},p_{\text{CM}},\bm{\delta}_{[27]},\bm{\delta}_{[10]})$. To study the effect of uncertainties due to the assumed fixed phase shifts, fitting was performed for three conditions with different sets of fixed parameters.
\begin{itemize}
\item[A] $\bm{\delta}_{[27]}$ was fixed at values taken from the $pp$ scattering. $\delta_{^3D_3},\delta_{^3D_2}, \delta_{^3D_1}$, and $\epsilon_1$ were fixed at 0.
\item[B] $\bm{\delta}_{[27]}$ was fixed at values from the ESC16 or NSC97f models. $\delta_{^3D_3},\delta_{^3D_2}, \delta_{^3D_1}$, and $\epsilon_1$ were fixed at 0.
\item[C] $\bm{\delta}_{[27]}$ was fixed at values from the ESC16 or NSC97f models. $\delta_{^3D_3},\delta_{^3D_2}, \delta_{^3D_1}$, and $\epsilon_1$ were fixed at the values from ESC16 or NSC97f.
\end{itemize}
By comparing conditions A and B, the effect of uncertainty in $\bm{\delta}_{[27]}$ was studied. From conditions B and C, the effect of uncertainty in the other parameters, $\delta_{^3D_3},\delta_{^3D_2}, \delta_{^3D_1}$, and $\epsilon_1$, was evaluated.
Although the sign of $\delta_{^3S_1}$ is expected to be negative, as predicted by recent theoretical models including ESC16, numerical fittings with a positive $\delta_{^3S_1}$ are possible using a different set of phase-shift parameters, such as NSC97f. To investigate the negative and positive $\delta_{^3S_1}$ cases, two different sets of fixed parameters were obtained from ESC16 and NSC97f, respectively. The fixed phase shifts are presented in Table \ref{tab:phaseshift}.
\begin{table}[!h]
\begin{center}
\caption[$\bm{\delta}_{[27]}$ and $\bm{\delta}_{[10]}$ for the $pp$ scattering and $\Sigma^+ p$ scattering in ESC16 and NSC97f.]{$\bm{\delta}_{[27]}$ and $\bm{\delta}_{[10]}$ for the $pp$ scattering and $\Sigma^+ p$ scattering in ESC16 \cite{Nagels:2019} and NSC97f \cite{Rijken:1999}, respectively. The units of  $p_\Sigma$ and $p_{\text{CM}}$ are [GeV/$c$]. $E_{\text{lab}}^{pp}$ (unit: [MeV]) represents the kinetic energy of the beam proton in the $pp$ scattering in which $p_{\text{CM}}$ is equal to that of the $\Sigma^+ p$ scattering. The units of phase shifts are [$^\circ$].}
\begin{tabular}{c|ccc|ccc|ccc}\hline
\label{tab:phaseshift} 
 &  & low & & & mid &  &  & high & \\ \hline
 &$pp$ & ESC16 & NSC97f &$pp$  & ESC16 & NSC97f &$pp$  & ESC16 & NSC97f  \\ \hline \hline
$p_\Sigma$ & 0.496 & 0.50 & 0.50 & 0.59 & 0.60 & 0.60 & 0.71 &0.70 &0.70 \\
$p_{\text{CM}}$& 0.214 & 0.216 &0.216 & 0.253 & 0.257 & 0.257 & 0.303 & 0.297 & 0.297 \\
$E_{\text{lab}}^{pp}$  & 87.6 & -- &-- & 122.1 & -- & -- &173.7 & -- & --\\ \hline \hline
$\delta_{^1S_0}$ & 27.9 &19.1 & 20.2 &19.5 &10.8 &11.8 &10.4 &2.80 &3.71 \\
$\delta_{^3P_2}$& 9.92 &6.76 &6.44 &12.7 &8.50 &8.02 &14.9 &9.82 &9.04 \\
$\delta_{^3P_1}$& $-12.2$ &$-13.2$ &$-13.3$ &$-15.5$ &$-16.9$ &$-17.1$ &$-19.5$ &$-20.8$ &$-21.0$ \\
$\delta_{^3P_0}$& 10.5 &7.19 &8.10 &7.29 &3.59 &4.49 &2.15 &$-0.92$ &$-0.23$ \\
$\delta_{^1D_2}$& 3.24 &3.38 &3.25 &4.71 &4.99 &5.02 &6.41 &6.61 &6.99 \\ \hline
$\delta_{^3S_1}$ & -- & $(-27.9)$ &(21.9)  & -- & $(-32.6)$ &(28.5) & -- & $(-36.6)$ &(35.01) \\ 
$\delta_{^1P_1}$ & -- &(8.33) &(12.2) & -- & (8.45) & (13.71) & -- & (7.46) &(13.7) \\
$\delta_{^3D_3}$ & -- &1.14 & 1.42 & -- &1.59 &2.29 & -- &1.93 &3.18 \\
$\delta_{^3D_2}$ & -- & $-3.53$ &$-3.23$ & -- &$-4.87$ &$-4.23$ & -- &$-6.42$ &$-5.31$ \\
$\delta_{^3D_1}$ & -- &1.35 &1.48 & -- &0.69 &1.30 & -- &$-0.70$ & 0.41\\
$\epsilon_1$ & -- &$-5.04$ &$-1.65$ & -- &$-5.24$ &0.11 & -- &$-5.14$ &1.87 \\ \hline \hline
\end{tabular}
\end{center}
\end{table}

\begin{figure}[!h]
\centerline{\subfloat[$\delta_{^3S_1}<0$ case]{\includegraphics[width=3.0in]{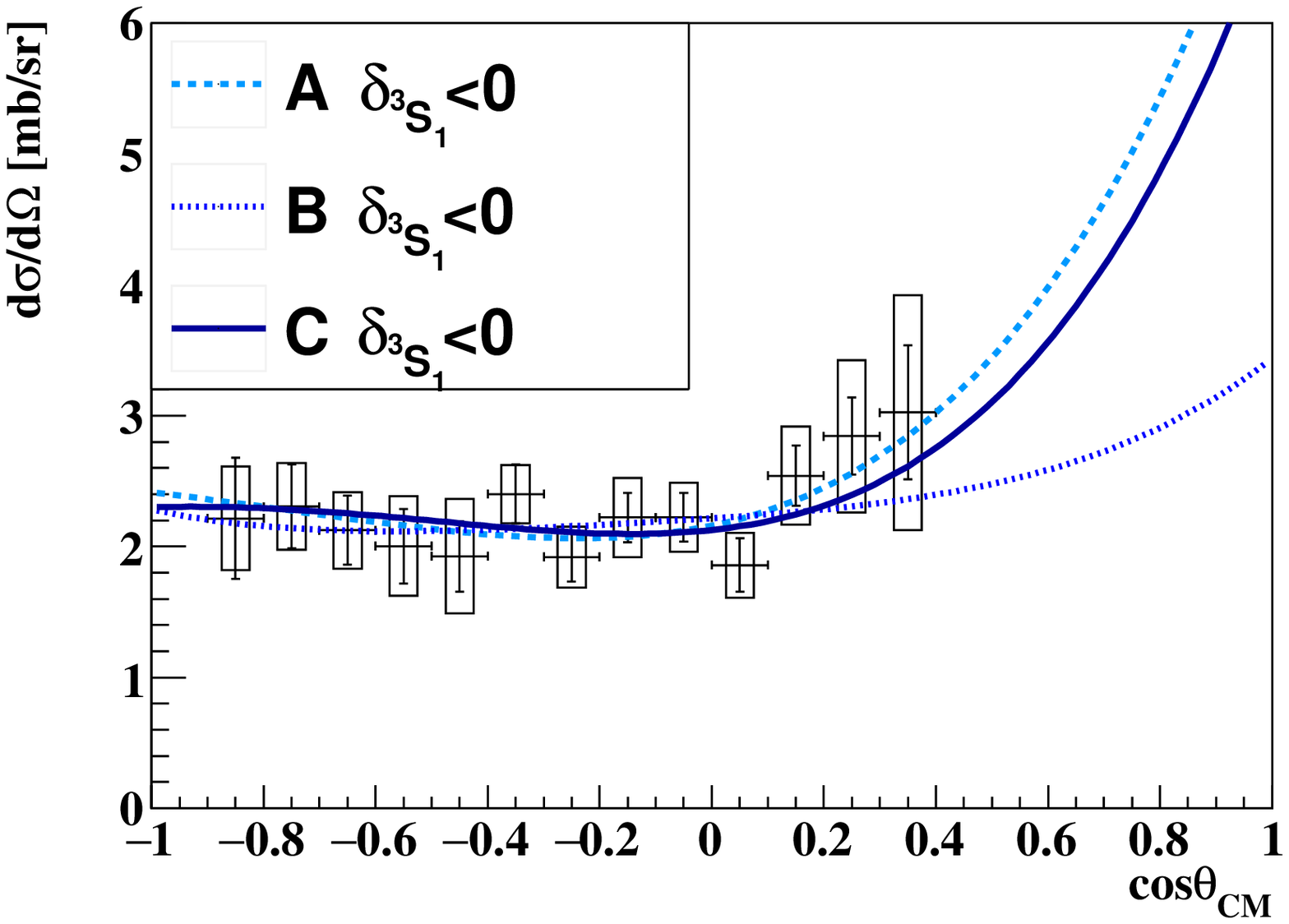}
\label{PAlowm}}
\hfil
\subfloat[$\delta_{^3S_1}>0$ case]{\includegraphics[width=3.0in]{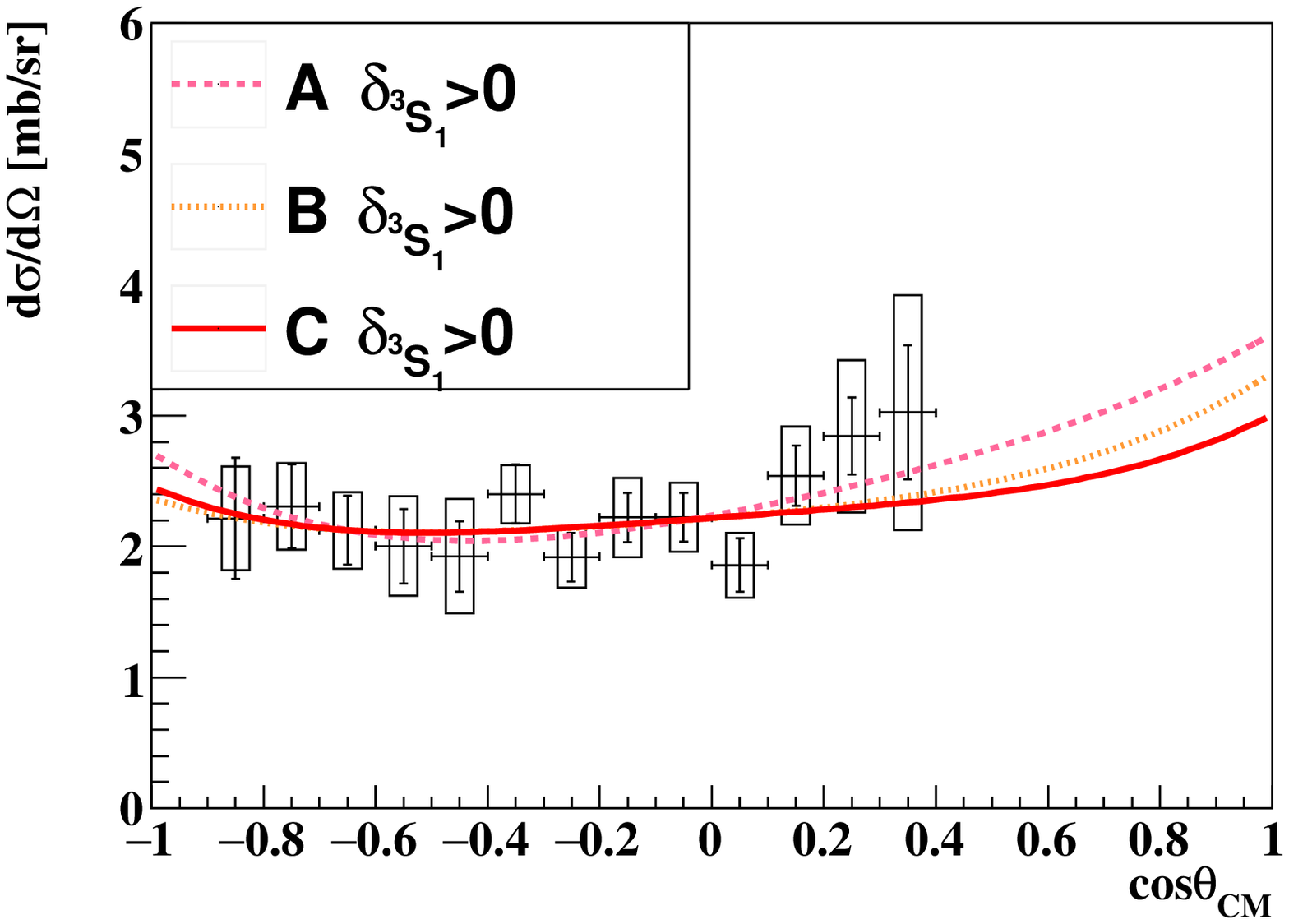}
\label{PAlowp}}}
\caption{\red{Differential cross section as a function of the scattering angle together with the calculated angular distribution in the phase-shift analysis} for the low-momentum region ($0.44<p_\Sigma [\text{GeV}/c] <0.55 $) \red{in} (a) the negative $\delta_{^3S_1}$ case  and (b) the positive case. Three lines in each graph show the fitting results with three different fitting conditions A, B, and C, as described in the text. The typical $\chi^2/$ndf is $4.4/11$.}
\label{PAlow}      
\end{figure}

\begin{figure}[!h]
\centerline{\subfloat[$\delta_{^3S_1}<0$ case]{\includegraphics[width=3.0in]{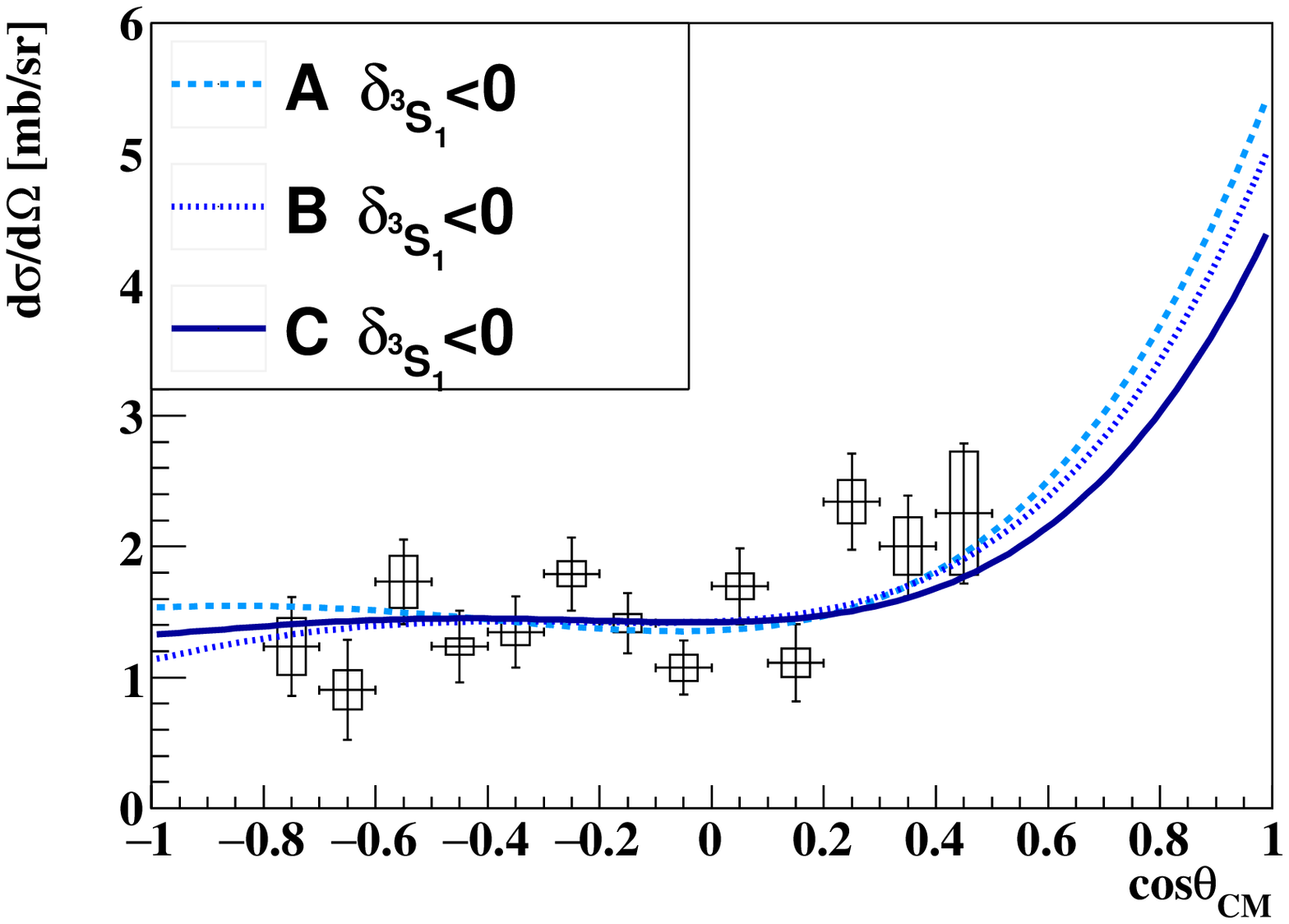}
\label{PAmidm}}
\hfil
\subfloat[$\delta_{^3S_1}>0$ case]{\includegraphics[width=3.0in]{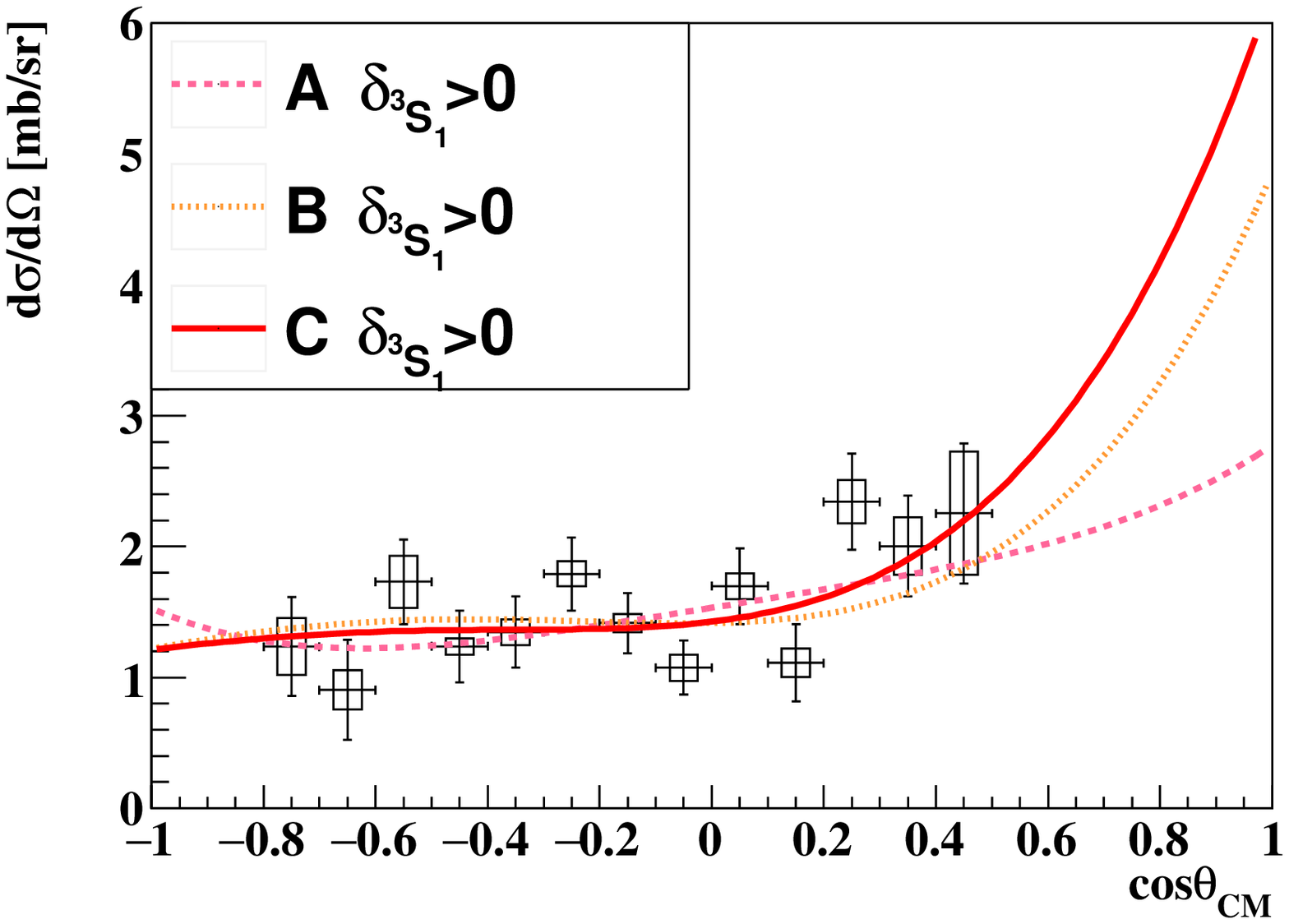}
\label{PAmidp}}}
\caption{\red{Differential cross section as a function of the scattering angle together with the calculated angular distribution in the phase-shift analysis} for the middle-momentum region ($0.55<p_\Sigma [\text{GeV}/c] <0.65 $). The fitting conditions are the same as in \Fig{PAlow}. The typical $\chi^2/$ndf is $14.0/11$.}
\label{PAmid}      
\end{figure}

\begin{figure}[!h]
\centerline{\subfloat[$\delta_{^3S_1}<0$ case]{\includegraphics[width=3.0in]{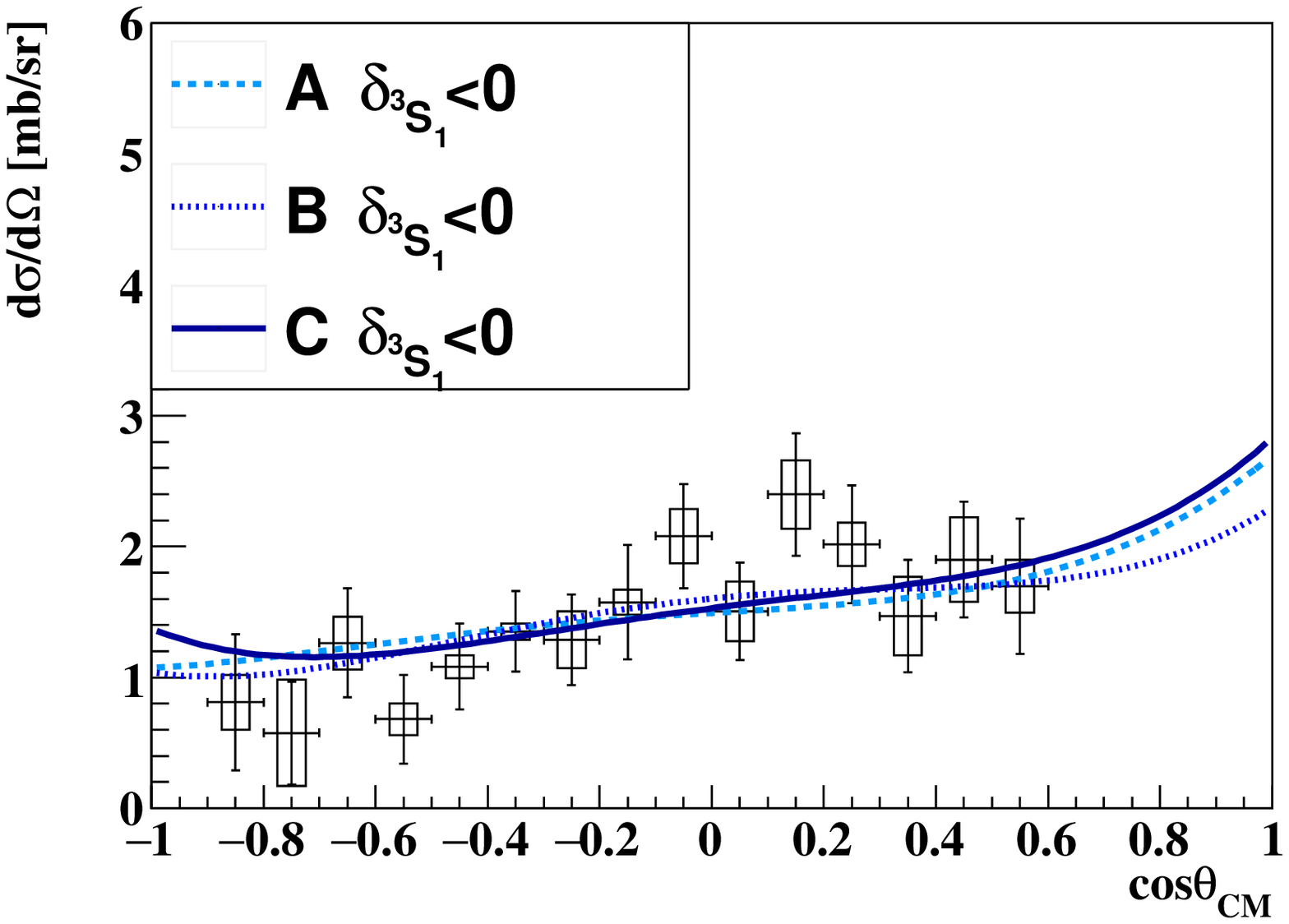}
\label{PAhighm}}
\hfil
\subfloat[$\delta_{^3S_1}>0$ case]{\includegraphics[width=3.0in]{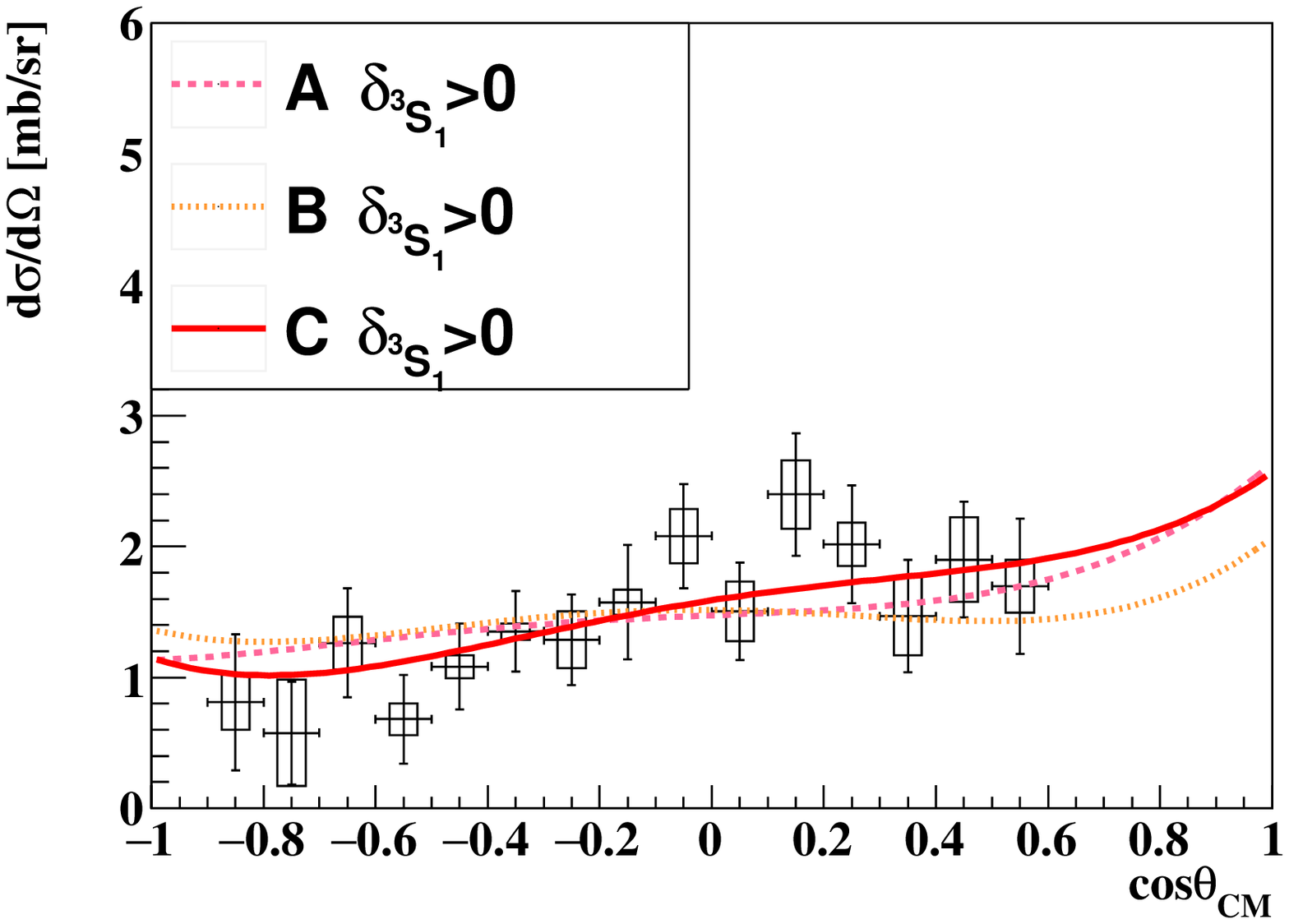}
\label{PAhighp}}}
\caption{\red{Differential cross section as a function of the scattering angle together with the calculated angular distribution in the phase-shift analysis} for the high-momentum region ($0.65<p_\Sigma [\text{GeV}/c]<0.80 $). The fitting conditions are the same as in \Fig{PAlow}. The typical $\chi^2/$ndf is $11.0/13$.}
\label{PAhigh}      
\end{figure}

\begin{figure}[!h]
\begin{center}
\centerline{\subfloat[$\delta_{^3S_1}$]{
\includegraphics[width=3in]{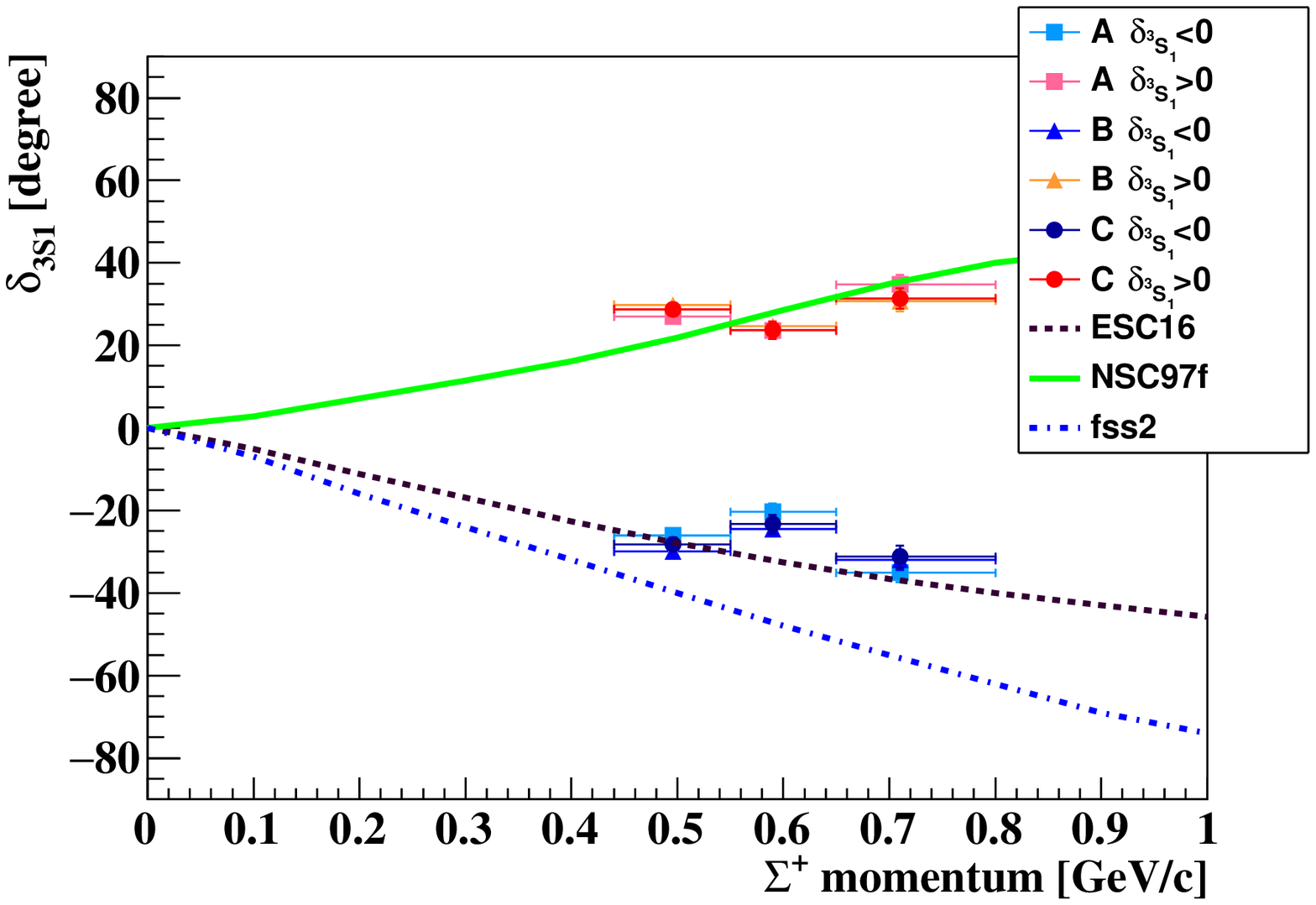}
\label{delta3S1}
}
\subfloat[$\delta_{^1P_1}$]{
\includegraphics[width=3in]{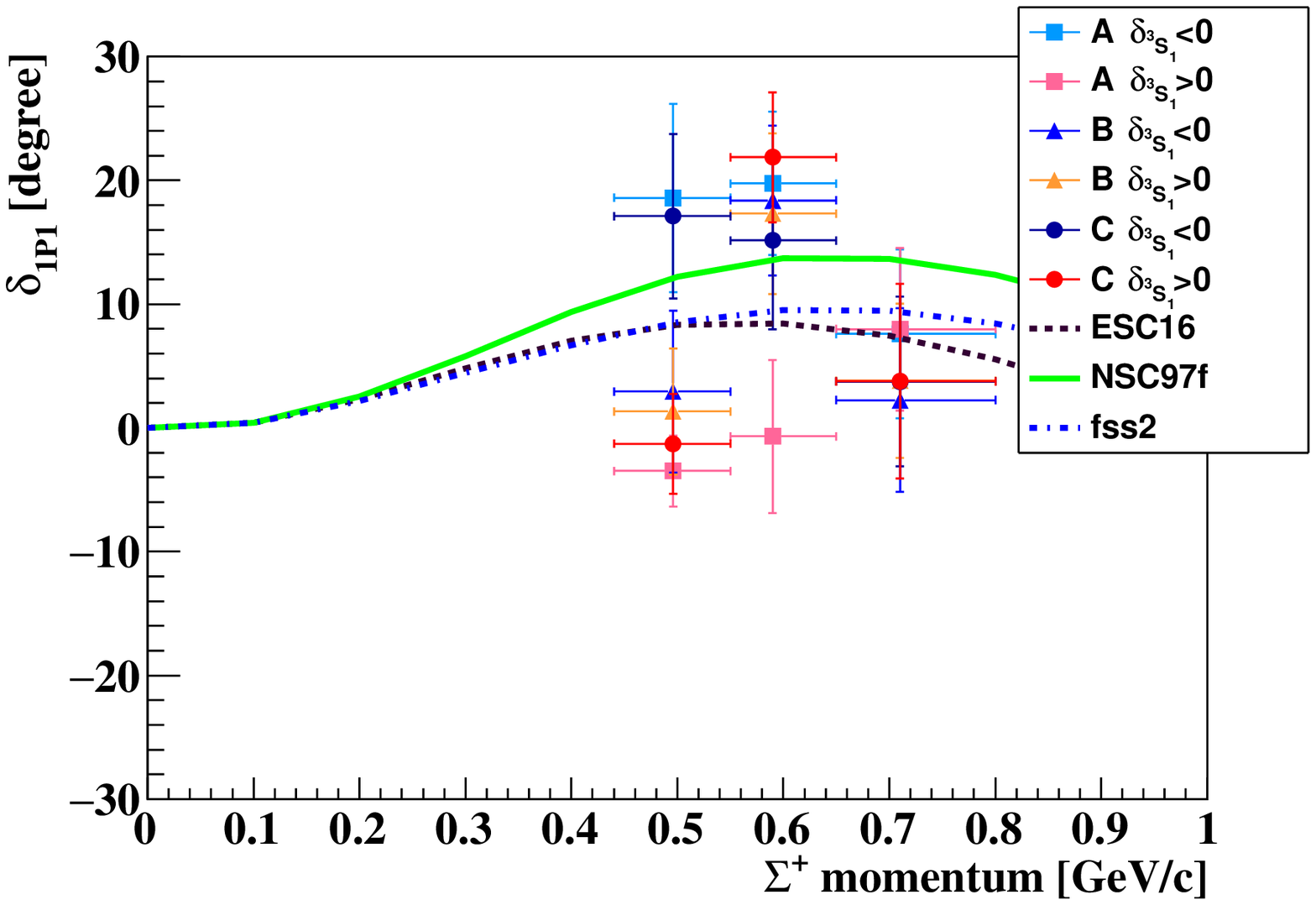}
\label{delta1P1}
}}
\end{center}
\caption{\red{Obtained phase shifts $\delta_{^3S_1}$ and $\delta_{^1P_1}$ as a function of the incident momentum.} The black-dashed, green-solid, and blue-dotted lines represent the calculated phase shifts of ESC16 \cite{Nagels:2019}, NSC97f  \cite{Rijken:1999}, and fss2 \cite{Fujiwara:2007}, respectively.}
\label{PAsummary}       
\end{figure}

The fitting results \red{in} the three momentum regions are shown in \Fig{PAlow}, \ref{PAmid}, and \ref{PAhigh}. In all momentum regions, reasonably reduced $\chi^2$ values of approximately one were obtained. The momentum dependencies of the obtained $\delta_{^3S_1}$ and $\delta_{^1P_1}$ values are plotted in \Fig{PAsummary}. The absolute values of $\delta_{^3S_1}$ for the low-, middle-, and high-momentum regions were $(28.3\pm1.5\pm 2.1)^\circ, (23.4\pm 2.0\pm 3.0 )^\circ, $ and $(32.5\pm 2.5 \pm 2.5) ^\circ$, respectively. The former error comes from the fitting error and the latter shows the effect of the different sets of the fixed parameters. If the sign is assumed to be negative, the momentum dependence of $\delta_{^3S_1}$ is consistent with the ESC models, suggesting that the repulsive force is moderate as in the ESC models, as discussed in subsection \ref{comparetext}. In contrast, the obtained $\delta_{^1P_1}$ values deviate considerably in the range of $-5^\circ<\delta_{^1P_1}<25^\circ$ depending on the conditions. Although the results of $\delta_{^1P_1}$ are ambiguous, they may support the predictions of fss2, ESC, and NSC97f, in which the interaction of the $^1P_1$ state in the $\Sigma^+ p$ system is weakly attractive. 

\section{Summary}
Revealing the nature of flavor SU(3) multiplets is important for a systematic understanding of $BB$ interactions. Among them, 10-plet is predicted to be considerably repulsive, owing to the Pauli effect at the quark level, which is closely related to the origin of the repulsive core in the nuclear force. The $\Sigma^+ p$ channel is one of the best channels for studying the repulsive nature of the 10-plet. 
With this motivation, we performed a novel high-statistics $\Sigma^+ p$ scattering experiment at J-PARC (J-PARC E40).

The experiment was performed at the K1.8 beam line in the J-PARC Hadron Experimental Facility. 
$\Sigma^+$ particles were produced via the $\pi^+ p \to K^+ \Sigma^+$ reaction in the $\text{LH}_2$ target. The $\Sigma^+ p$ scattering events caused by running $\Sigma^+$ in the $\text{LH}_2$ target were identified by a kinematical consistency check for the recoil proton detected with the CATCH detector system. Approximately 2400 $\Sigma^+ p$ elastic scattering events were identified from $4.9 \times 10^7$ tagged $\Sigma^+$ particles in the momentum range 0.44 -- 0.80~GeV/$c$.

The differential cross sections of $\Sigma^+ p$ scattering were derived for the three separate momentum regions. Their uncertainties were typically less than 20\% with an angular step of $\Delta\cos\theta_{\text{CM}}=0.1$. The data quality was significantly improved as compared to previous experiments.
The angular dependencies of the obtained differential cross sections are relatively isotropic for the present angular regions of $-0.8<\cos\theta_{\text{CM}}<0.6$, particularly \red{in the low-momentum region.} The obtained values of the differential cross sections are approximately 2~mb/sr or less, which are not as large as those predicted by the fss2 and FSS models based on the QCM in the short-range region \cite{Fujiwara:2007}. Predictions from the Nijmegen ESC models \cite{Rijken:2010}\cite{Nagels:2019}, which include the moderate repulsive force according to the Pomeron effect, are in close proximity to the data, although sizable discrepancies still exist between the data and ESC08. The $\chi$EFT model predicts much larger cross sections, particularly in the higher-momentum region. We expect that our data will be used to specify the LECs for $P$ waves in $\chi$EFT models \cite{Haidenbauer:2013}\cite{Haidenbauer:2020}.

Owing to the precise data points and simple representation of the $\Sigma^+ p$ system, with respect to the multiplets of the $BB$ interaction, we derived the phase shifts of the $^3S_1$ and $^1 P_1$ channels for the first time, by performing a phase-shift analysis for the obtained differential cross sections.
The absolute values of $\delta_{^3S_1}$ range from $20^\circ$ to $35^\circ$ in the present momentum range. If the sign is assumed to be negative, the momentum dependence of $\delta_{^3S_1}$ is consistent with the ESC models, which predict a relatively moderate repulsive force. Because the $^3 S_1$ channel is expected to be related to the quark Pauli effect, the obtained $\delta_{^3S_1}$ will impose a strong constraint on the size of the repulsive force.

\section*{Acknowledgments}
We would like to thank the staff of the J-PARC accelerator and the Hadron Experimental Facility for their support in providing the beam during beam time. Detector tests at CYRIC and ELPH at Tohoku University were also important during the preparation period of the experiment. We additionally thank the staff at CYRIC and ELPH for their support during the experiments.
We would like to express our gratitude to Y. Fujiwara for theoretical support from the initial stages of the experimental design, who suggested extracting the phase shift of the $^3S_1$ state from the differential cross sections. Additionally, we thank T. A. Rijken and J. Haidenbauer for their theoretical calculations.
We appreciate the computational and network resources provided by KEKCC and SINET.
This work was supported by JSPS KAKENHI Grant Number 23684011, 15H00838, 15H05442, 15H02079 and 18H03693. 
This work was also supported by Grants-in-Aid Number 24105003 and 18H05403 for Scientific Research from the Ministry of Education, Culture, Science and Technology (MEXT), Japan.


%



\let\doi\relax

\bibliography{bibliography}
\bibliographystyle{ptephy_wdoi} 

\appendix
\section{Appendix A: Specific expressions of differential cross section as a function of phase shifts}
\label{PAformula}
\red{Similarly to} the $NN$ scattering case \cite{NNformula}, a wave function for the $\Sigma^+ p$ scattering can asymptotically be written as follows:
\begin{equation}
\psi_{m'}^{s'}(\bm{r}) \sim e^{ikz}\xi^{s'}_{m'} + \frac{e^{ikr}}{r} \sum_{s,m} \xi^{s}_{m} M^{s, s'}_{m,m'}(\theta,\phi)
\end{equation}
where $k$ represents the wavenumber of relative motion in the CM system defined as $k=p_{\text{CM}}/\hbar$, $\xi^s_m$ denotes the spin state with spin quantum number $s$ and the projection on the quantization axis $m$. $M^{s, s'}_{m,m'}$ are the matrix elements of the spin-$1/2$ spin-$1/2$ scattering amplitude with a polar angle $\theta$ and azimuthal angle $\phi$. By the partial-wave decomposition, the matrix element becomes
\begin{align}
 M^{s, s'}_{m,m'}(\theta,\phi)&=\sum_L \sum_{J=|L-s|}^{L+s} \sum_{L'=|J-s'|}^{J+s'} \sqrt{4\pi(2L'+1)}Y_L^{m'-m}(\theta,\phi) \notag \\
&\times  C_{L\times s}(J,m',m'-m,m) C_{L'\times s'} (J,m',0,m') i^{L'-L} \frac{\bra{L,s}S^{\text{mat}}-1 \ket{L',s'}}{2ik}
\end{align}
where the $C_{L\times s}$ are Clebsch-Gordan coefficients defined as $C_{L \times s}(J,m_J,m_L,m_s)=\braket{Lm_Lsm_s|LsJm_J}$, $Y_L^m$ is a spherical harmonic, and $S^\text{mat}$ is the scattering matrix.
The differential cross section for scattering of an unpolarized incident particle on an unpolarized target $I_0$ is expressed by the matrix elements as follows:
\begin{equation}
I_0=\frac{1}{4}|M^{0,0}_{0,0}|^2+\frac{1}{2}|M^{1,1}_{1,1}|^2+\frac{1}{4}|M^{1,1}_{0,0}|^2+\frac{1}{2}|M^{1,1}_{0,1}|^2+\frac{1}{2}|M^{1,1}_{1,0}|^2+\frac{1}{2}|M^{1,1}_{1,-1}|^2 .
\end{equation}
The explicit formulae of the matrix elements as a function of the partial wave amplitudes $h$'s for a general angular momentum $L$ can be found in \cite{Hoshizaki}. The specific expressions up to D wave ($L\le 2$) are described as follows:
\begin{align}
M^{0,0}_{0,0}&=h_{^1S_0}+3h_{^1P_1}\cos\theta+5h_{^1D_2}\times \Bigl{(}\frac{3\cos^2\theta-1}{2}\Bigr{)}, \\
M^{1,1}_{1,1}&=\Bigl{(}h_{^3S_1}-\frac{\sqrt{2}}{2}h^{^3S_1-^3D_1}\Bigr{)}+\Bigl{(}\frac{3}{2}h_{^3P_2}+\frac{3}{2}h_{^3P_1}\Bigr{)}\cos\theta \notag \\
&\quad+\Bigl{(}2h_{^3D_3}+\frac{5}{2}h_{^3D_2}+\frac{1}{2}h_{^3D_1}-\frac{\sqrt{2}}{2}h^{^3S_1-^3D_1} \Bigr{)}\times\frac{3\cos^2\theta-1}{2}, \\
M^{1,1}_{0,0}&=(h_{^3S_1}+\sqrt{2}h^{^3S_1-^3D_1}) +(2h_{^3P_2}+h_{^3P_0})\cos\theta \notag \\
&\quad+(3h_{^3D_3}+2h_{^3D_1}+\sqrt{2}h^{^3S_1-^3D_1})\times \frac{3\cos^2\theta-1}{2} ,\\
M^{1,1}_{0,1}&=\Bigl{(}-\frac{3}{2\sqrt{2}}h_{^3P_2}+\frac{3}{2\sqrt{2}}h_{^3P_1} \Bigr{)}\times(-\sin\theta)\notag \\
&\quad+\Bigl{(}-\frac{4}{3\sqrt{2}}h_{^3D_3}+\frac{5}{6\sqrt{2}}h_{^3D_2}+\frac{1}{2\sqrt{2}}h_{^3D_1}-\frac{1}{\sqrt{2}}h^{^3S_1-^3D_1}\Bigr{)}\times(-3\cos\theta\sin\theta), \\
M^{1,1}_{1,0}&=\Bigl{(}\frac{1}{\sqrt{2}}h_{^3P_2}-\frac{1}{\sqrt{2}}h_{^3P_0} \Bigr{)}\times(-\sin\theta) \notag \\
&\quad+\Bigl{(}\frac{1}{\sqrt{2}}h_{^3D_3}-\frac{1}{\sqrt{2}}h_{^3D_1} -\frac{1}{\sqrt{2}}h^{^3S_1-^3D_1}\Bigr{)}\times(-3\cos\theta\sin\theta), \\
M^{1,1}_{1,-1}&=\Bigl{(}\frac{1}{6}h_{^3D_3}-\frac{5}{12}h_{^3D_2}+\frac{1}{4}h_{^3D_1}-\frac{1}{2\sqrt{2}}h^{^3S_1-^3D_1}\Bigr{)}\times(3\sin^2\theta),
\end{align}
where partial wave amplitudes $h$'s were defined as
\begin{align}
h_{^{2s+1}L_J}&=\begin{cases}
\frac{1}{2ik}(\cos(2\bar{\epsilon}_1)\exp(2i\bar{\delta}_{^{2s+1}L_J})-1) & (^3S_1 \text{ and } ^3D_1 \text{ cases})\\
\frac{1}{2ik}(\exp(2i \bar{\delta}_{^{2s+1}L_J})-1) & (\text{otherwise})
\end{cases}\\
h^{^3S_1-^3D_1}&=\frac{1}{2k}\sin(2\bar{\epsilon}_1)\exp(i\bar{\delta}_{^3S_1}+i\bar{\delta}_{^3D_1}).
\end{align} 
$\bar{\delta}_{^{2s+1}L_J}$ and $\bar{\epsilon}_1$ are the bar-phase shifts and mixing parameter for the $^3S_1-^3D_1$ mixing, which are different from the commonly used nuclear bar-phase shifts separated from Coulomb effects in a precise sense. Because the energies of the $\Sigma^+ p$ scattering are sufficiently high and the data for very-forward angle is absent, the Coulomb effects might be negligible. Therefore, the bar-phase shifts were equated with the nuclear bar-phase shifts and called merely ``phase shifts $\delta$'' in this study.

\section{Appendix B: Angular dependence of the $\Delta E(\Sigma^+ p)$ distribution}
\label{Fittext}
\begin{figure}[!h]
\begin{center}
\includegraphics[width=4.5in]{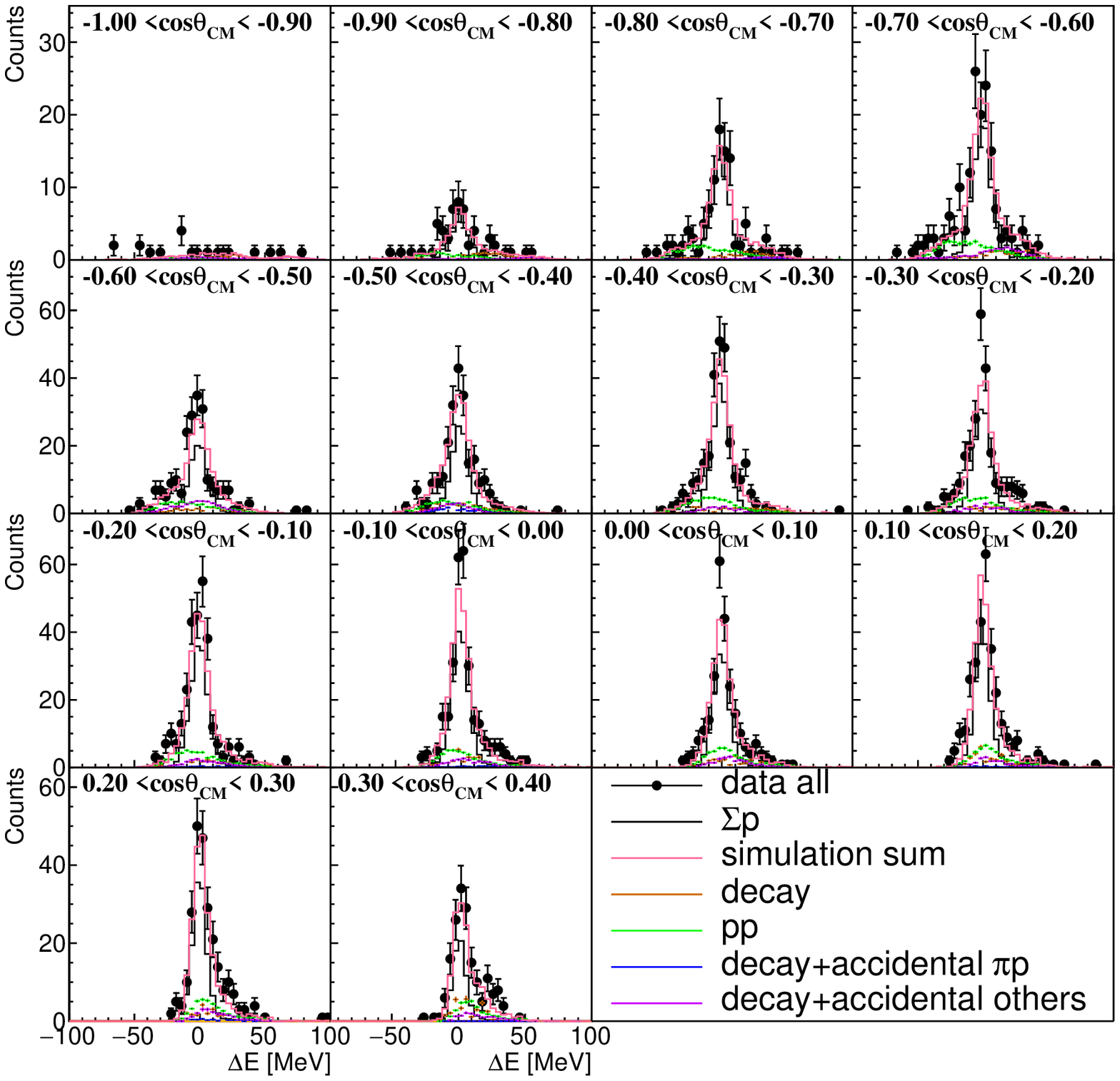}
\end{center}
\caption{$\Delta E$ spectra at each scattering angle of $\Sigma^+$ in the low-momentum region ($0.44<p_\Sigma [\text{GeV}/c]<0.55$). \red{The data points} with error bars show the experimental data. Simulated spectra for the assumed reactions are also shown and \red{the red histogram} showing the sum of these spectra.}
\label{fitLow}       
\end{figure}

\begin{figure}[!h]
\begin{center}
\includegraphics[width=4.5in]{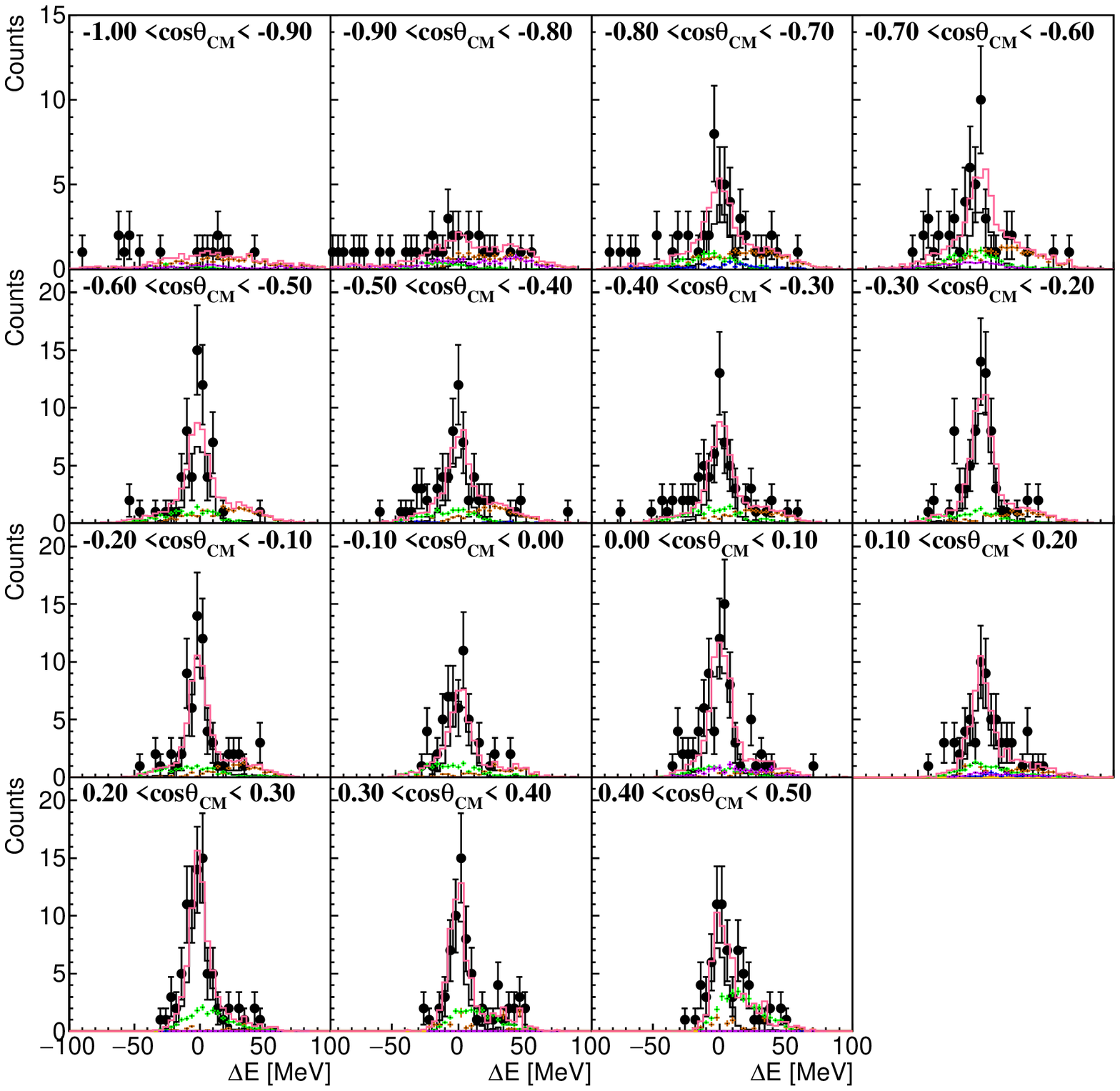}
\end{center}
\caption{$\Delta E$ spectra at each scattering angle of $\Sigma^+$ in the middle-momentum region ($0.55<p_\Sigma [\text{GeV}/c]<0.65$). \red{The legends are the same as those in Fig. \ref{fitLow}}.}
\label{fitMid}       
\end{figure}

\begin{figure}[!h]
\begin{center}
\includegraphics[width=4.5in]{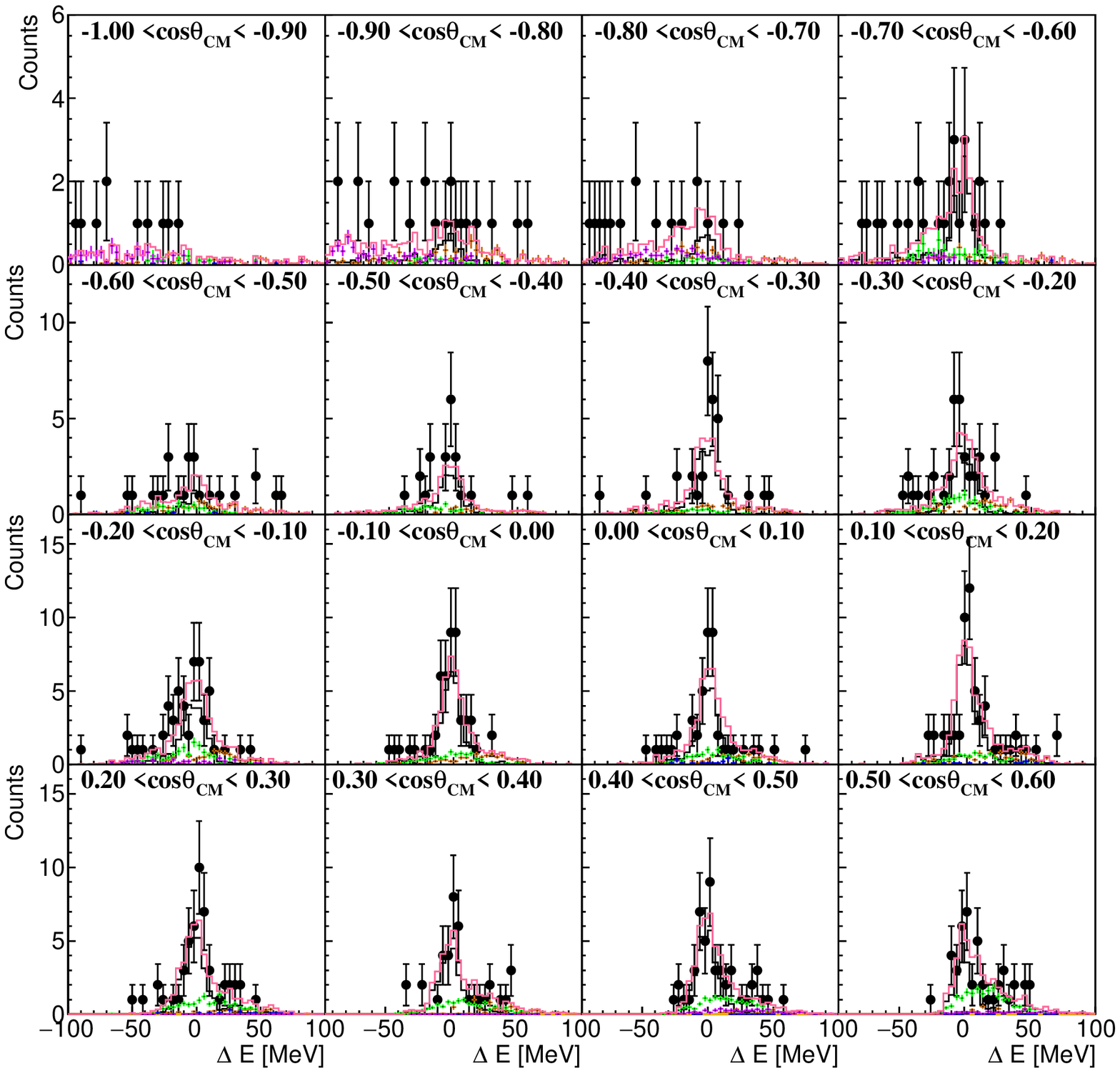}
\end{center}
\caption{$\Delta E$ spectra at each scattering angle of $\Sigma^+$ in the high-momentum region ($0.65<p_\Sigma [\text{GeV}/c]<0.80$). \red{The legends are the same as those in Fig. \ref{fitLow}}.}
\label{fitHigh}       
\end{figure}

In order to show the statistical significance of the $\Sigma^+ p$ scattering events, the fitting results of the $\Delta E$ spectra for each scattering angle and momentum region of $\Sigma^+$ are shown in \Fig{fitLow}, \ref{fitMid}, and \ref{fitHigh}. The $\Delta E (\Sigma^+p)$ spectrum can be reproduced by the sum of the simulated spectra.


\section{Appendix C: Table of the differential cross sections}
The values of the differential cross sections and their \red{uncertainties} are summarized in Tables \ref{tab:DCSlow}, \ref{tab:DCSmid}, and \ref{tab:DCShigh}.
\label{Tablesdiff}
\begin{table}[!h]
\begin{center}
\caption[Data table of differential cross sections for  $\Sigma^+ p$ scattering in the low momentum region ( $0.44<p_\Sigma<0.55$ GeV/c).]{Data table of \red{differential cross sections for $\Sigma^+ p$ scattering} in the low-momentum region ($0.44<p_\Sigma [\text{GeV}/c]<0.55$). The systematic error was estimated as a quadratic sum of the error from \red{the background estimation (BG), averaged efficiency (eff), and $\Sigma^+$ total flight length (L)}.}
\begin{tabular}{cccc|ccc}\hline
\label{tab:DCSlow} 
$\cos \theta_{\text{CM}}$ & $\red{d\sigma/d\Omega}$ & stat. & syst. (Total) & syst. (BG) & syst. (eff) & syst. (L) \\
 & [mb/sr] & [mb/sr] & [mb/sr] & [mb/sr] & [mb/sr] & [mb/sr]\\ \hline
$-0.85 \pm$ 0.05 &2.22	&$\pm0.46$	&$\pm0.40$	&$\pm0.23$	&$\pm0.32$	&$\pm0.04$ \\
$-0.75 \pm$ 0.05 &2.31	&$\pm0.32$	&$\pm0.33$	&$\pm0.17$	&$\pm0.28$	&$\pm0.05$ \\
$-0.65 \pm$ 0.05 &2.12	&$\pm0.26$	&$\pm0.29$	&$\pm0.18$	&$\pm0.22$	&$\pm0.04$ \\
$-0.55 \pm$ 0.05 &2.00	&$\pm0.29$	&$\pm0.38$	&$\pm0.34$	&$\pm0.17$	&$\pm0.04$ \\
$-0.45 \pm$ 0.05 &1.93	&$\pm0.27$	&$\pm0.43$	&$\pm0.41$	&$\pm0.13$	&$\pm0.04$ \\
$-0.35 \pm$ 0.05 &2.40	&$\pm0.22$	&$\pm0.22$	&$\pm0.15$	&$\pm0.15$	&$\pm0.05$ \\
$-0.25 \pm$ 0.05 &1.92	&$\pm0.19$	&$\pm0.23$	&$\pm0.18$	&$\pm0.15$	&$\pm0.04$ \\
$-0.15 \pm$ 0.05 &2.22	&$\pm0.19$	&$\pm0.31$	&$\pm0.25$	&$\pm0.17$	&$\pm0.04$ \\
$-0.05 \pm$ 0.05 &2.22	&$\pm0.19$	&$\pm0.26$	&$\pm0.16$	&$\pm0.21$	&$\pm0.04$ \\
$  0.05 \pm$ 0.05 &1.86	&$\pm0.21$	&$\pm0.25$	&$\pm0.15$	&$\pm0.20$	&$\pm0.04$ \\
$  0.15 \pm$ 0.05 &2.54	&$\pm0.23$	&$\pm0.37$	&$\pm0.12$	&$\pm0.35$	&$\pm0.05$ \\
$  0.25 \pm$ 0.05 &2.84	&$\pm0.29$	&$\pm0.58$	&$\pm0.17$	&$\pm0.55$	&$\pm0.06$ \\
$  0.35 \pm$ 0.05 &3.02	&$\pm0.51$	&$\pm0.90$	&$\pm0.24$	&$\pm0.86$	&$\pm0.06$ \\ \hline
\end{tabular}
\end{center}
\end{table}
\begin{table}[!h]
\begin{center}
\caption[Data table of the differential cross sections of $\Sigma^+ p$ elastic scattering for the middle momentum region ( $0.55<p_\Sigma<0.65$ GeV/c).]{Data table of \red{differential cross sections for $\Sigma^+ p$ scattering} in the middle-momentum region ($0.55<p_\Sigma [\text{GeV}/c]<0.65$). }
\begin{tabular}{cccc|ccc}\hline
\label{tab:DCSmid} 
$\cos \theta_{\text{CM}}$ & $\red{d\sigma/d\Omega}$ & stat. & syst. (Total) & syst. (BG) & syst. (eff) & syst. (L) \\
 & [mb/sr] & [mb/sr] & [mb/sr] & [mb/sr] & [mb/sr] & [mb/sr]\\ \hline
$-0.75 \pm$ 0.05 &1.24	&$\pm0.38$	&$\pm0.22$	&$\pm0.17$	&$\pm0.13$	&$\pm0.02$ \\
$-0.65 \pm$ 0.05 &0.90	&$\pm0.38$	&$\pm0.15$	&$\pm0.13$	&$\pm0.07$	&$\pm0.02$ \\
$-0.55 \pm$ 0.05 &1.73	&$\pm0.32$	&$\pm0.20$	&$\pm0.18$	&$\pm0.08$	&$\pm0.03$ \\
$-0.45 \pm$ 0.05 &1.24	&$\pm0.28$	&$\pm0.06$	&$\pm0.04$	&$\pm0.04$	&$\pm0.02$ \\
$-0.35 \pm$ 0.05 &1.35	&$\pm0.27$	&$\pm0.10$	&$\pm0.08$	&$\pm0.05$	&$\pm0.03$ \\
$-0.25 \pm$ 0.05 &1.79	&$\pm0.28$	&$\pm0.09$	&$\pm0.07$	&$\pm0.05$	&$\pm0.04$ \\
$-0.15 \pm$ 0.05 &1.42	&$\pm0.23$	&$\pm0.07$	&$\pm0.06$	&$\pm0.02$	&$\pm0.03$ \\
$-0.05 \pm$ 0.05 &1.07	&$\pm0.21$	&$\pm0.10$	&$\pm0.10$	&$\pm0.02$	&$\pm0.02$ \\
$  0.05 \pm$ 0.05 &1.70	&$\pm0.29$	&$\pm0.10$	&$\pm0.08$	&$\pm0.05$	&$\pm0.03$ \\
$  0.15 \pm$ 0.05 &1.11	&$\pm0.29$	&$\pm0.11$	&$\pm0.09$	&$\pm0.05$	&$\pm0.02$ \\
$  0.25 \pm$ 0.05 &2.34	&$\pm0.37$	&$\pm0.16$	&$\pm0.05$	&$\pm0.15$	&$\pm0.05$ \\
$  0.35 \pm$ 0.05 &2.00	&$\pm0.39$	&$\pm0.22$	&$\pm0.11$	&$\pm0.19$	&$\pm0.04$ \\
$  0.45 \pm$ 0.05 &2.25	&$\pm0.53$	&$\pm0.47$	&$\pm0.23$	&$\pm0.40$	&$\pm0.05$ \\ \hline
\end{tabular}
\end{center}
\end{table}
\begin{table}[!h]
\begin{center}
\caption[Data table of the differential cross section of the $\Sigma^+ p$ elastic scattering for the high momentum region ( $0.65<p_\Sigma<0.80$ GeV/c).]{Data table of \red{differential cross sections for $\Sigma^+ p$ scattering} in the high-momentum region ($0.65<p_\Sigma [\text{GeV}/c]<0.80$). }
\begin{tabular}{cccc|ccc}\hline
\label{tab:DCShigh} 
$\cos \theta_{\text{CM}}$ & $\red{d\sigma/d\Omega}$ & stat. & syst. (Total) & syst. (BG) & syst. (eff) & syst. (L) \\
 & [mb/sr] & [mb/sr] & [mb/sr] & [mb/sr] & [mb/sr] & [mb/sr]\\ \hline
$-0.85 \pm$ 0.05 &0.81	&$\pm0.52$	&$\pm0.21$	&$\pm0.19$	&$\pm0.09$	&$\pm0.01$ \\
$-0.75 \pm$ 0.05 &0.58	&$\pm0.39$	&$\pm0.41$	&$\pm0.40$	&$\pm0.04$	&$\pm0.01$ \\
$-0.65 \pm$ 0.05 &1.26	&$\pm0.42$	&$\pm0.20$	&$\pm0.19$	&$\pm0.06$	&$\pm0.02$ \\
$-0.55 \pm$ 0.05 &0.68	&$\pm0.34$	&$\pm0.12$	&$\pm0.12$	&$\pm0.02$	&$\pm0.01$ \\
$-0.45 \pm$ 0.05 &1.08	&$\pm0.33$	&$\pm0.09$	&$\pm0.08$	&$\pm0.02$	&$\pm0.02$ \\
$-0.35 \pm$ 0.05 &1.35	&$\pm0.31$	&$\pm0.06$	&$\pm0.05$	&$\pm0.02$	&$\pm0.02$ \\
$-0.25 \pm$ 0.05 &1.29	&$\pm0.35$	&$\pm0.22$	&$\pm0.21$	&$\pm0.02$	&$\pm0.02$ \\
$-0.15 \pm$ 0.05 &1.58	&$\pm0.44$	&$\pm0.10$	&$\pm0.09$	&$\pm0.01$	&$\pm0.02$ \\
$-0.05 \pm$ 0.05 &2.08	&$\pm0.40$	&$\pm0.21$	&$\pm0.20$	&$\pm0.06$	&$\pm0.03$ \\
$  0.05 \pm$ 0.05 &1.50	&$\pm0.37$	&$\pm0.23$	&$\pm0.22$	&$\pm0.06$	&$\pm0.02$ \\
$  0.15 \pm$ 0.05 &2.40	&$\pm0.47$	&$\pm0.26$	&$\pm0.23$	&$\pm0.10$	&$\pm0.04$ \\
$  0.25 \pm$ 0.05 &2.02	&$\pm0.45$	&$\pm0.17$	&$\pm0.15$	&$\pm0.06$	&$\pm0.03$ \\
$  0.35 \pm$ 0.05 &1.47	&$\pm0.43$	&$\pm0.30$	&$\pm0.29$	&$\pm0.06$	&$\pm0.02$ \\
$  0.45 \pm$ 0.05 &1.90	&$\pm0.44$	&$\pm0.32$	&$\pm0.30$	&$\pm0.11$	&$\pm0.03$ \\
$  0.55 \pm$ 0.05 &1.70	&$\pm0.52$	&$\pm0.20$	&$\pm0.09$	&$\pm0.18$	&$\pm0.03$ \\ \hline
\end{tabular}
\end{center}
\end{table}

\end{document}